\documentclass[12pt,a4paper]{article}
\usepackage{preamble}

\newcommand{\RMSD}{\text{RMSD}_{\text{fit}}}

\begin{document}

%%%%%%%%%%%%%%%%%%%%%%%%%%%%%%%%%%%%%%%%%%%%%%%%%%%%%%%%%%%%%%%%%%%%%%
%%%%%%%%%%%%%%%%%%%%%%%%%%%%%%%%%%%%%%%%%%%%%%%%%%%%%%%%%%%%%%%%%%%%%%
% Titlepage
\begin{titlepage}

\numberwithin{equation}{section}
\begin{flushright}
\small
% Saclay preprint - not needed anymore before arXiv, done afterwards 

\normalsize
\end{flushright}
%\vspace{0.4 cm}

\begin{center}

{\LARGE \textbf{Black Hole Photon Rings\\ Beyond General Relativity}}
\\

\medskip

\vspace{1 cm} {\large Seppe Staelens$^{1}$, Daniel R. Mayerson$^1$, Fabio Bacchini$^{2,3}$, Bart Ripperda$^{4,5,6}$, Lorenzo K\"uchler$^{1,7}$}\\

\vspace{1cm}
 
 {$^1$ Institute for Theoretical Physics, KU Leuven, Celestijnenlaan 200D, B-3001 Leuven, Belgium}
 
 {$^2$ Centre for mathematical Plasma Astrophysics, KU Leuven, Celestijnenlaan 200B, B-3001 Leuven, Belgium}
 
 {$^3$ Royal Belgian Institute for Space Aeronomy, Solar-Terrestrial Centre of Excellence, Ringlaan 3, 1180 Uccle, Belgium}
 
 {$^4$ School of Natural Sciences, Institute for Advanced Study, 1 Einstein Drive, Princeton, NJ 08540, USA}

 {$^5$ NASA Hubble Fellowship Program Einstein Fellow}
 
 {$^6$ Center for Computational Astrophysics, Flatiron Institute, 162 Fifth Avenue, New York, NY 10010, USA}

 {$^7$ Universit\'e Libre de Bruxelles and International Solvay Institutes, C.P. 231, B-1050 Bruxelles, Belgium}

\vspace{1cm}

seppe.staelens@student.kuleuven.be, daniel.mayerson@kuleuven.be, fabio.bacchini@kuleuven.be, bartripperda@ias.edu, lorenzo.kuchler@ulb.be

\vspace{1cm}

\textbf{Abstract}
\end{center}
\noindent 
We investigate whether photon ring observations in black hole imaging are able to distinguish between the Kerr black hole in general relativity and alternative black holes that deviate from Kerr.
Certain aspects of photon rings have been argued to be robust observables in Very-Long-Baseline Interferometry (VLBI) black hole observations which carry imprints of the underlying spacetime. The photon ring shape, as well as its Lyapunov exponent (which encodes the narrowing of successive photon subrings), are detailed probes of the underlying geometry; measurements thereof have been argued to provide a strong null test of general relativity and the Kerr metric.  However, a more complicated question is whether such observations of the photon ring properties can distinguish between Kerr and alternative black holes.
We provide a first answer to this question by calculating photon rings of the Johannsen, Rasheed-Larsen, and Manko-Novikov black holes. We find that large deviations from Kerr and large observer inclinations are needed to obtain measurable differences in the photon ring shape. In other words, the Kerr photon ring shape appears to be the universal shape even for deviating black holes at low inclinations. On the other hand, the Lyapunov exponent shows more marked variations for deviations from the Kerr metric. Our analysis lays out the groundwork to determine deviations from the Kerr spacetime in photon rings that are potentially detectable by future observing missions.

\vspace{2em}

%\newpage

%%%%%%%%%%%%%%%%%%%%%%%%%%%%%%%%%%%%%%%%%%%%%%%%%%%%%%%%%%%%%%%%%%%%%%
%%%%%%%%%%%%%%%%%%%%%%%%%%%%%%%%%%%%%%%%%%%%%%%%%%%%%%%%%%%%%%%%%%%%%%
% Table of contents
\setcounter{tocdepth}{2}
\tableofcontents

\end{titlepage}
\newpage
%%%%%%%%%%%%%%%%%%%%%%
%%%%%%%%%%%%%%%%%%%%%%
%%%%%%%%%%%%%%%%%%%%%%
\section{Introduction and Summary}

The Event Horizon Telescope (EHT) has recently given us a first glimpse at the black holes in the centers of the M87 galaxy \cite{EHT2019a} and our Milky Way, Sgr A* \cite{EventHorizonTelescope:2022wkp}, and how they influence the emission from the surrounding plasma with their strong gravity. 

The EHT images show a bright ring of light surrounding a darker depression. This depression is sometimes called the black hole ``shadow'': the absence of light in the image due to the black hole absorbing any light that travels too close to it. The shadow's size and shape encodes properties of the black hole such as its mass and angular momentum, and so measuring it precisely could in principle give constraints on these black hole properties.
However, the black hole shadow is \emph{not} a direct observable \cite{Gralla:2019xty,Gralla:2020pra}: the dark depression can be somewhat obscured in the actual image due to light coming from matter that orbits and surrounds the black hole (e.g.\ an accretion disc). The black hole shadow is also a highly degenerate object, so its measurement does not uniquely fix a black hole's parameters but only identifies patches of the allowed parameter space. Moreover, the shadow is not even a truly universal quantity, since its size depends on the particular emission that illuminates the black hole \cite{Gralla:2019xty}.\footnote{\label{fn:shadow} The shadow is also a term that is sometimes used rather ambiguously \cite{Gralla:2019xty}. (We thank S. Gralla for pointing this out.) To avoid confusion, we will use the unambiguous term ``critical curve'' in our paper to describe the curve on the viewer screen corresponding to exactly bound photon orbits; see Section \ref{sec: analytic shadow}.}

In general, the physics that produces emission from the region nearby a black hole is extremely complicated. The plasma comprising the black hole's accretion disc and its jet (if present) emits photons that eventually may reach our telescopes. There are many uncertainties and unknowns in the details of the plasma emission that determine many of the features of the final image that is observed.
To extract detailed information about the black hole geometry, the key is to identify observables that are largely independent of the details of the emitting plasma, but nevertheless sensitive to the geometry in which the photons travel.

Aspects of the black hole \emph{photon ring} have been argued to provide such universal observables that are largely independent of emission details but sensitive to geometry \cite{Johnson_2020}. The photon ring is precisely the bright ring of light in black hole images \cite{Johnson_2020,Broderick:2022tfu,Lockhart:2021xel,Lockhart:2022rui}.
In fact, if viewed with high enough resolution and observed wavelength, this photon ring can be seen to consist of an infinite series of \emph{subrings}. The $n$-th subring consists of photons that have travelled $n$ half-orbits around the black hole. Each consecutive subring is brighter, but also exponentially narrower, than the previous one. The $n$-th photon subring for $n$ ``large enough'' should be largely independent of the emitting plasma details \cite{Johnson_2020}, providing an exciting window into the spacetime geometry of the  black hole.

Gralla, Lupsasca, and Marrone (GLM) \cite{Gralla:2020srx} argued that a hypothetical but feasible Earth-orbit satellite Very-Long-Baseline Interferometry (VLBI) observation of M87* could resolve its $n=2$ photon ring. This particular subring is an observational ``sweet spot'': the lower-order ($n<2$) rings still show dependence on the plasma emission details; $n=2$ is thought to be the first subring to be largely independent of the plasma, but still feasible to observe \cite{Johnson_2020, Gralla:2020srx}.
GLM showed that the \emph{shape} of this $n=2$ photon ring is fixed for the Kerr black hole, and could be measured extremely precisely by such an experiment, providing a strong null test of general relativity. 
 
The intensities of successive photon subrings --- or equivalently, their widths --- decrease exponentially according to the photon ring \emph{Lyapunov exponent} $\gamma$, which is intimately tied to the details of the geometry. These successive ring intensities provide a strong and universal signature on long interferometric baselines \cite{Johnson_2020}.
 Measuring successive subring intensities or widths for M87* would then be an intricate probe of the Kerr geometry; however, such detailed ring measurements would require an extremely long baseline which is well beyond the reach even of the feasible Earth-orbit missions proposed by GLM.
 Another approach, taken by the authors of \cite{hadar2021photon}, is to look at autocorrelations in brightness fluctuations of the ring emission. This does not require the ring to be resolved, and these peaks encode the black hole parameters through (among others) the Lyapunov exponent.

 \medskip

The discussion of the $n=2$ photon ring shape, first by GLM \cite{Gralla:2020srx} and later expanded by these authors and others (see e.g.\ \cite{Paugnat:2022qzy,Vincent:2022fwj,Cardenas-Avendano:2022csp}), focuses on the interferometric signal produced by the shape of the $n=2$ photon ring \emph{for Kerr}. Similarly, the Lyapunov exponent has been discussed at length \emph{for Kerr} \cite{Johnson_2020}. These discussions clearly show that measurement of these quantities can provide a precise null test of the Kerr metric and thus of general relativity. However, it remains unclear how much these quantities can be expected to \emph{differ} in corrections or alternatives to Kerr. The fundamental question thus stands: \emph{Can photon ring observables be used to efficiently distinguish between Kerr and deviations from Kerr?} This question has largely been left unanswered, and we aim to provide a first preliminary answer with this paper.

We will consider three different black hole geometries that have a smooth Kerr limit.
This allows us to study the dependence of both the photon ring shape and Lyapunov exponent on the Kerr-deviating parameters of these black holes.
Specifically, we consider the Johannsen \cite{johannsen2013regular}, Rasheed-Larsen \cite{Rasheed:1995zv,larsen2000rotating}, and Manko-Novikov \cite{manko1992novikov} black holes. Each of these has different characteristics and so provides complementary ways of deviating from Kerr. The Johannsen metric is a parametrized extension of Kerr that is constructed to remain stationary, axisymmetric, and integrable (i.e.\ admits a Carter-like constant), but it is not necessarily a solution to any theory with matter \cite{johannsen2013regular}. The Rasheed-Larsen metric is a black hole with electric and magnetic charges which moreover breaks the equatorial symmetry (top-down reflection symmetry) that Kerr enjoys;\footnote{See e.g.\ \cite{Chen:2020aix,Bah:2021jno,Chen:2022lct,Fransen:2022jtw} for earlier discussions on how breaking equatorial symmetry can lead to observable signals in black hole imaging or gravitational waves.} it is a solution to Einstein gravity coupled to a dilaton and gauge field. The Rasheed-Larsen solution also preserves integrability for null geodesics. Finally, the Manko-Novikov solution is a metric which breaks both equatorial symmetry and geodesic integrability; it is a vacuum solution of the Einstein equations which implies part of its horizon is singular in accordance with black hole uniqueness theorems.

All three of the considered black holes can have large spins $J/M^2 \sim 1$, which is important to launch powerful jets \cite{Blandford:1977ds}. In addition, we will present each of these black holes' gravitational multipoles \cite{Thorne:1980ru} --- these are asymptotic, gauge-invariant observables that quantify the (asymptotic) deviation with respect to Kerr.

We use our ray tracer \texttt{FOORT} (Flexible Object-Oriented Ray Tracer) \cite{FOORT} to numerically integrate null geodesics in these black hole spacetimes and extract photon ring data. Synthetic image creation with numerical ray tracing is a standard technique that is used by the EHT collaboration \cite{EHT2019a,EventHorizonTelescope:2022wkp} and others \cite{Gralla:2020srx}, although we do not require realistic photon emission profiles for our purposes.
We will show four-color screen ``images'' of these black holes throughout the paper. For such images, we follow each geodesic backward in time from the camera until it reaches an outer ``celestial sphere'', and assign a color to the geodesic according to the quadrant of the celestial sphere it originated from. To determine which photon subring a geodesic belongs to, we additionally keep track of the number of passes through the equatorial plane that the geodesic makes.

 \medskip

We summarize our main results immediately below in Section \ref{sec:summary}. In Section \ref{sec:BHandshadows}, we briefly review the three black holes we consider --- the Johannsen, Rasheed-Larsen, and Manko-Novikov black holes --- and briefly discuss their critical curves. Section \ref{sec:shape} focuses on the GLM-proposed measurement of the $n=2$ photon ring shape \cite{Gralla:2020srx} of these black holes, and how it can differ from the Kerr shape. Section \ref{sec:lyap} investigates the Lyapunov exponent for these black holes in an analytical way. 

Many details of calculations have been deferred to appendices for readability. Appendix \ref{app: MaNo} contains more information on the Manko-Novikov black hole; Appendix \ref{app: auxiliary calcs} contains detailed calculations of the critical curves and Lyapunov exponents. Appendix \ref{app:FOORT} describes our ray-tracing code and method of identifying the photon rings, while Appendix \ref{app:moreshapemethod} contains additional information about the ray-traced photon rings and their shapes. Finally, Appendix \ref{app: num Lyap} describes our ray-tracing numerical approximations to the Lyapunov exponent.

\subsection{Summary \& conclusions}\label{sec:summary}
We consider the Johannsen, Rasheed-Larsen, and Manko-Novikov black holes (introduced in Section \ref{sec:BHandshadows}). In particular, we have calculated and analyzed the shape of their $n=2$ photon ring and/or critical curve (Section \ref{sec:shape}) and the Lyapunov exponent for the Johannsen and Rasheed-Larsen black holes (Section \ref{sec:lyap}).

\medskip

The shape of the Kerr black hole (for most of its parameter space) is a \emph{circlipse} \cite{Gralla:2020srx,Gralla:2020yvo}. One of our main findings in Section \ref{sec:shape} is that this shape is a rather robust feature of \emph{all} black holes we consider, as long as the inclination $\theta_0$ --- i.e.\ the angle between the observer and the black hole spin axis --- remains small. Roughly, the smaller the deviation from Kerr, the closer to edge-on (i.e.\ when the black hole spin axis is perpendicular to the line of sight) the viewer needs to be in order for deviations to the circlipse shape function to be noticeable. At around $\theta_0\approx 17^\circ$ (which is the derived viewing angle for M87* from observations of the large-scale jet \cite{CraigWalker:2018vam}), the deviation from Kerr necessary for a deviation from the circlipse shape to be discernible is rather large. For example, for the Johannsen metric, at $\theta_0\approx 17^\circ$, we require the dimensionless deviation parameter $\alpha_{22}\gtrsim 2.5$ for the photon ring shape to deviate noticeably from the circlipse (see Fig.~\ref{fig:Johshapefits}); this is already a rather large deviation from Kerr (see e.g Fig.~\ref{fig:Johbasicfig}). In addition, we find that the shape of the Rasheed-Larsen photon ring is always indistinguishable from the Kerr circlipse, at all inclinations.

Our conclusion is that the photon ring shape is only a good distinguishing observable that can separate Kerr from other black hole metrics when the observer inclination is relatively large and/or the deviation of these other metrics from Kerr is large (in which case we could have reasonably expected to see deviations in other observations already).

\medskip

The Lyapunov exponent is certainly a better distinguishing feature that can differentiate other black holes from Kerr, as we show in Section \ref{sec:lyap} (although it is less established how the Lyapunov exponent could be measured in practice). The Lyapunov exponent is in fact an entire function's worth of metric-specific quantities, and we show that this function is significantly influenced by every and any deviation to Kerr.

\medskip

Our analysis is but a first step in understanding what possible beyond-Kerr signatures could appear in the photon rings of black hole imaging with VLBI. There are many more detailed follow-up questions to ask along these lines.
Can we differentiate different black holes (either with photon ring shape or Lyapunov exponent) using Bayesian model selection?
Do the photon rings have measurably different features if the imaged object does not have a horizon, but instead is an ultracompact, horizonless object such as a fuzzball \cite{Bacchini:2021fig,Bah:2021jno,Mayerson:2020tpn}? 
We leave answering these questions for future work.

%We use units in which $G=c=1$.

\section{Black Holes \& Critical Curves}\label{sec:BHandshadows}

We focus on three different stationary and axisymmetric black holes that each have a smooth limit to Kerr. All three of these black holes have different physical properties which give them varying observational features to look for.

The Johannsen metric is a stationary, axisymmetric deviation of Kerr that preserves the integrability of geodesics (by ensuring a Carter-like constant exists); however, this metric is not the solution of any (known) theory of gravity coupled to matter. The Rasheed-Larsen metric can be thought of as a generalization of Kerr with electric and magnetic charges, distinct from Kerr-Newman by the presence of a non-trivial scalar in addition to the gauge field; furthermore, null geodesics remain integrable in this metric. Finally, the Manko-Novikov metric is constructed as a vacuum solution to general relativity by adding multipole ``bumps'' to Kerr; it (necessarily) has a singularity on part of its horizon, but has the interesting feature of breaking integrability for (all) geodesics.

In this Section, we review each of these three black holes in turn. As a point of comparison to Kerr, we give each black hole's gravitational multipole moments $M_\ell, S_\ell$ \cite{Thorne:1980ru,Bena:2020see,Bena:2020uup,Mayerson:2022ekj}. These multipoles are an infinite series of gauge-invariant quantities that characterize the metric\footnote{In an appropriate coordinate system, these can be read off from the asymptotic expansion of the metric, i.e. schematically $g_{tt}\sim \sum_\ell M_\ell/r^{\ell+1}$ and $g_{t\phi}/(r^2\sin^2\theta)\sim \sum_\ell S_\ell/r^{\ell+1}$ \cite{Thorne:1980ru,Bena:2020uup}.} --- and so for our purposes, characterize the black holes' (asymptotic) deviation from Kerr, which has multipoles $M_\ell + iS_\ell = M(ia)^\ell$.

We conclude this section by reviewing the calculation of the critical curve (or ``shadow boundary'') for the Johannsen and Rasheed-Larsen metric, and discuss the influence of the different metric parameters.
Four-color screen images of examples of these black holes are given in Section \ref{sec:PRresults}.

\subsection{Johannsen metric}\label{sec:joh}

The Johannsen metric is constructed as a stationary and axisymmetric generalization of the Kerr metric that keeps a Carter-like constant \cite{johannsen2013regular}. This ensures that the geodesic equations are still integrable, facilitating an analytic treatment. The metric is given by\footnote{We consider only the first non-zero correction in the deviation functions.}

\begin{empheq}[box=\fbox]{align}
    \dd s^2 =\,& -\Tilde{\Sigma}\frac{\Delta - a^2 A_2(r)^2 \sin^2\theta}{N(r)} \, \dd t^2  + \frac{\Tilde{\Sigma}}{\Delta A_5(r)}\, \dd r^2 + \Tilde{\Sigma}\, \dd \theta^2\nonumber\\ 
    & - 2 a\Tilde{\Sigma}\sin^2\theta \frac{(r^2+a^2)A_1(r)A_2(r) - \Delta}{N(r)}\, \dd t \, \dd \phi \label{eq: Johannsen metric}\\
    & + \Tilde{\Sigma} \sin^2 \theta \frac{(r^2+a^2)^2A_1(r)^2 - a^2 \Delta \sin^2\theta}{N(r)}\, \dd \phi^2\,,
    \nonumber
\end{empheq}
where
\be
\begin{aligned}
    A_1(r) & = 1+\alpha_{13}\left(\frac{M}{r}\right)^3\,,  & & &\Tilde{\Sigma}  & = r^2 + a^2\cos^2\theta + \epsilon_3 \frac{M^3}{r} \label{eq: dev functions Joh}
    \,,\\
    A_5(r) & = 1+\alpha_{52}\left(\frac{M}{r}\right)^2\,, & & &
    \Delta &  = r^2 - 2Mr + a^2 \,, \\
    A_2(r) & = 1+\alpha_{22}\left(\frac{M}{r}\right)^2\,, & & &
    N(r) & = \left[\left(r^2+a^2\right)A_1(r) - a^2A_2(r)\sin^2\theta\right]^2 \,.
\end{aligned}
\ee
This new metric depends on four additional parameters with respect to Kerr: $\alpha_{13}, \alpha_{22}, \alpha_{52}$, and $\epsilon_3$. We will set $\epsilon_3=0$ as this parameter is only of interest for massive particle trajectories.\footnote{This can be seen from the fact that $\tilde{\Sigma}$ couples only to $\mu$ in (\ref{eq: Ham Jac Joh}).}

The Johannsen metric is asymptotically flat and reduces to Kerr when all deviation parameters $\alpha_{13}, \alpha_{22}, \alpha_{52}, \epsilon_3$ vanish. The event horizon of this black hole is still the same as that of the Kerr metric, $r_H = M + \sqrt{M^2 - a^2}$. Demanding that the metric is regular outside of the horizon restricts the parameters as
\begin{align}
    \alpha_{52} & > -\frac{r_H^2}{M^2}\,, & 
    \epsilon_3 & > -\frac{r_H^3}{M^3} \,, &
    \alpha_{13} & > -\frac{r_H^3}{M^3}, \label{eq: bound a13}
\end{align}
together with the requirement that
\begin{equation}\label{eq: det req Joh}
    \alpha_{13} \neq \frac{a^2r(r^2+\alpha_{22}M^2)\sin^2\theta - r^3(r^2+a^2)}{M^3(r^2+a^2)}\,
\end{equation}
everywhere outside the horizon. In the case that only one of the deviation parameters is non-zero, as we will take in this paper, we find that this last condition reduces to
\begin{equation}\label{eq: single bounds a13 a22}
    \alpha_{22} < \frac{r_H^4}{a^2M^2} \qquad \text{or} \qquad 
    \alpha_{13} > - \frac{r_H^4}{2M^4}\,.
\end{equation}

The Johannsen black hole has precisely the same gravitational multipoles as Kerr:
\be 
\begin{aligned}
    M_{2n} &= M(-a^2)^n ,& M_{2n+1} &= 0,\\
    S_{2n+1} &= Ma(-a^2)^n, & S_{2n} &= 0,
\end{aligned}
\ee
which implies the black hole cannot be distinguished from Kerr by asymptotic observations of the metric alone.

\subsection{Rasheed-Larsen black hole}\label{sec:RL}

The Rasheed-Larsen black hole can be obtained by a Kaluza-Klein reduction of five-dimensional Einstein gravity \cite{Rasheed:1995zv,larsen2000rotating}. Contrary to the Johannsen metric, this black hole solution is obtained from an actual physical theory in four dimensions, with the Lagrangian
\begin{equation}\label{eq:lagrRL}
    L = \frac{1}{16\pi G_N}\int \dd^4 x \left[ R - 2\partial_\mu \Phi \partial^\mu \Phi - \frac14 e^{-2\sqrt{3}\Phi} F_{\mu\nu}F^{\mu\nu}\right]\,,
\end{equation}
where $R$ is the Ricci scalar, $\Phi$ is the dilaton field and $F_{\mu\nu}$ is associated with the electromagnetic gauge field $A_\mu$.

The Rasheed-Larsen black hole is a solution to (\ref{eq:lagrRL}) given by
\begin{equation}\label{eq: RL metric}
\boxed{
 \dd s^2 = -\frac{H_3}{\sqrt{H_1 H_2}}\left(\dd t + \mathbf{B}\right)^2 + \sqrt{H_1 H_2} \left(\frac{\dd r^2}{\Delta} +\dd\theta^2 + \frac{\Delta}{H_3}\sin^2\theta \dd \phi^2 \right)  \,.} 
\end{equation}
The functions $H_1, H_2, H_3, \Delta$, and 1-form $\mathbf{B}$ are given by
\begin{align}
    H_1  = &\, r^2 + a^2 \cos^2\theta + r(p-2m) + \frac{p}{p+q}\frac{(p-2m)(q-2m)}{2} \\
  \nonumber  & - \frac{p}{2m(p+q)}\sqrt{(q^2 - 4m^2)(p^2-4m^2)} \,a \cos\theta\,, \\
    H_2 = &\,  r^2 + a^2 \cos^2\theta + r(q-2m) + \frac{q}{p+q}\frac{(p-2m)(q-2m)}{2}\\
   \nonumber &  +\frac{q}{2m(p+q)}\sqrt{(q^2 - 4m^2)(p^2-4m^2)}\, a \cos\theta\,, \\
    H_3 = &\,  r^2 + a^2 \cos^2\theta - 2mr\,, \\
    \Delta = &\,  r^2 + a^2 - 2mr\,, \\
    \mathbf{B} = &\, \sqrt{pq}\frac{(pq+4m^2)r - m(p-2m)(q-2m)}{2m(p+q)H_3} a\sin^2\theta \,\dd \phi\,. \label{eq: B RL}
\end{align}
The solution depends on four parameters $a, m, p, q$. These are related to the physical quantities of the metric: mass ($M$), angular momentum ($J$), and electric \& magnetic charges ($Q$ \& $P$) as
\begin{align}
    M & = \frac{p+q}{4}\, \label{eq: mass RL} \,,&
    J & = \frac{\sqrt{pq}(pq+4m^2)}{4m(p+q)}\,a\,, \\
    Q^2 & = \frac{q(q^2-4m^2)}{4(p+q)} \,, &
    P^2 & = \frac{p(p^2-4m^2)}{4(p+q)} \,.
\end{align}

The event horizon of the Rasheed-Larsen black hole is located at
\begin{equation}
    r_H  = m + \sqrt{m^2 - a^2}\,,
\end{equation}
which implies the bound $a^2\leq m^2$. The other parameters need to satisfy $p,q \geq 2m$, where equality corresponds to the vanishing of the electric and/or magnetic charges. The case $p=q=2m$ corresponds to the chargeless Kerr limit (in which the mass is given by $m$).

Interestingly, the Rasheed-Larsen black hole only admits integrable geodesic equations for \textit{null} geodesics due to the existence of a conformal Killing tensor \cite{Keeler:2012mq}; see also Appendix \ref{app: aux calc RL}.

The multipoles of the Rasheed-Larsen black hole were calculated in \cite{Bena:2020see,Bena:2020uup}:
\be \label{eq:RLmultipoles} M_\ell = \sum_{k=0}^\ell \binom{\ell}{k} \tilde M_k \left( -\frac{\tilde M_1}{\tilde M_0}\right)^{\ell-k},\qquad S_\ell = \sum_{k=0}^\ell \binom{\ell}{k} \tilde S_k \left( -\frac{\tilde M_1}{\tilde M_0}\right)^{\ell-k},\ee
where
\be
\begin{aligned}
\tilde M_{2n} &= \left[\frac{p+q}{4}\right](-a^2)^n, & \tilde M_{2n+1} &= \left[\frac{a}{8m}\frac{p-q}{p+q}\sqrt{(p^2 -4m^2)(q^2-4m^2)}\right](-a^2)^n,\\
\tilde S_{2n} &= 0, & \tilde S_{2n+1} &= \left[ \frac{a}{4m} \frac{\sqrt{p q} (pq+4m^2)}{p+q}\right](-a^2)^n.
\end{aligned}
\ee
Note that when $a\neq 0, p,q > 2m$ and $p\neq q$, the Rasheed-Larsen metric has $S_{2n},M_{2n+1} \neq 0$, which indicates a breaking of equatorial symmetry that is not present in Kerr. When $p=q=2m$, the metric reduces to Kerr; however, note that the equal charge case $p=q>2m$ does \emph{not} correspond to Kerr-Newman \cite{larsen2000rotating}.

\subsection{Manko-Novikov black hole}\label{sec:MN}

The Manko-Novikov metric is constructed as a solution to the vacuum Einstein equations in four dimensions \cite{manko1992novikov}; it can be thought of as deforming the Kerr black hole by multipole ``bumps''. Since it is a vacuum solution, the uniqueness theorems imply it must be singular; in particular, the event horizon has a singularity (only) at the equator.

The Manko-Novikov metric can be written in Boyer-Lindquist coordinates as
\begin{equation}\label{eq: MaNo metric}\boxed{
    \dd s^2 = -f\left(\dd t - \omega \dd \phi\right)^2 + \frac{e^{2\gamma} \rho^2}{f \Delta}\dd r^2 + \frac{e^{2\gamma} \rho^2}{f}\dd \theta^2 + \frac{\Delta \sin^2 \theta}{f} \dd \phi^2\,.
    }
\end{equation}
The functions $\rho,\Delta$ are given by
\be \rho^2 = (r-M)^2 - k^2\cos^2\theta, \qquad \Delta = r^2 - 2Mr + a^2,\ee
where $k = \sqrt{M^2-a^2}$. The event horizon is at
\be r_H = M + \sqrt{M^2-a^2}.\ee
The functions $\gamma, f, \omega$ depend on the mass and rotation parameters $M, a$ as well as an octopole-bump parameter $\alpha_3$ in a highly non-trivial way --- see Appendix \ref{app: MaNo}. When $\alpha_3=0$, the metric is simply Kerr. (The generic Manko-Novikov black hole depends on an infinite series of such parameters, one for each multipole that can be added.)

As opposed to the Johannsen and Rasheed-Larsen black holes presented above, this metric no longer permits separability of the (null) geodesic equations. This breaking of integrability leads to chaotic phenomena in its image, and the critical curve (if it is even well-defined) cannot be determined analytically. (See also below in Section \ref{sec:MNshape}.)

The first few non-zero multipoles of the Manko-Novikov metric are
\be
\begin{aligned}
M_0 &= M, & M_2 &= -Ma^2, & M_3 &= -\alpha_3 (M^2-a^2)^2, & M_4 = Ma^4,\\
S_1 &= Ma,  & S_3 &= - Ma^3, & S_4 &= -2a \left[\alpha_3 (M^2-a^2)^2\right].
\end{aligned}
\ee
From these expressions, it is clear that the dimensionless parameter $\alpha_3$ controls the deviations from Kerr by turning on the equatorial symmetry-breaking multipole moments $M_3$ and $S_4$ (and will also feature in higher-order odd-parity multipole moments).

\subsection{Black hole critical curves}\label{sec: analytic shadow}

In an image of a black hole, there will always be a central area with little to no brightness. In the case of a theoretical uniformly illuminated black hole, this is an entirely black shape that is sometimes called the \emph{shadow} of the black hole (although see footnote \ref{fn:shadow} above). The outer boundary of this shadow is the \emph{critical curve}, and is determined by the exactly bound photons of the black hole. The critical curve can also be thought of as the $n\rightarrow\infty$ limit of the photon rings.
This critical curve is typically somewhat ``larger'' than the black hole itself --- e.g.\ the (circular) Schwarzschild black hole critical curve has radius $\sqrt{27}M$ from the point of view of an observer at infinity \cite{bardeen1973timelike}, whereas the horizon radius is $2M$.

The critical curve can be analytically derived for metrics where null geodesics are separable. In this section, we review this method for stationary and axisymmetric separable metrics (following \cite{bardeen1973timelike}) and their application to the critical curves for the Johannsen and Rasheed-Larsen black holes (first studied in resp.\ \cite{johannsen2013photon} and \cite{mirzaev2022observational}). The Manko-Novikov metric does not have separable null geodesics and so does not allow such a treatment.

\medskip

Consider a photon following a null geodesic with 4-momentum $p_\mu$, satisfying $p_\mu p^\mu = 0$. Stationarity and axisymmetry of the metric imply that the photon energy $E\equiv - p_t$ and azimuthal angular momentum $L \equiv p_\phi$ are conserved quantities. An observer at distance $r_0$ can set up a frame of reference given by an orthonormal tetrad $\{e_{(\mu)}\}$, where $(\mu)$ is a \textit{label} and not an index.  The photon's velocity as measured by this observer is then $p^{(t)} = -e_{(t)}^\mu p_\mu$ and $p^{(i)} = e_{(i)}^\mu p_\mu$ for $i \in \{r, \theta,\phi\}$. When the observer is at a large distance $r_0\rightarrow\infty$ the photon is characterized alternatively by the impact parameters
\begin{align}
    \alpha & = \lim_{r_0\to \infty} - r_0 \frac{p^{(\phi)}}{p^{(t)}}\,, &
    \beta & = \lim_{r_0\to \infty}  r_0 \frac{p^{(\theta)}}{p^{(t)}}\,.
\end{align}
For both black holes of interest, the impact parameters are given by
\begin{align}
    \alpha & = -\frac{\lambda}{\sin\theta_0}\,, \label{eq: alpha beta} &
    \beta & = p_\theta\big|_{\theta_0}\,, 
\end{align}
where $\lambda\equiv L/E$ is the energy-rescaled angular momentum\footnote{For null geodesics, only the energy-rescaled quantities are relevant.}, and $\theta_0$ is the observer's angle of inclination with respect to the black hole's angular momentum. The separable null geodesic equations demand that
\be p_r^2 = \R(r) , \qquad p_\theta^2 = \Theta(\theta),\ee
for a particular \emph{angular potential} $\Theta(\theta)$ and \emph{radial potential} $\R(r)$. The angular potentials for the Johannsen and Rasheed-Larsen black holes are (see Appendix \ref{app: auxiliary calcs})
\begin{align}
    \Theta_{\text{Joh}}(\theta) & = \eta + a^2 \cos^2\theta - \lambda^2 \cot^2 \theta \,, \label{eq: Theta Joh}\\
    \Theta_{\text{RL}}(\theta) & = \Theta_{\text{Joh}}(\theta) - a \frac{(p-q)\sqrt{(p^2-4m^2)(q^2-4m^2)}}{2m(p+q)}\cos\theta \, \label{eq: Theta RL}
\end{align}
whereas the radial potentials are
\begin{align}
    \R_{\text{Joh}}(r) \equiv &&A_5^2(r)\Delta^2 p_r^2 & = A_5(r)\left[A_1(r) (a^2+r^2)  - A_2(r)a\lambda \right]^2 - \Delta \chi A_5(r)\,, \label{eq: R Joh} \\
    \R_{\text{RL}}(r) \equiv &&  \Delta^2 p_r^2 & = -\Delta\left[\chi + T + T_r(r) + 2\lambda f(r)\right]+a^2\lambda^2\,. \label{eq: R RL}
\end{align}
These potentials depend on a constant $\eta$, which can be physically interpreted as the square of the angular momentum in the $\theta$-direction of the photon when it passes through the equatorial plane $\theta = \pi/2$. Therefore, the geodesics that we consider have $\eta \geq 0$, with equality holding for geodesics that are confined to the equatorial plane; $\eta$ is precisely related to the Carter-like constant for these separable black holes. The precise definition of $\eta$ as well as the auxiliary functions $T, T_r, f$ are given in Appendix \ref{app: auxiliary calcs}.

Photons can only arrive at the distant observer if they travel on geodesics that are not bound. Geodesics are characterized by their constants of motion, meaning that the impact parameters (\ref{eq: alpha beta}) only make sense for $\lambda, \eta$ associated with unbound geodesics. Constants of motion related to geodesics that fall into the black hole correspond to impact parameters at which no photon will be detected. No light will arrive on the observer screen for this set of impact parameters. The \emph{critical curve} is the boundary of this region: it is defined by the impact parameters for \textit{bound geodesics}. In the parameter space of impact parameters, these separate the geodesics that fall into the black hole from those that escape to infinity.

A bound null geodesic orbits at a fixed radial coordinate for radii within a certain interval.\footnote{This was shown for Kerr in \cite{gralla2020null}, and we assume this holds for the Johannsen and Rasheed-Larsen spacetimes as well. Given that the resulting critical curves define convex, connected regions, there is no evidence our analysis misses any other bound orbits.} The radial potential for a geodesic on a bound orbit of radius $r_B$ must satisfy
\begin{equation}\label{eq: cond bound geod}
    \R(r_B) = \R'(r_B) = 0\,.
\end{equation}
These equations can be solved to determine $\eta$ and $\lambda$ in terms of $r_B$, see Appendix \ref{app: aux calc Joh}. 

The range of allowed values for the radial coordinate is determined by demanding that the angular potential $\Theta(\theta)$ is positive somewhere in the interval $[0,\pi]$, i.e.\ there are values for $\theta$ at which the momentum in the $\theta$-direction is real. The angular potential of bound geodesics depends on the radius of the bound orbit $r_B$ through the constants of motion $\eta,\lambda$.

Note that an observer located at an inclination $\theta_0$ can only observe photons that travel on geodesics for which $\Theta(\theta_0) \geq 0$ (see (\ref{eq: alpha beta})). In the case of the Johannsen (or Kerr) metric, the range of bound orbit radii $r_B$ for which this holds is largest for an observer in the equatorial plane. The range of $r_B$ for off-equatorial observers is contained within this maximal range. This means that an equatorial observer probes the largest part of the black hole spacetime. In case of the Rasheed-Larsen metric, this is no longer the case, as the spacetime is no longer symmetric with respect to $\theta = \frac{\pi}{2}$. However, there is still an observer inclination angle $\theta'_0$ which corresponds to observing a maximal range in $r_B$. For further discussion, see Section \ref{sec:lyap-inclangle}.

\medskip

\begin{figure}[htbp]
    \centering
    \includegraphics[width = \textwidth]{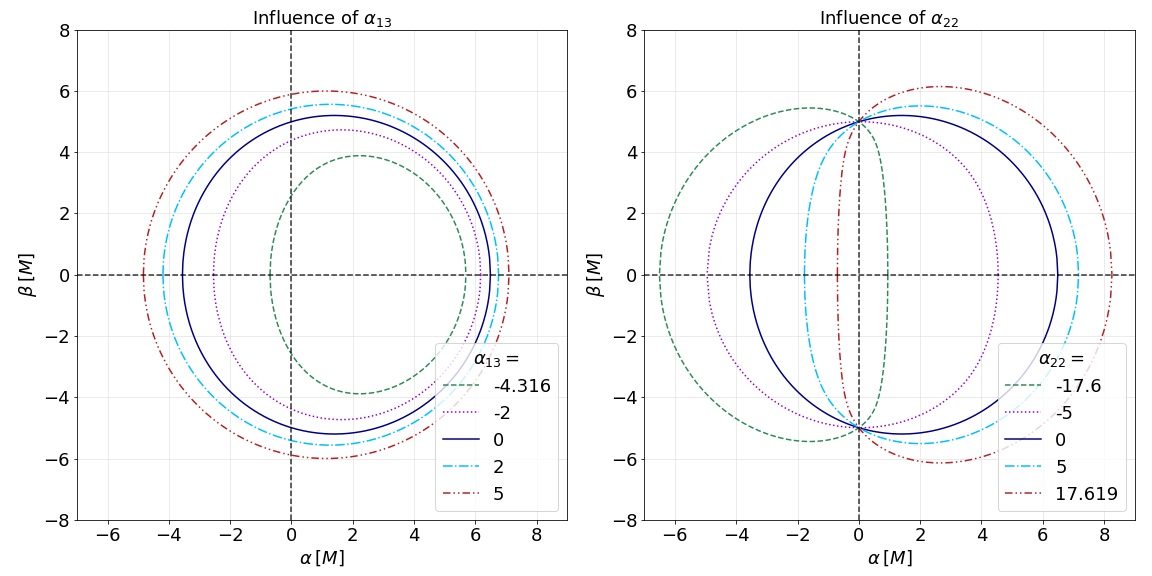}
    \caption{Influence of changing the different parameters of the Johannsen metric on the black hole critical curve. The spin parameter is $a/M=0.7$. Only one deviation parameter in each figure is non-zero; $\alpha_{ij}=0$ corresponds to the Kerr black hole. The parameter $\alpha_{52}$ has no effect on the critical curve so is not shown. The observer is located in the equatorial plane.}
    \label{fig: Sh Joh}
\end{figure}

\begin{figure}[htbp]
    \centering
    \includegraphics[width = \textwidth]{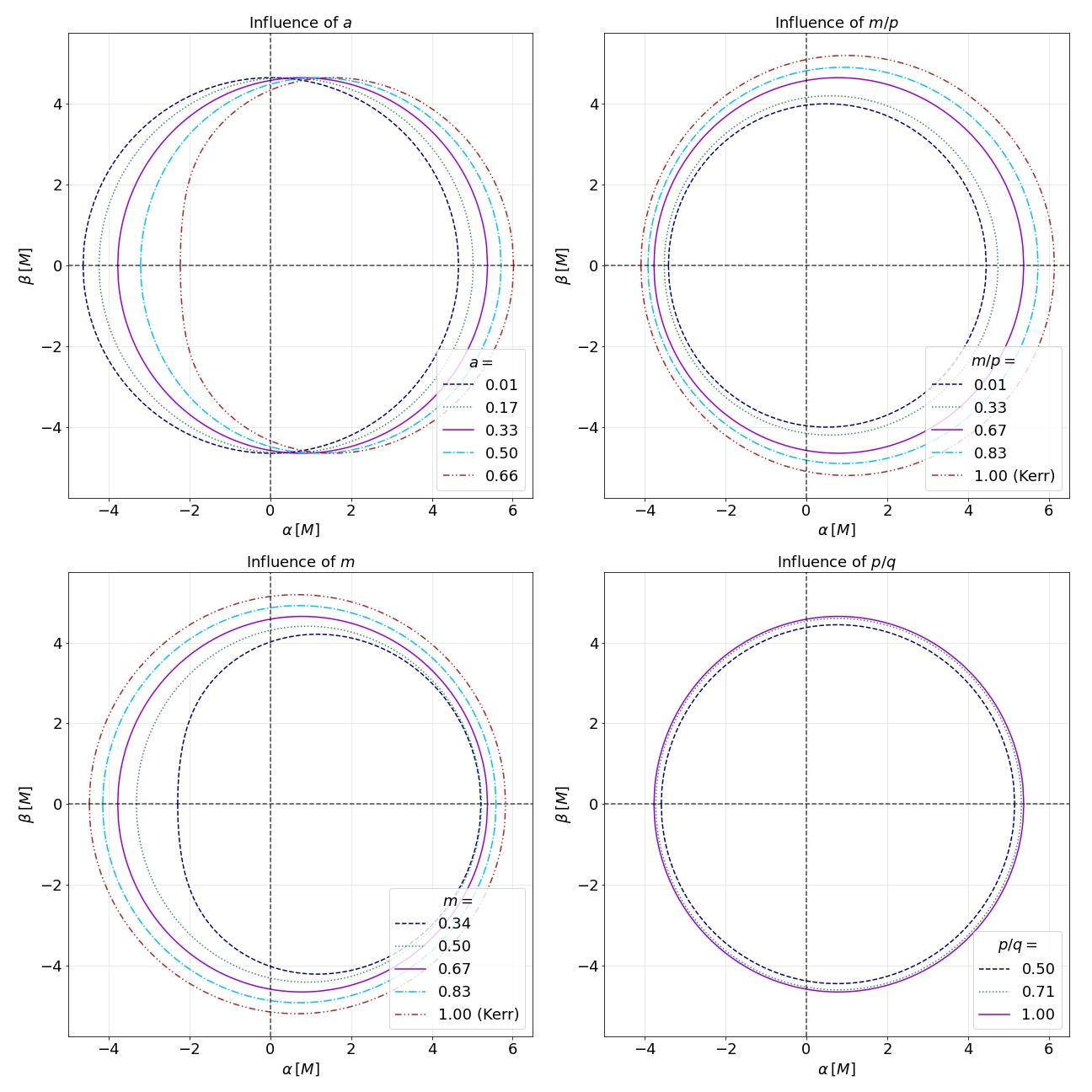}
    \caption{Influence of changing the different parameters of the Rasheed-Larsen metric on the black hole critical curve, as measured by an equatorial observer. The standard parameters are $a/M=1/3$, $m/M = 2/3$, $p/M=q/M=2$. In the two left panels we change $a$ and $m$, and in the top right panel we change $m/p$ while keeping $a/m$ and $p$ fixed. The bottom right panel shows the influence of changing $p/q$, keeping $a,m$ fixed. The two parameter values that correspond to Kerr black holes are indicated.}
    \label{fig: Sh RL}
\end{figure}

We present examples of critical curves for the Johannsen and Rasheed-Larsen black holes in Figures \ref{fig: Sh Joh} and \ref{fig: Sh RL}. We have also verified their correctness by comparing them with ray-traced images.

For the Johannsen metric (Fig.~\ref{fig: Sh Joh}), the effect of changing the spin parameter is the same as for Kerr --- increasing spin compresses the critical curve on one side. The parameter $\alpha_{52}$ can be seen not to have any influence on the critical curve, so we fix $\alpha_{52}=0$. The other two deviation parameters have an observable effect; $\alpha_{13}$ mostly changes the overall size of the critical curve, whereas $\alpha_{22}$ compresses the critical curve in one direction, much like the effect of increasing the spin parameter. We also show the critical curve for parameter values close to the bounds (\ref{eq: single bounds a13 a22}).

Rasheed-Larsen black hole critical curves are plotted in Fig.~\ref{fig: Sh RL}. We again see the typical one-sided compression of the critical curve when the spin parameter $a$ is increased (see upper left panel). The parameter $m$ influences the overall size, whereas increasing the charge while keeping the mass fixed decreases the size of the critical curve (see upper right panel) --- as usual for charged black holes. Finally, changing the ratio $p/q$ between the two kinds of charge parameters has a small influence on the size of the shadow as well, and the results are symmetric in $p,q$.

\clearpage

\section{Photon Ring Shape}\label{sec:shape}
Although the shape and size of the critical curve may be intrinsically linked to the specific geometry of the black hole, they are not true observables, as mentioned above in the Introduction. However, the successive photon rings do approach the critical curve --- closer for every successive $n$. Therefore, from about $n\gtrsim 2$, the photon ring curves approximate the critical curve very well.

As mentioned in the Introduction, GLM argued \cite{Gralla:2020srx} that measuring the shape of the $n=2$ ring can provide a precise null test of the Kerr metric. In particular, the projected diameter of the shape of the Kerr critical curve (and the $n=2$ photon ring) is to a very good approximation fit by a so-called \emph{circlipse} \cite{Gralla:2020srx, Gralla:2020yvo}, which can be thought of as a superposition of a circle and an ellipse. GLM projected that measurements of the $n=2$ ring could be fit to a circlipse to an incredible $0.04\%$ precision at relatively low inclinations $\theta_0\leq 30^\circ$.
The precise method of extracting the shape of the photon ring from the actual interferometric measurements of the image visibility amplitude (i.e.\ the Fourier transform of the actual image) is explained in \cite{Gralla:2020srx} (see also \cite{Paugnat:2022qzy}).

 Below in Section  \ref{sec:PRmethod} we introduce the precise definition of projected diameter (i.e.\ the measurable component of the photon ring shape), the circlipse function, and our method of generating and analyzing the photon ring shapes.

 In Section \ref{sec:PRresults}, we analyze the critical curve and $n=2$ photon ring shape at low inclinations for the three different black hole metrics at hand, and discuss the circlipse fit for their projected diameters.

\subsection{Methodology}\label{sec:PRmethod}
We summarize our method of generating the $n=2$ ring and critical curve projected diameter $d(\varphi)$ here.
In all cases except the Manko-Novikov metric, the critical curve  can be analytically calculated; see Section \ref{sec: analytic shadow}.
We use numerical ray tracing of null geodesics in black hole backgrounds to generate a collection of Cartesian coordinates $(x,y)$ for the $n=2$ photon ring. We use the ray tracer \texttt{FOORT} \cite{FOORT}; see Appendix \ref{app:FOORT} for additional information on \texttt{FOORT} and Appendix \ref{app:moreshapemethod} for more details on the extraction of the $n=2$ ring.

\subsubsection{Projected diameter}
Given a collection of points $(x,y)$ describing a curve (i.e.\ the $n=2$ ring or the critical curve), we can parametrize every point on this curve by the standard polar angle $\theta=\tan^{-1}(y/x)$, giving us a parametrized shape $(x(\theta),y(\theta))$. We then define the angle $\varphi$ through the normal at every point on the curve as \cite{Gralla:2020yvo}
\be \label{eq:phifromtheta} \tan \varphi(\theta) = - \frac{x'(\theta)}{y'(\theta)},\ee
giving the projected position function as
\be\label{eq:deffphi} f(\varphi) = x(\varphi) \cos\varphi + y(\varphi)\sin(\varphi),\ee
and finally the projected diameter is simply
\be \label{eq:defdphi} d(\varphi) = \frac12\left(f(\varphi) + f(\varphi + \pi)\right).\ee
The angle $\varphi$ is periodic with a range from $0$ to $2\pi$ and the projected diameter satisfies $d(\varphi+\pi)=d(\varphi)$. By convention, $\varphi=0$ and $\varphi=\pi$ are on the equatorial plane, so an \emph{equatorially symmetric} closed curve will also satisfy
\be d(\varphi) = d(\pi-\varphi).\ee

Following \cite{Gralla:2020srx}, we will consider datapoints for $d(\varphi)$ at the equidistant 35 points $\varphi=5^\circ, \cdots, 175^\circ$.\footnote{In \cite{Gralla:2020srx}, also $\varphi=0^\circ$ is used; we exclude this point as it is too sensitive to numerical errors inherent in our method.} To mimic observational noise, we also consider a ``noisy'' version of this dataset, where we artificially add a random Gaussian noise to the 35 datapoints with deviation $\sigma$ given by:
\be \label{eq:mocknoise} \sigma = 4.45026\times 10^{-4}\times \mu,\ee
where $\mu=\langle d(\varphi)\rangle$ is the average projected diameter over the 35 datapoints. This is the same noise as the forecast of \cite{Gralla:2020srx} exhibits (where $\sigma = 0.017\, \mu \text{as}$ and $\mu\approx 38.2\, \mu\text{as}$).
In principle, one should infer the photon ring diameter at a given angle $\varphi$ from the periodicity of the visibility amplitude on long baselines, in which case the periodicity would give a definite value for the ring diameter lying within the range of diameters contained within the lensing band.\footnote{We thank A. Lupsasca for emphasizing this point to us.} The mock ``noise'' that we are adding to our data points can be thought as mimicking the intrinsic error in this interferometric measurement that comes from the finite ring thickness; our assumption is that this error is comparable in scale to that found in \cite{Gralla:2020srx}.

The resulting $d(\varphi)$ data set can then be fit to the \emph{circlipse} function \cite{Gralla:2020srx,Gralla:2020yvo}
\be \label{eq:circlipsedef} d_\text{circlipse}(\varphi) = R_0 + \sqrt{R_1^2 \cos^2(\varphi-\varphi_0) + R_2^2\sin^2(\varphi-\varphi_0)}.\ee
This function depends on the three physical parameters $R_0,R_1,R_2$, and an offset angle $\varphi_0$ which is degenerate with the orientation of the camera \cite{Gralla:2020srx}. As measure for the precision of the fit, we consider, as in \cite{Gralla:2020srx}, the normalized root-mean square deviation of the fit
\be \RMSD:= \frac{\sqrt{\langle (d_\text{obs}-d_\text{fit})^2 \rangle}}{\langle d_\text{fit} \rangle} ,\ee
where $d_\text{fit}$ is the best-fit circlipse of (\ref{eq:circlipsedef}) for the datapoints $d_\text{obs}$.

\begin{figure}[htpb]\centering

\begin{subfigure}{0.45\textwidth}
\includegraphics[width=\textwidth]{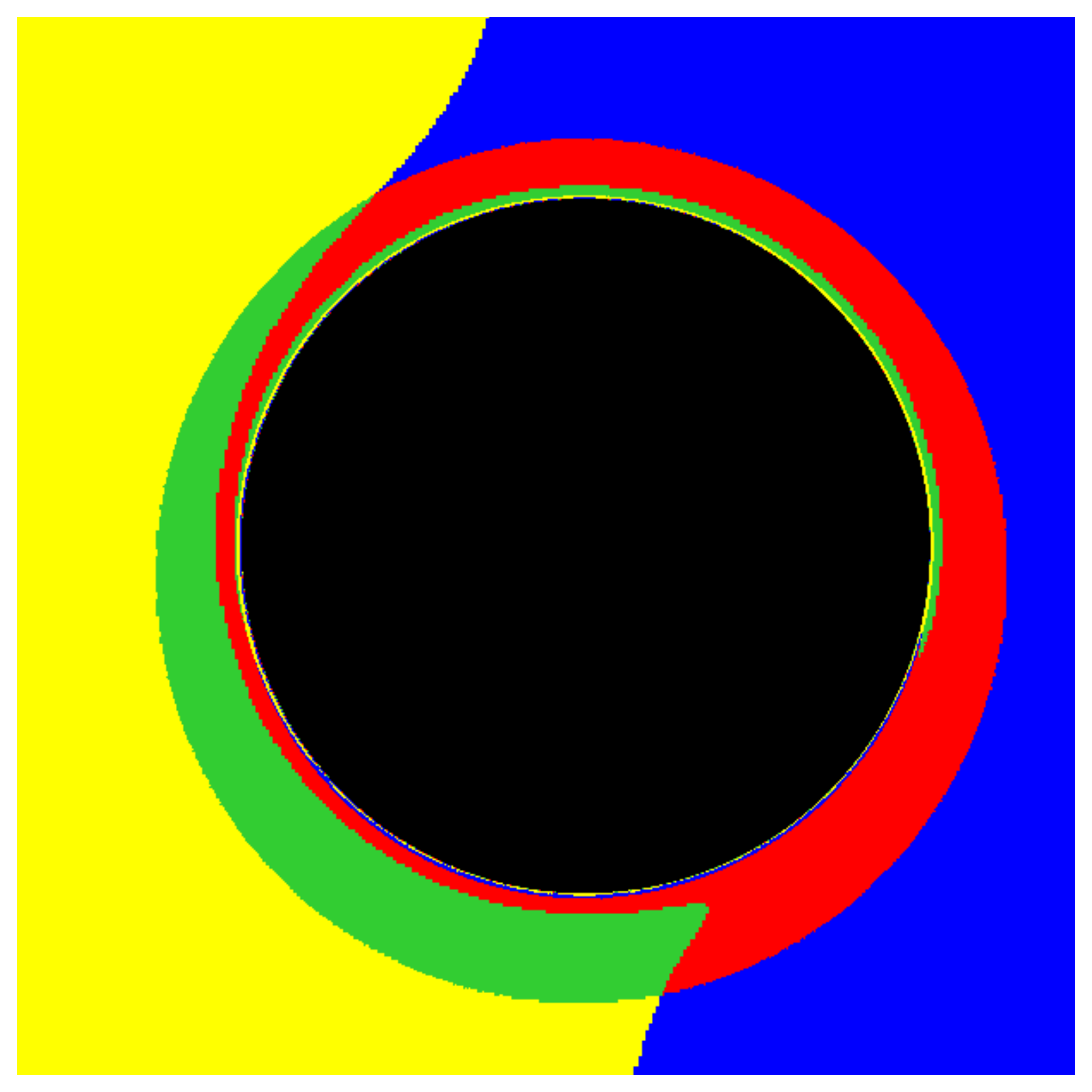}
\caption{Four-color screen image}
\end{subfigure}
\begin{subfigure}{0.45\textwidth}
\includegraphics[width=\textwidth]{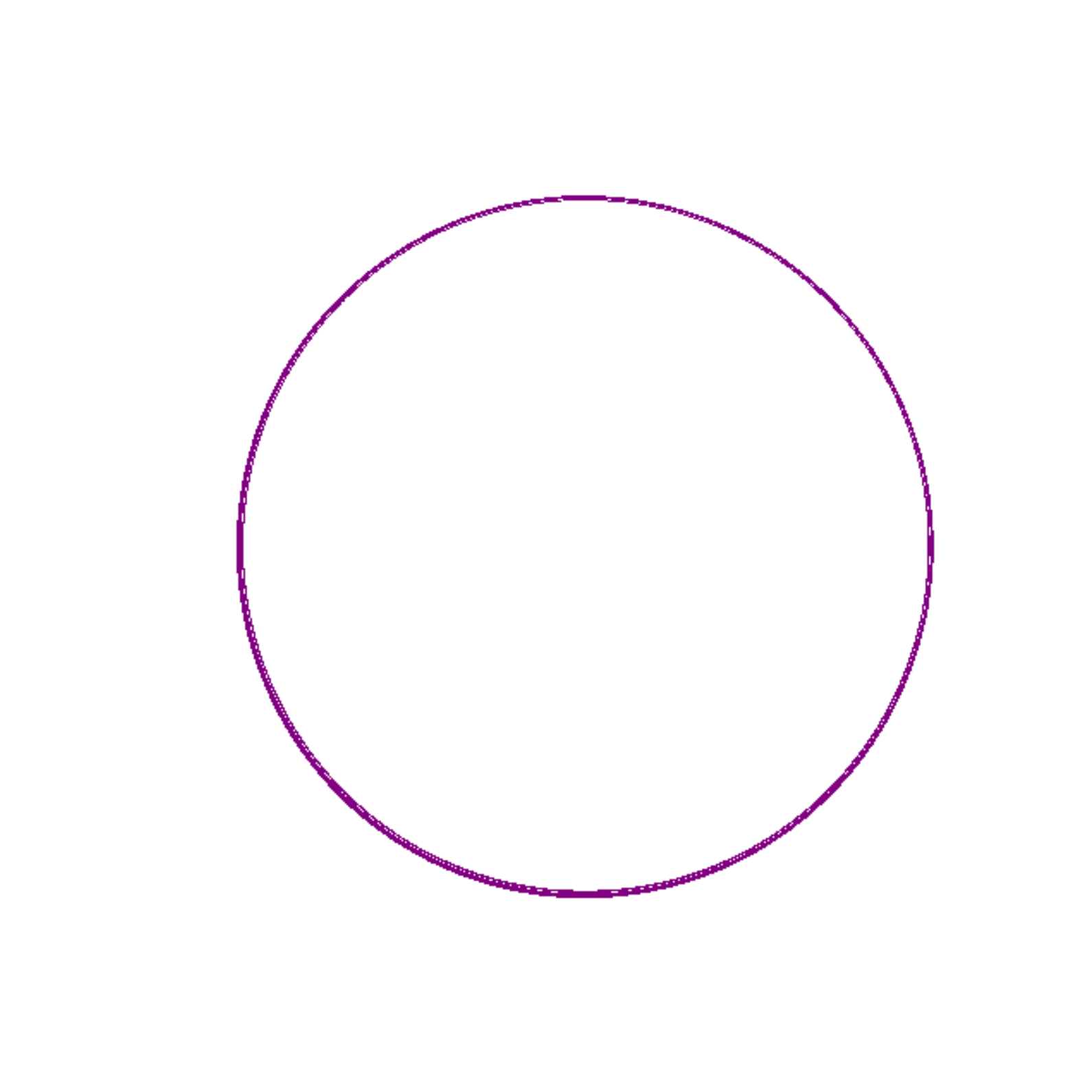}
\caption{The complete $n=2$ ring.}
\end{subfigure}
\begin{subfigure}{0.45\textwidth}
\includegraphics[width=\textwidth]{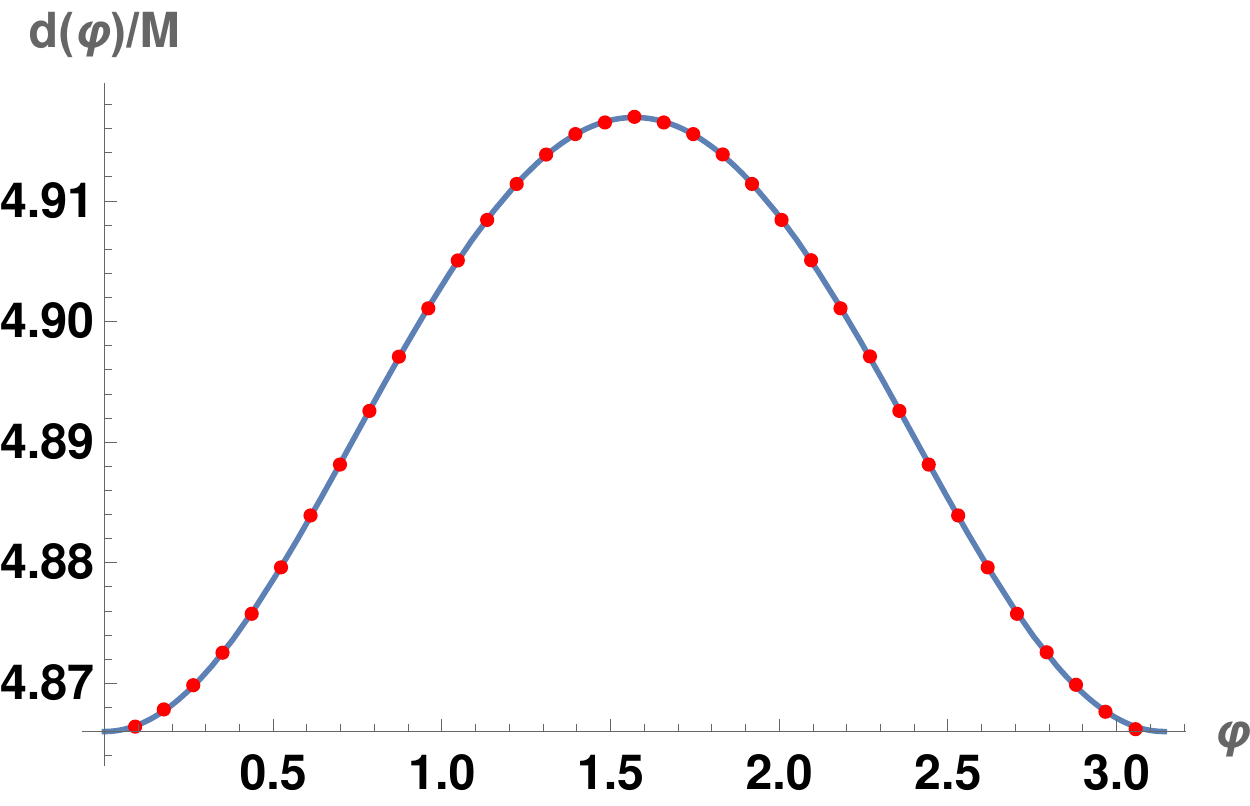}
\caption{Fit with original data}
\end{subfigure}
\begin{subfigure}{0.45\textwidth}
\includegraphics[width=\textwidth]{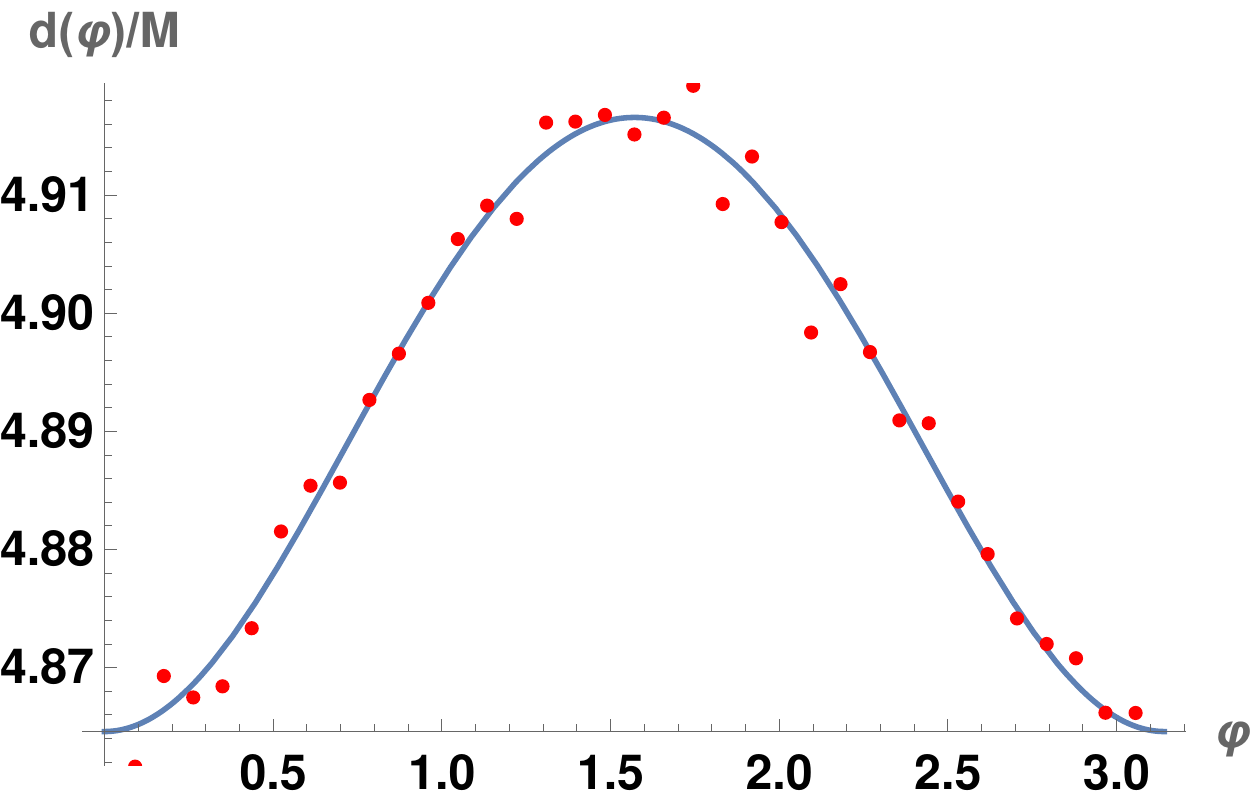}
\caption{Fit with noisy data}
\end{subfigure}
\caption{Our procedure of determining the $n=2$ ring or critical curve shape for Kerr with $a/M = 0.94$ and $\theta_0=17^\circ$. The critical curve is fit in (c) and (d), but the fit for the $n=2$ data is analogous. The RMSD of the original data fit is $\RMSD = 0.002\%$ and for the mock noisy data fit is $\RMSD = 0.046\%$.}
\label{fig:Kerrexample}
\end{figure}

An example of our procedure is shown in Fig.~\ref{fig:Kerrexample} for Kerr with $a/M=0.94$, viewed at $\theta_0=17^\circ$ inclination. According to \cite{Gralla:2020srx}, $\RMSD$ for the best-fit circlipse model to the actual forecasted observational data for the $n=2$ photon ring is 0.04\%. Our procedure generates data for the $n=2$ ring and critical curve that has $\RMSD = 0.002\%$, and the corresponding mock noisy data has $\RMSD = 0.046\%$, which indeed matches the forecasted precision of \cite{Gralla:2020srx}. Note that an analogous fit with $\RMSD\approx 0.04\%$ for the noisy data is obtained at all inclinations $\theta_0\leq 35^\circ$.

At higher inclinations, the circlipse function still fits the Kerr critical curve projected diameter very well \cite{Gralla:2020yvo}. The RMSD errors for the best-fit circlipse increase only mildly --- for $a/M=0.99$ at $\theta_0=90^\circ$, we find $\RMSD = 0.073\%, 0.081\%$ for resp.\ the original and mock-noisy data, again very much in agreement with the fit errors found in \cite{Gralla:2020yvo} (i.e.\ $\RMSD = 0.06\%$).

\subsection{Results}\label{sec:PRresults}
Here, we present an analysis of the critical curves and/or $n=2$ photon ring projected diameters for the Johannsen, Rasheed-Larsen, and Manko-Novikov black holes introduced above. We will present results for low inclinations $\theta_0\lesssim 30^\circ$, although we will often focus on $\theta_0 = 17^\circ$ (i.e.\ the observationally inferred inclination for M87*).

\subsubsection{Johannsen}
The only parameter that affects the photon ring or critical curve \emph{shape} measurably is the parameter $\alpha_{22}$,\footnote{The \emph{size} of the critical curve or photon ring is affected significantly by $\alpha_{13}$ as well, but this does not show up in an analysis of only the shape. The parameter $\alpha_{52}$ does not influence the critical curve at all. See Section \ref{sec: analytic shadow}.} so we focus only on changing this parameter.

For all the Johannsen metrics we consider, there is no measurable difference in the shape analysis between critical curve and curve of the $n=2$ photon ring; for simplicity, we focus on the critical curve as it can be determined analytically (see Section \ref{sec: analytic shadow}.)

\begin{figure}[htpb]\centering
\begin{subfigure}{0.48\textwidth}
\includegraphics[width=\textwidth]{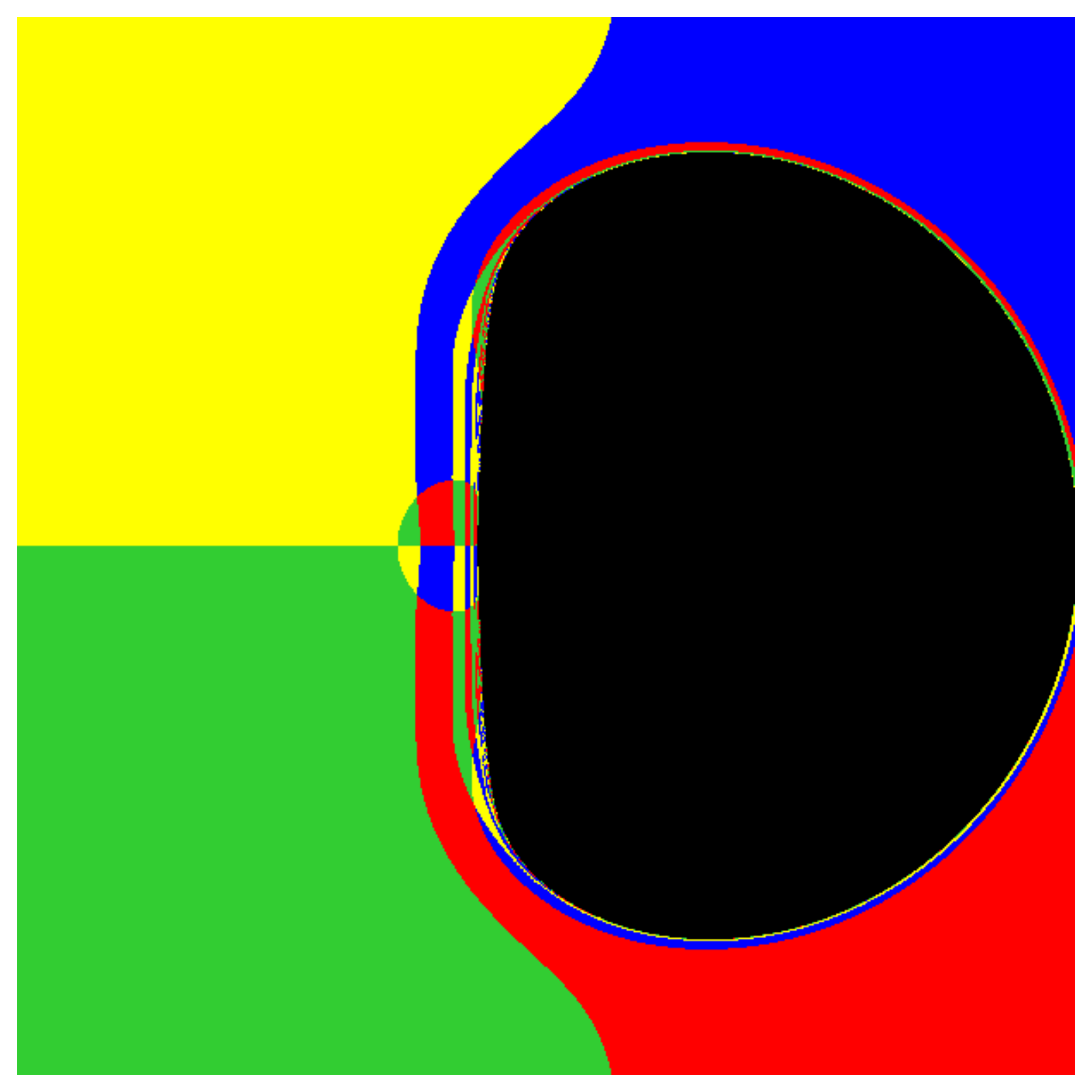}
\caption{$\theta_0=90^\circ$}
\end{subfigure}
\begin{subfigure}{0.48\textwidth}
\includegraphics[width=\textwidth]{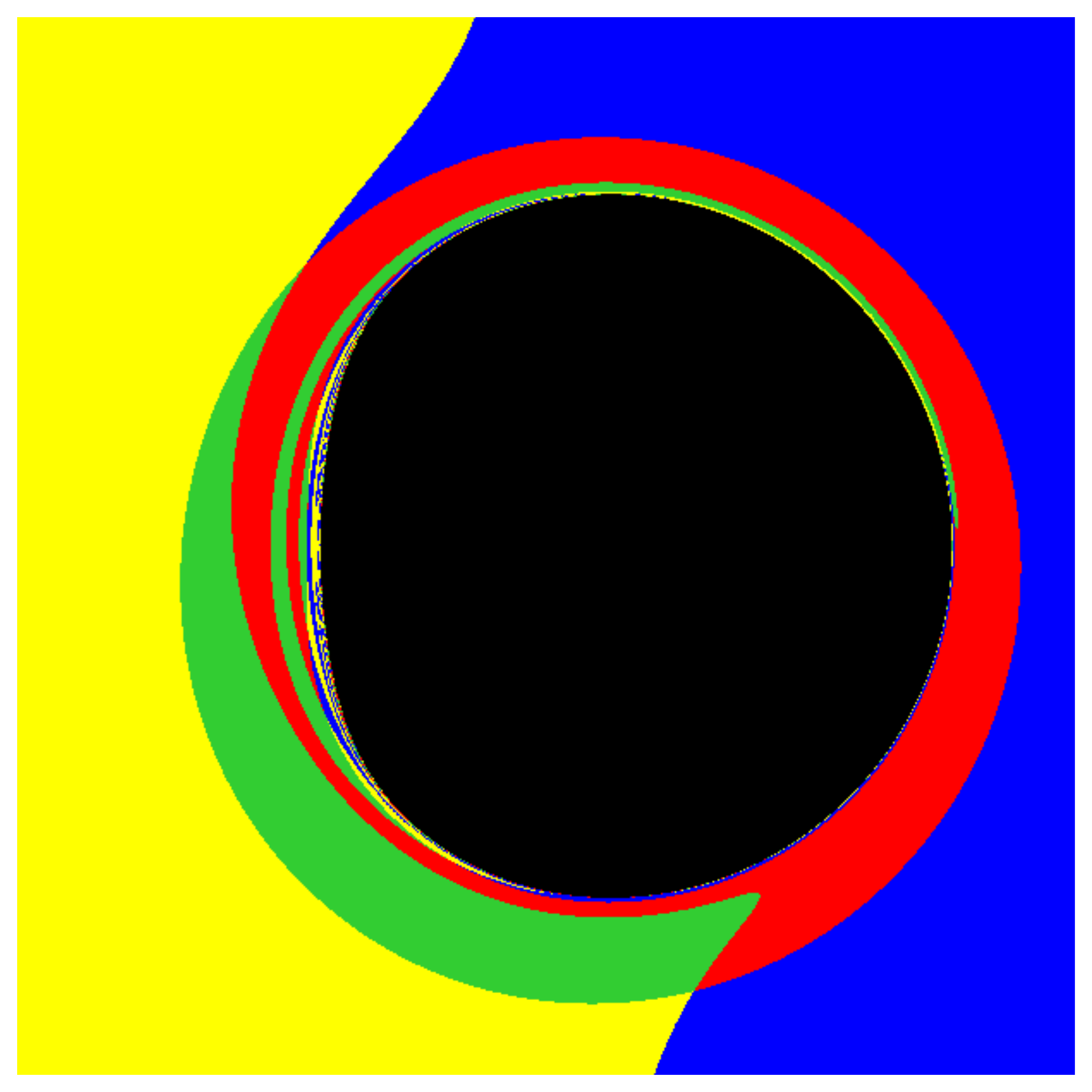}
\caption{$\theta_0=15^\circ$}
\end{subfigure}
\caption{Four-color screen image of the Johannsen black hole with $a/M=0.94$ and $\alpha_{22} = 3.65$, at inclinations $\theta_0=90^\circ$ and $\theta_0=15^\circ$.}
\label{fig:Johbasicfig}
\end{figure}

Consider $a/M=0.94$, $\alpha_{13}=\alpha_{52}=\epsilon_3=0$. When $\alpha_{22}=0$, this is simply a Kerr black hole shown in Fig.~\ref{fig:Kerrexample}. We show the four-color screen image for $\alpha_{22} = 3.65$ in Fig.~\ref{fig:Johbasicfig}; this is close to the upper bound (\ref{eq: single bounds a13 a22}), which is $\alpha_{22}\leq 3.66172$.

\begin{figure}[htpb]\centering
\begin{subfigure}{0.48\textwidth}
\includegraphics[width=\textwidth]{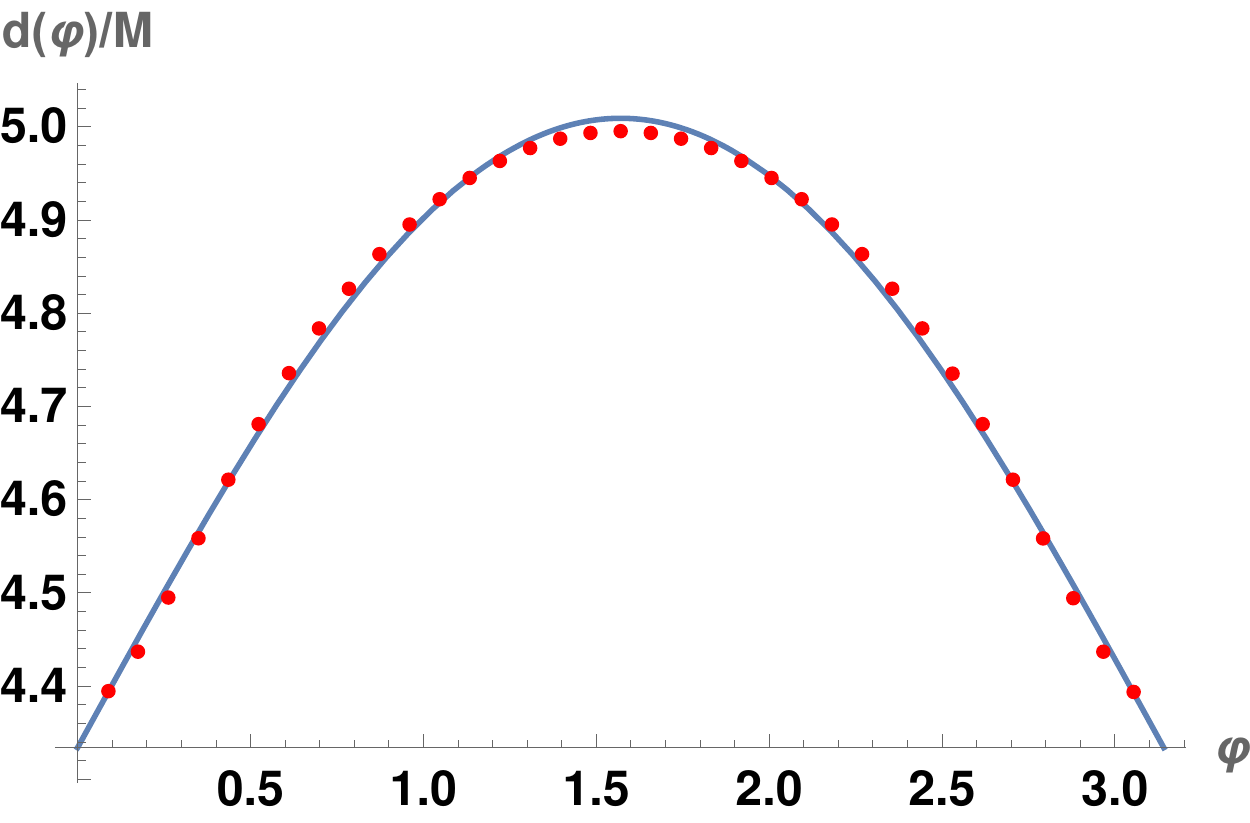}
\caption{$d(\varphi)$ fit original data}
\end{subfigure}
\begin{subfigure}{0.48\textwidth}
\includegraphics[width=\textwidth]{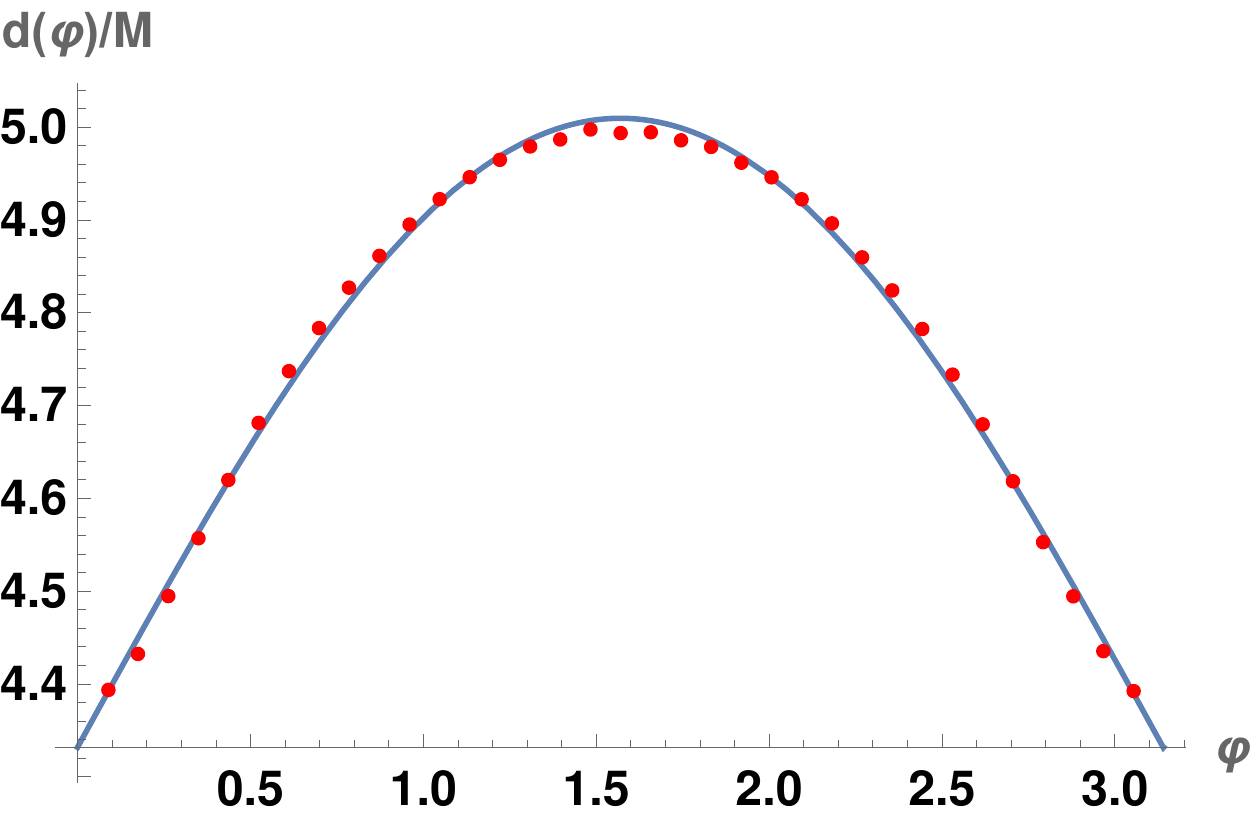}
\caption{$d(\varphi)$ fit noisy data}
\end{subfigure}
\begin{subfigure}{0.48\textwidth}
\includegraphics[width=\textwidth]{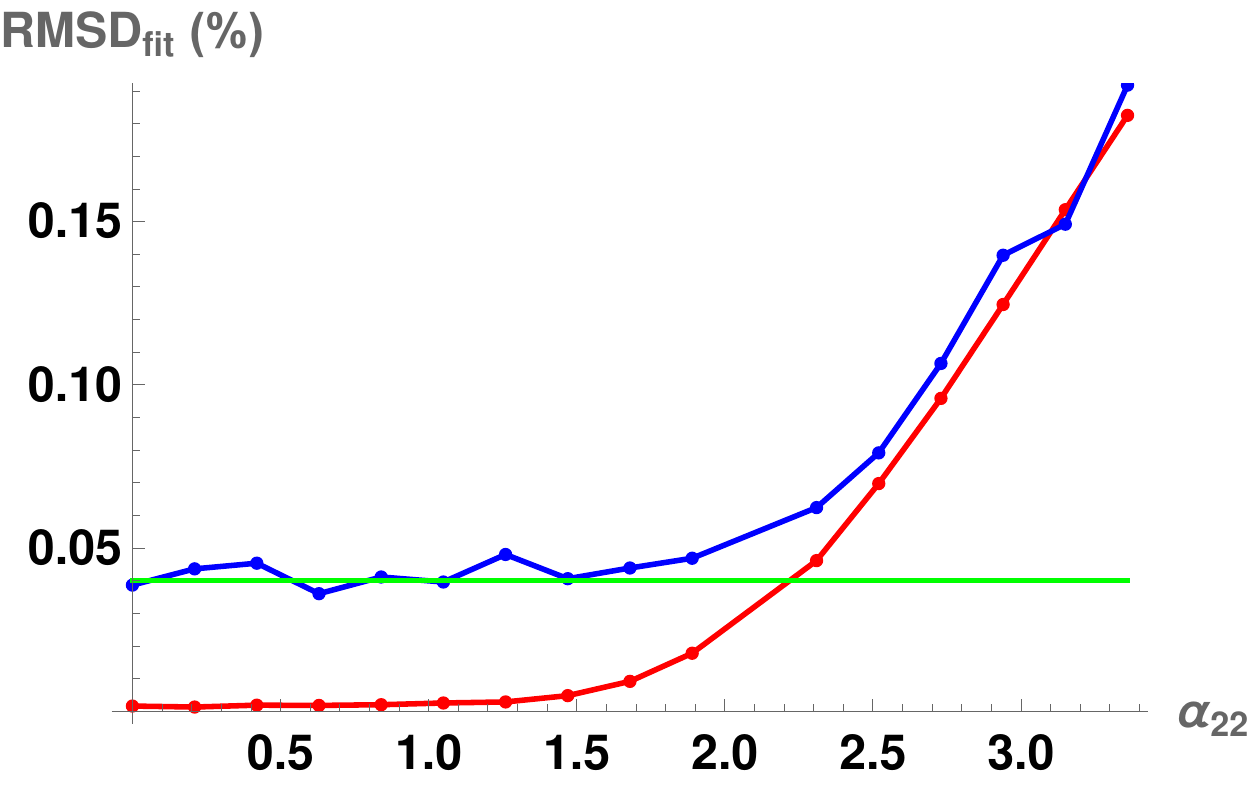}
\caption{$\theta_0=17^\circ$ fixed, varying $\alpha_{22}$.}
\end{subfigure}
\begin{subfigure}{0.48\textwidth}
\includegraphics[width=\textwidth]{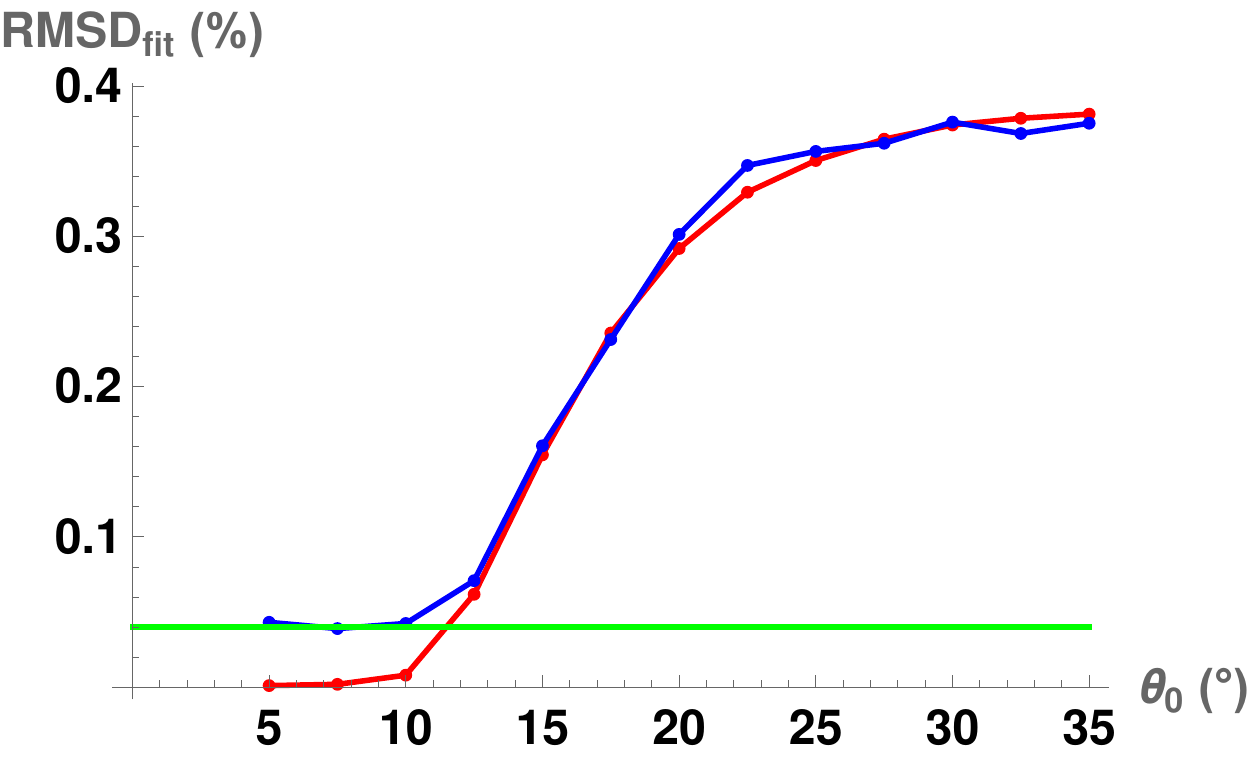}
\caption{$\alpha_{22}=3.65$ fixed, varying $\theta_0$.}
\end{subfigure}
\caption{The circlipse fit for the Johanssen black hole with $a/M=0.94$. For $\alpha_{22}=3.65$, the top two panels illustrate the circlipse fit for the projected diameter $d(\varphi)$ for the original data (upper-left) and mock-noisy data (upper-right); the line gives the best-fit circlipse function. The lower panels show the fit uncertainty $\RMSD$ for the circlipse fit for varying $\alpha_{22}$ and fixed $\theta_0=17^\circ$ (lower-left), and for fixed $\alpha_{22}=3.65$ and varying $\theta_0$ (lower-right). The red dots and line indicate the circlipse fit RMSD for the original data; the blue dots and line indicate the RMSD for the mock-noisy data. The green line indicates the pure Kerr baseline of $\RMSD=0.04\%$.}
\label{fig:Johshapefits}
\end{figure}

The projected diameter for the critical curve of this Johannsen black hole (with $\alpha_{22}=3.65$) at inclination $\theta_0=15^\circ$ is shown in the upper panels of Fig.~\ref{fig:Johshapefits}. The circlipse fit is now significantly worse than for Kerr, with $\RMSD= 0.22\%$ for both the original and mock-noisy data.

It is illustrative to show the evolution of the fit error $\RMSD$ of the best-fit circlipse function for different values of $\alpha_{22}$ at fixed inclination $\theta_0=17^\circ$, and also the evolution for different inclinations for fixed $\alpha_{22}=3.65$, see the lower panels of Fig.~\ref{fig:Johshapefits}. The Kerr circlipse remains a good fit for the projected diameter for low inclinations and low $\alpha_{22}$. The higher the inclination, the smaller the value of $\alpha_{22}$ that is necessary to achieve an RMSD that is higher than that of Kerr. Around $\theta_0 = 35^\circ$ (for $\alpha_{22}=3.65$), the RMSD is still only about ten times worse than for Kerr, i.e.\ only $\RMSD \sim 0.4\%$. It is clear that the projected diameter for the Johannsen metric generally only deviates modestly from the circlipse function.

\subsubsection{Rasheed-Larsen}
\begin{figure}[htpb]\centering
\begin{subfigure}{0.48\textwidth}
 \includegraphics[width=\textwidth]{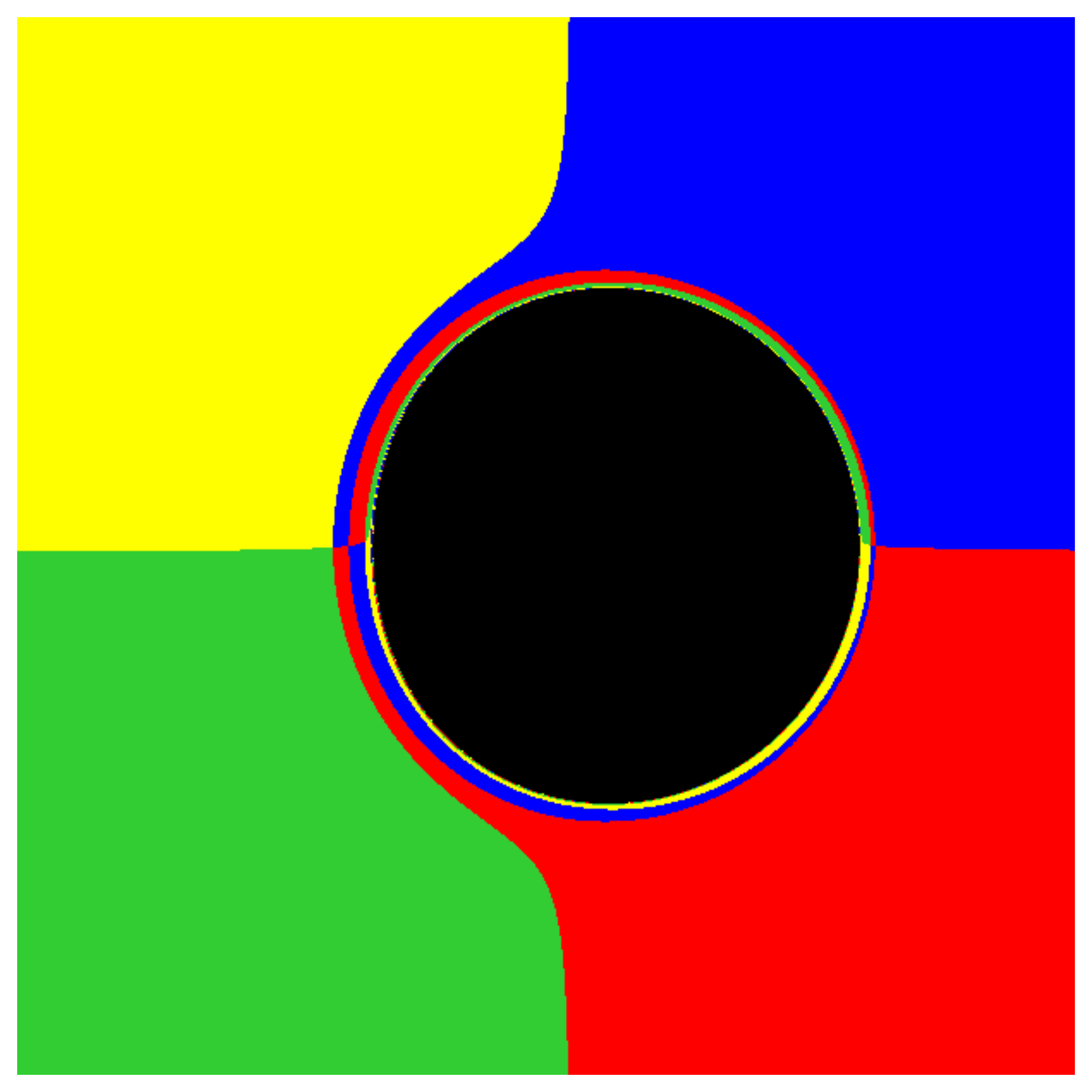}
 \caption{$\theta_0=90^\circ$}
 \end{subfigure}
 \begin{subfigure}{0.48\textwidth}
 \includegraphics[width=\textwidth]{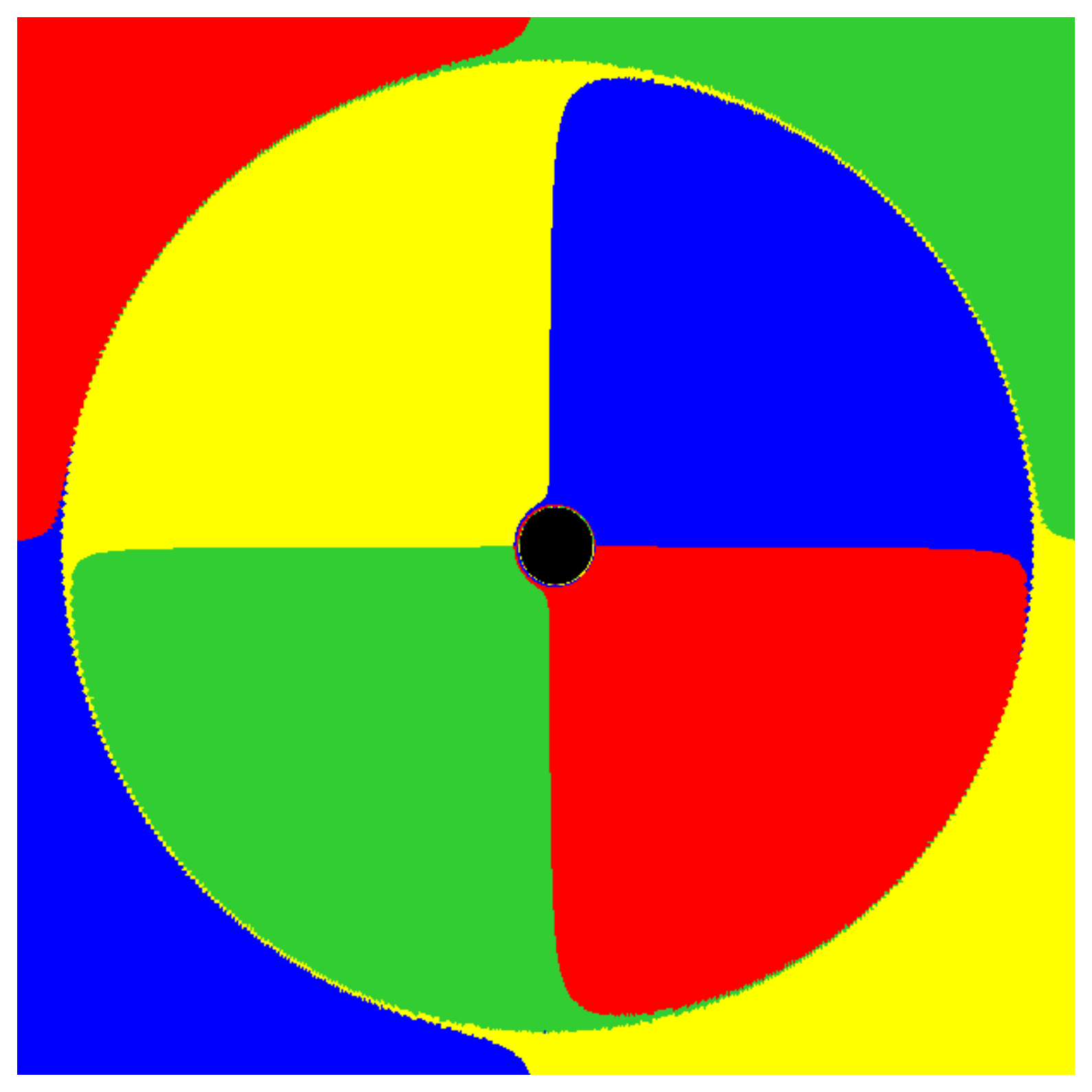}
 \caption{$\theta_0=90^\circ$ (zoomed out)}
 \end{subfigure}
 \begin{subfigure}{0.48\textwidth}
  \includegraphics[width=\textwidth]{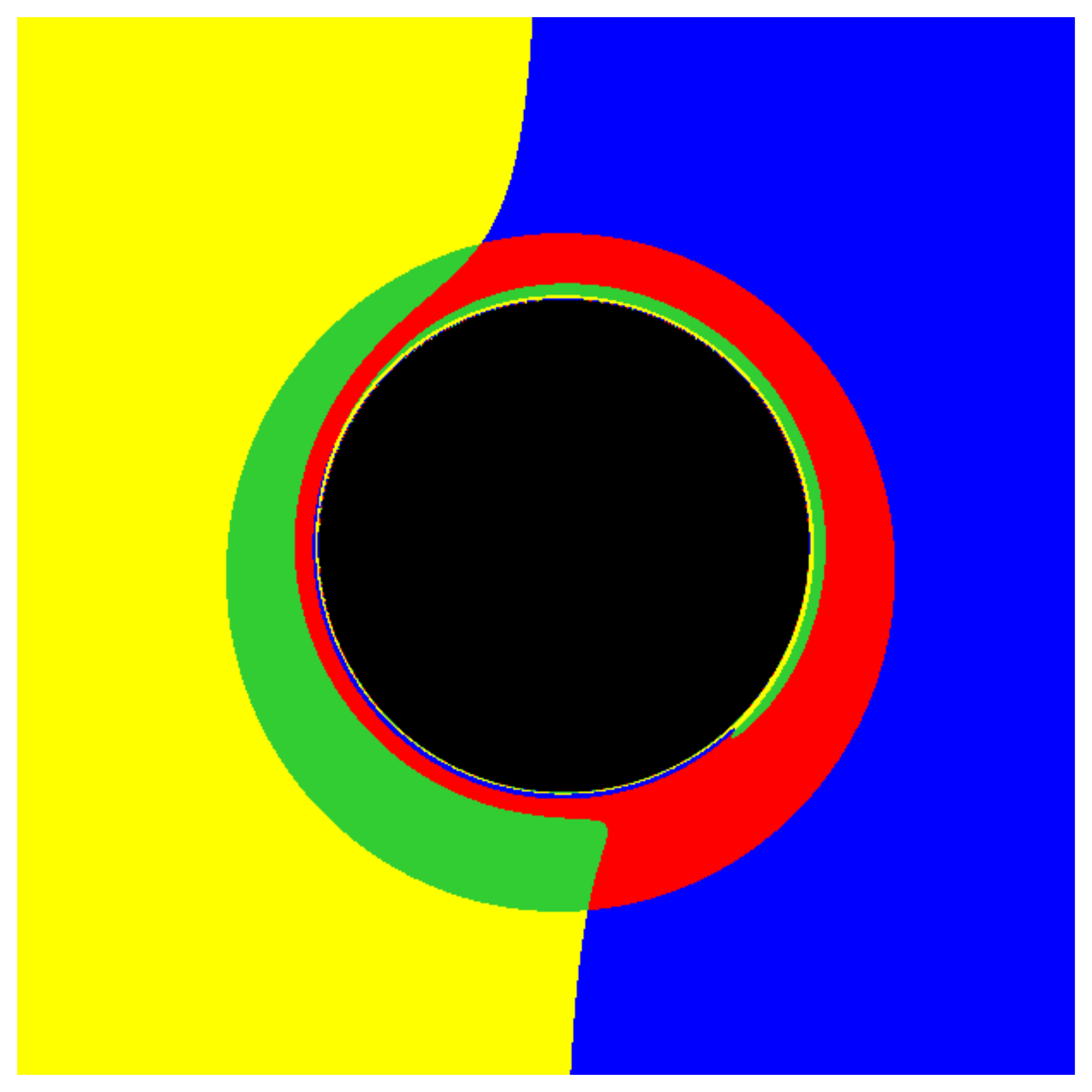}
  \caption{$\theta_0=15^\circ$}
  \end{subfigure}
\caption{Four-color screen images of the Rasheed-Larsen black hole with parameters given by (\ref{eq:RLmaxasymparams}), at inclinations $\theta_0=90^\circ$ and $\theta_0=15^\circ$. The critical curve is equatorially symmetric, but the image breaks this symmetry. This asymmetry is very slight and is most visible in the zoomed-out image at $\theta_0=90^\circ$ (with screen size $(100M)^2$ instead of $(15M)^2$).}
\label{fig:RLbasicfig}
\end{figure}

We consider the following  configuration (given in units of the black hole's mass):
\be \label{eq:RLmaxasymparams}
\begin{aligned}
m&=0.36M,&  a&=0.30M,\\
p&=3.06M, & q&=0.94M,
\end{aligned}\ee
which corresponds to the physical charges
\be  J=0.3M^2,\quad P=1.3M,\quad Q=0.15M.\ee
The metric breaks equatorial symmetry; in particular the lowest-order odd-parity multipoles (that vanish for Kerr) are (see (\ref{eq:RLmultipoles}))
\be \frac{S_2}{M^3} = -0.06, \quad \frac{M_3}{M^4} = 0.02 .\ee
This choice of parameters comes close to maximizing the equatorial symmetry-breaking multipole $S_2/M^3$, so it represents in some sense a (near-)maximal deviation from Kerr that can be achieved with the Rasheed-Larsen black hole.  However, note that since null geodesics are still integrable, the critical curve is still equatorially symmetric \cite{Cunha:2018uzc}; see Fig.~\ref{fig:RLbasicfig}.

The configuration (\ref{eq:RLmaxasymparams}) is also used in forthcoming general-relativistic magnetohydrodynamic (GRMHD) simulations of plasma accretion onto a Rasheed-Larsen black hole, and the resulting photon rings based on the emission from the fluid \cite{RLGRMHD}; our analysis provides a useful point of comparison to the more detailed analysis therein.

\begin{figure}[htpb]\centering
\includegraphics[width=0.45\textwidth]{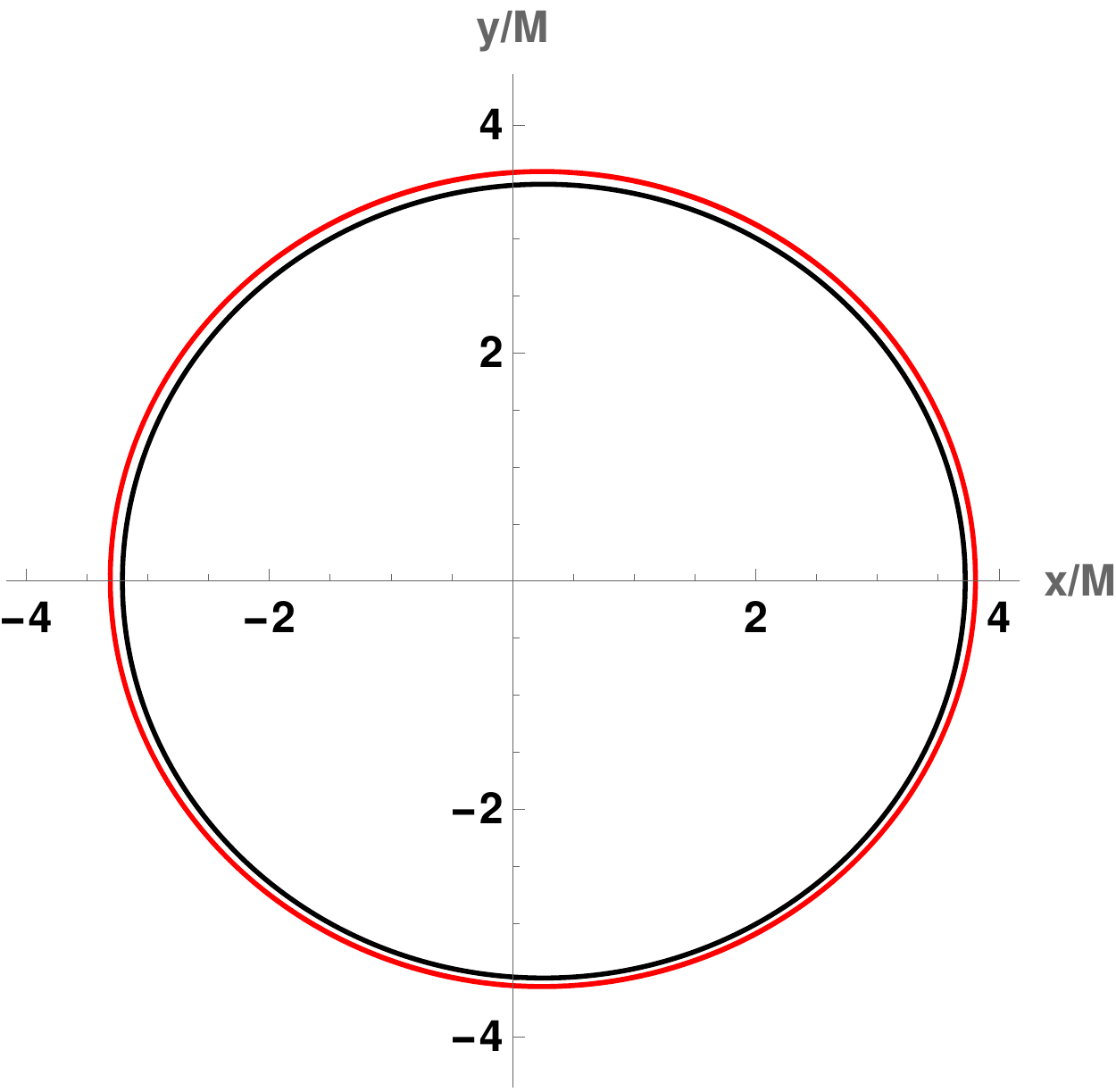}
\caption{The critical curve (black) and $n=2$ ring (red) of the Rasheed-Larsen black hole with parameters (\ref{eq:RLmaxasymparams}) at inclination $\theta_0=15^\circ.$}
\label{fig:RLshadowring}
\end{figure}

Since the metric is not equatorially symmetric, but the critical curve still is, an interesting question is whether the breaking of equatorial symmetry can be apparent in the $n=2$ ring. With the naked eye, there certainly does not seem to be any hint of this symmetry breaking in the $n=2$ ring, see Fig.~\ref{fig:RLshadowring}.
Our analysis below confirms the lack of signature from the symmetry breaking quantitatively; the $n=2$ ring is already too ``close'' to the critical curve to show any measurable deviation to the critical curve's equatorial symmetry.

\begin{figure}[htpb]\centering

\begin{subfigure}{0.48\textwidth}
\includegraphics[width=\textwidth]{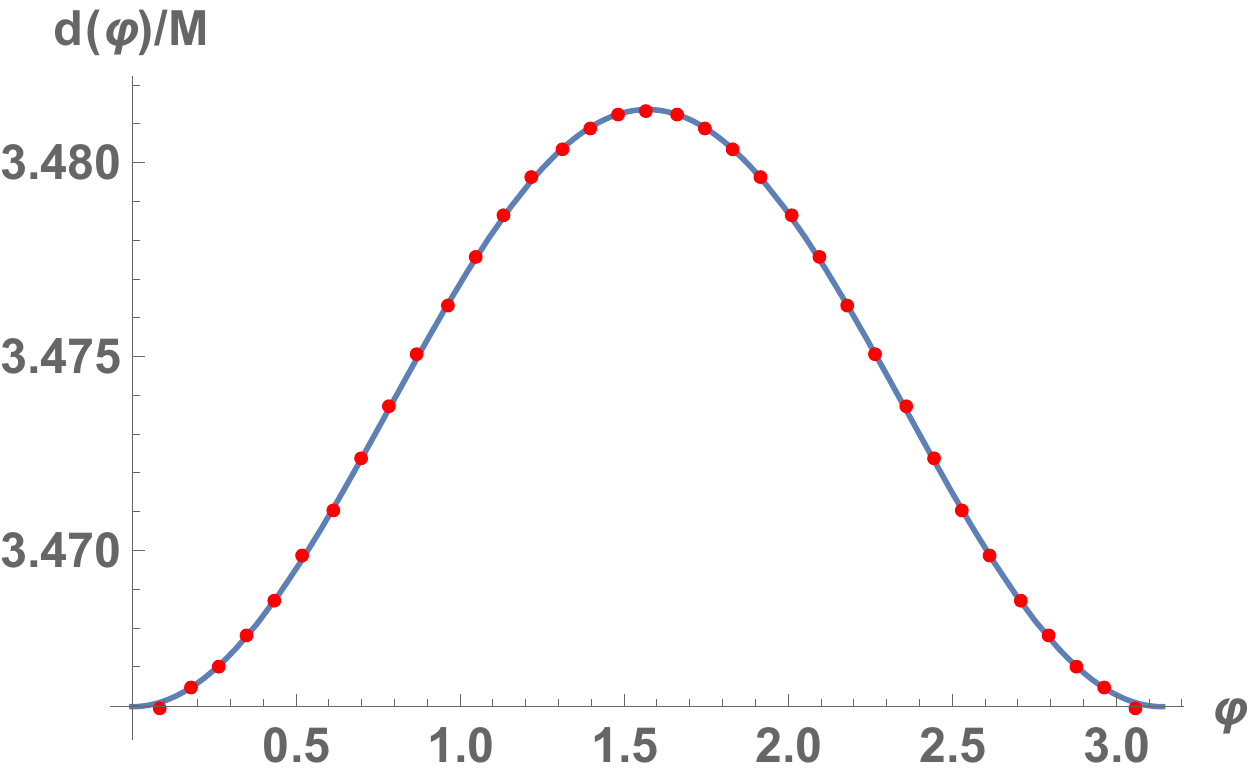}
\caption{Original critical curve data fit}
\end{subfigure}
\begin{subfigure}{0.48\textwidth}
\includegraphics[width=\textwidth]{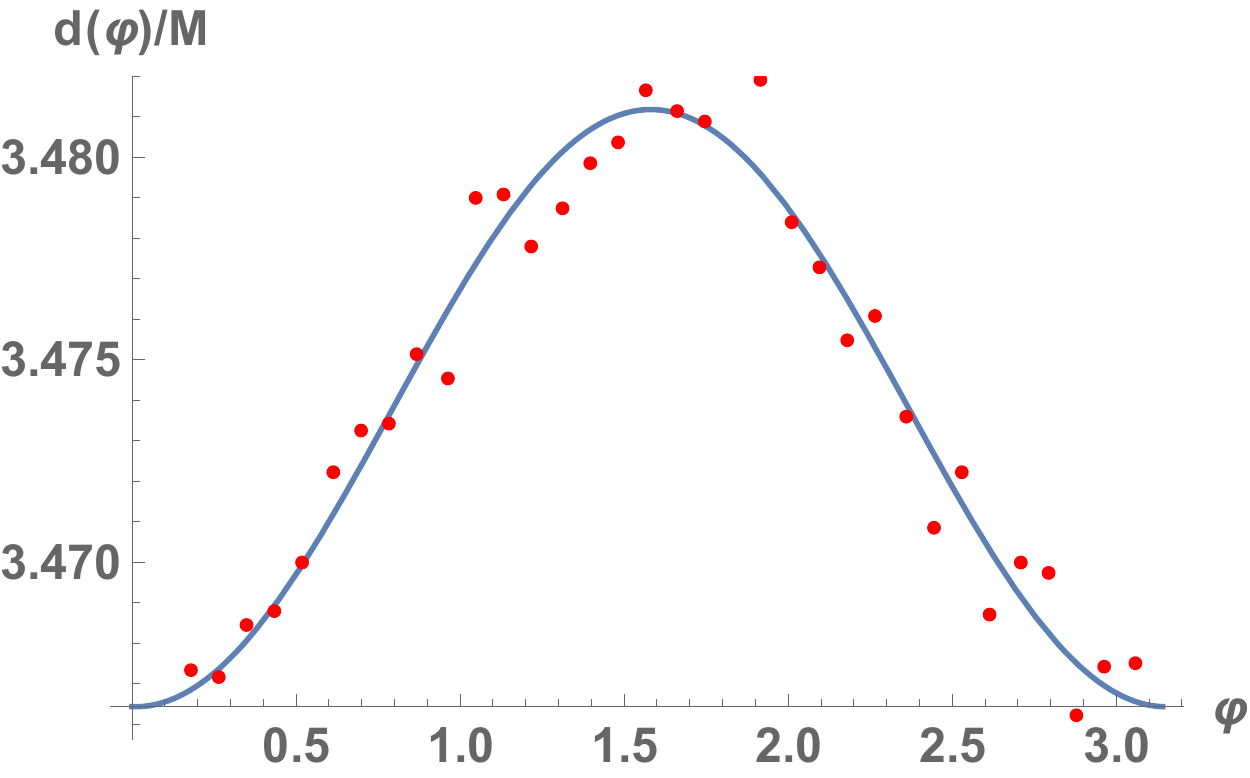}
\caption{Mock-noisy critical curve data fit}

\end{subfigure}
\begin{subfigure}{0.48\textwidth}
\includegraphics[width=\textwidth]{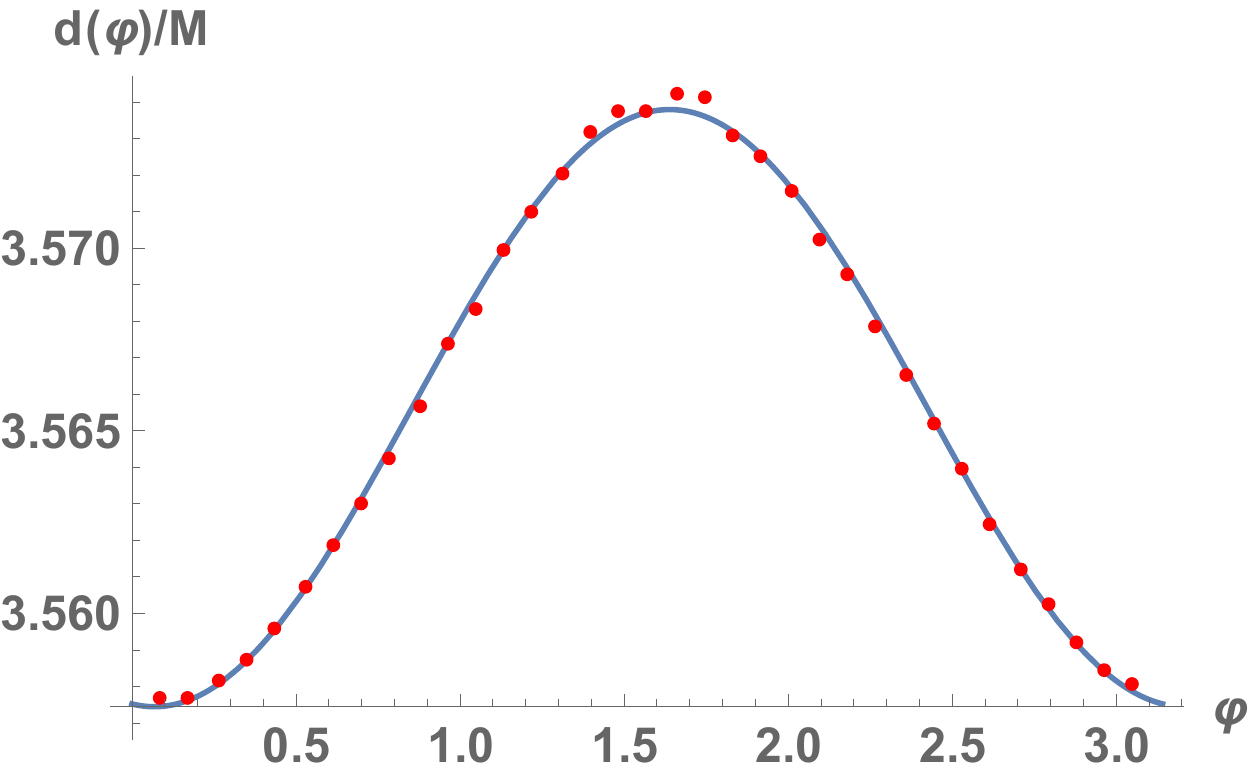}
\caption{Original $n=2$ ring data fit}
\end{subfigure}
\begin{subfigure}{0.48\textwidth}
\includegraphics[width=\textwidth]{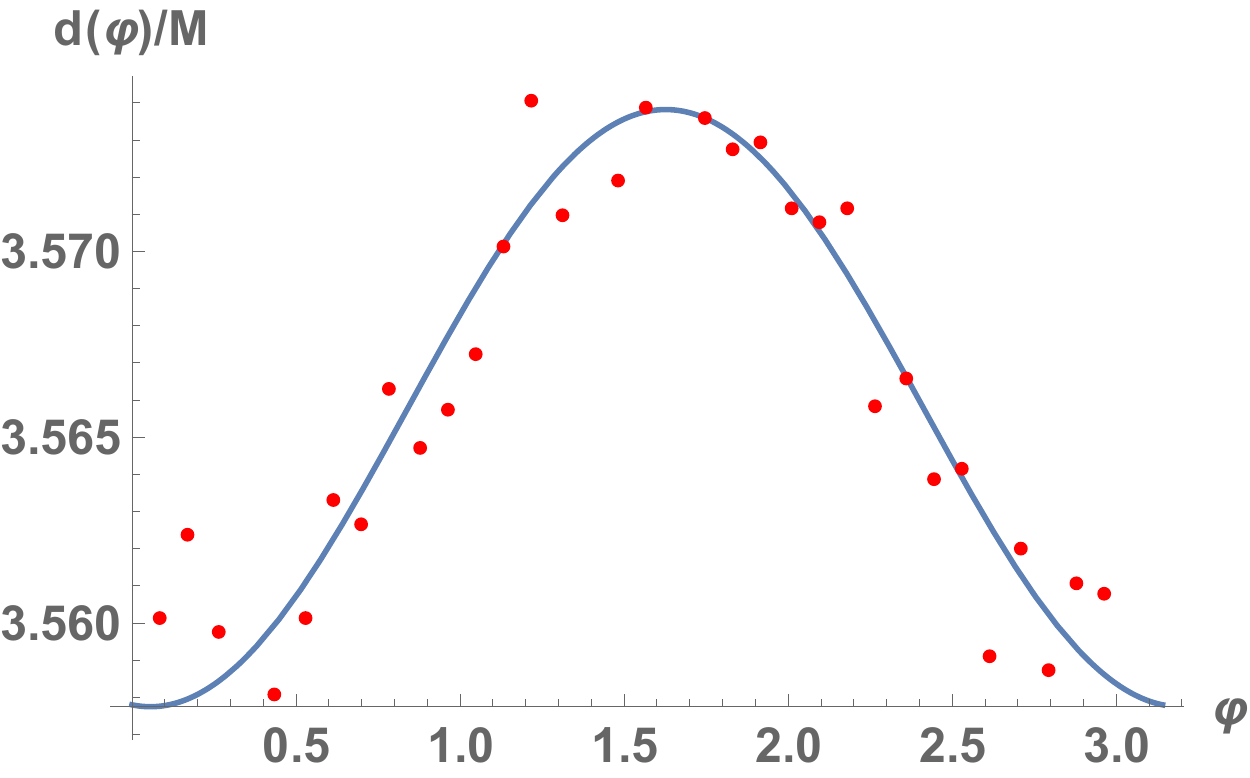}
\caption{Mock-noisy $n=2$ ring data fit}

\end{subfigure}
\begin{subfigure}{0.48\textwidth}
\includegraphics[width=\textwidth]{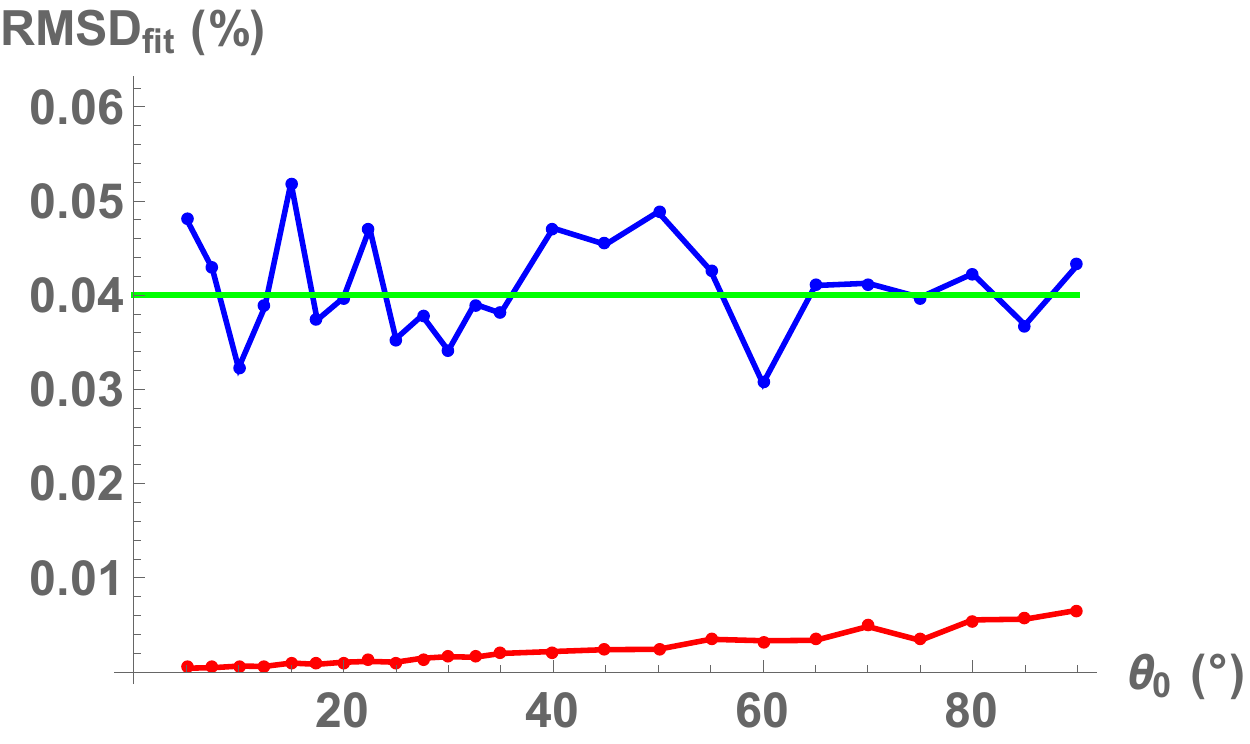}
\caption{Critical curve fit RMSDs}
\end{subfigure}
\begin{subfigure}{0.48\textwidth}
\includegraphics[width=\textwidth]{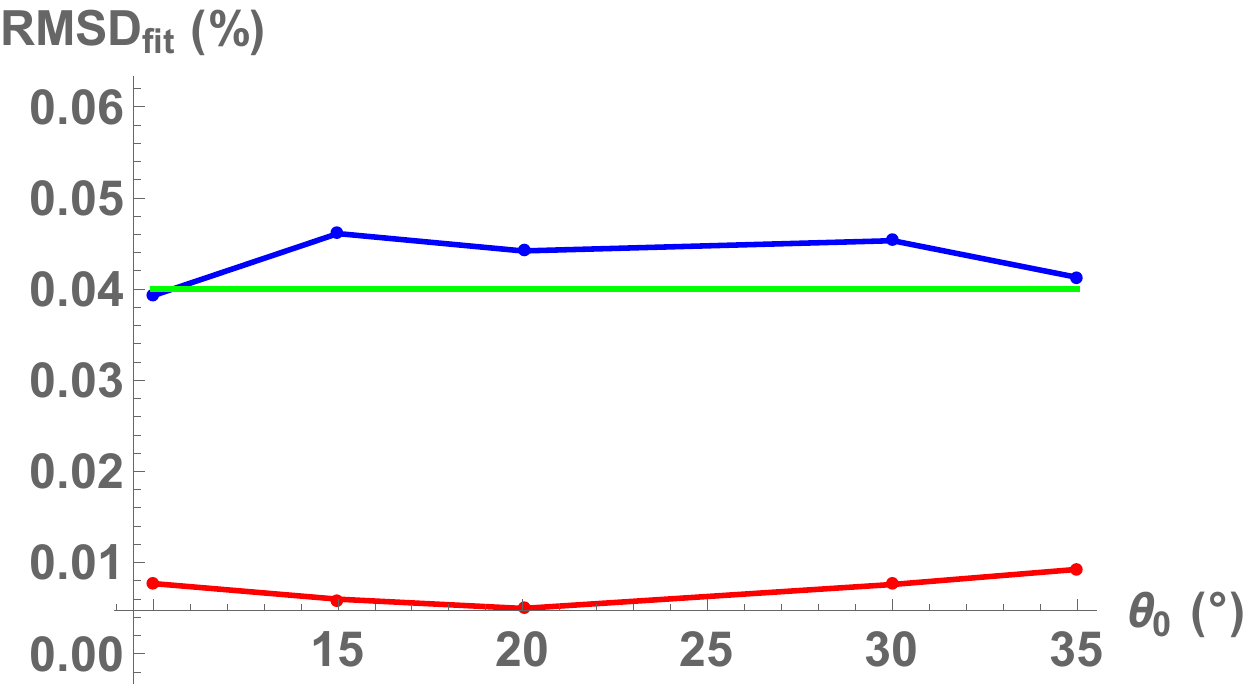}
\caption{Ring fit RMSDs}
\end{subfigure}
\caption{The circlipse fit illustrated for the Rasheed-Larsen black hole with parameters (\ref{eq:RLmaxasymparams}). We fit the critical curve (left panels) and $n=2$ ring (right panels). The top four panels show the projected diameters and circlipse fit at $\theta_0=15^\circ$. The bottom two panels show the fit uncertainty $\RMSD$ at varying inclination $\theta_0$. The red dots and line indicate the circlipse fit RMSD for the original data; the blue dots and line indicate the RMSD for the mock-noisy data. The green line indicates the pure Kerr baseline of $\RMSD=0.04\%$. The critical curve is fit all the way to $\theta_0=90^\circ$ (bottom-left panel), showing that the Rasheed-Larsen black hole always approaches the circlipse shape as well as Kerr does.}
\label{fig:RLshapefits}
\end{figure}

The projected diameters for the critical curve and $n=2$ photon ring at inclination $\theta_0=15^\circ$ are fit very well by the circlipse curve function (\ref{eq:circlipsedef}); see the top four panels of Fig.~\ref{fig:RLshapefits}. The corresponding RMSDs are $\RMSD = 0.0015\%$ and $\RMSD = 0.031\%$, respectively for the original and mock-noisy data of the critical curve, and $\RMSD = 0.006\%$ and $\RMSD = 0.051\%$ for the original and mock-noisy data of the $n=2$ ring.

Clearly, at $\theta_0=15^\circ$, the critical curve and $n=2$ ring cannot distinguish between the Rasheed-Larsen and Kerr black holes. In the bottom two panels of Fig.~\ref{fig:RLshapefits}, we show that this is also true at other inclinations.\footnote{Note that Fig.~\ref{fig:RLshapefits} shows the critical curve at all inclinations; we only show the results for the $n=2$ ring at inclination $\theta_0\leq 35^\circ$ to avoid the possible numerical issues described in Appendix \ref{app: rings asymm spacetime}.} It is simply not possible to distinguish between Kerr or Rasheed-Larsen using only the shape of the critical curve or $n=2$ ring.

\subsubsection{Manko-Novikov}\label{sec:MNshape}

In the analysis of the Manko-Novikov black hole, we will consider $a/M = 0.94$ and turn on the dimensionless asymmetry parameter to $\alpha_3 = 13$.

\begin{figure}[htpb]\centering

\begin{subfigure}{0.48\textwidth}
\includegraphics[width=\textwidth]{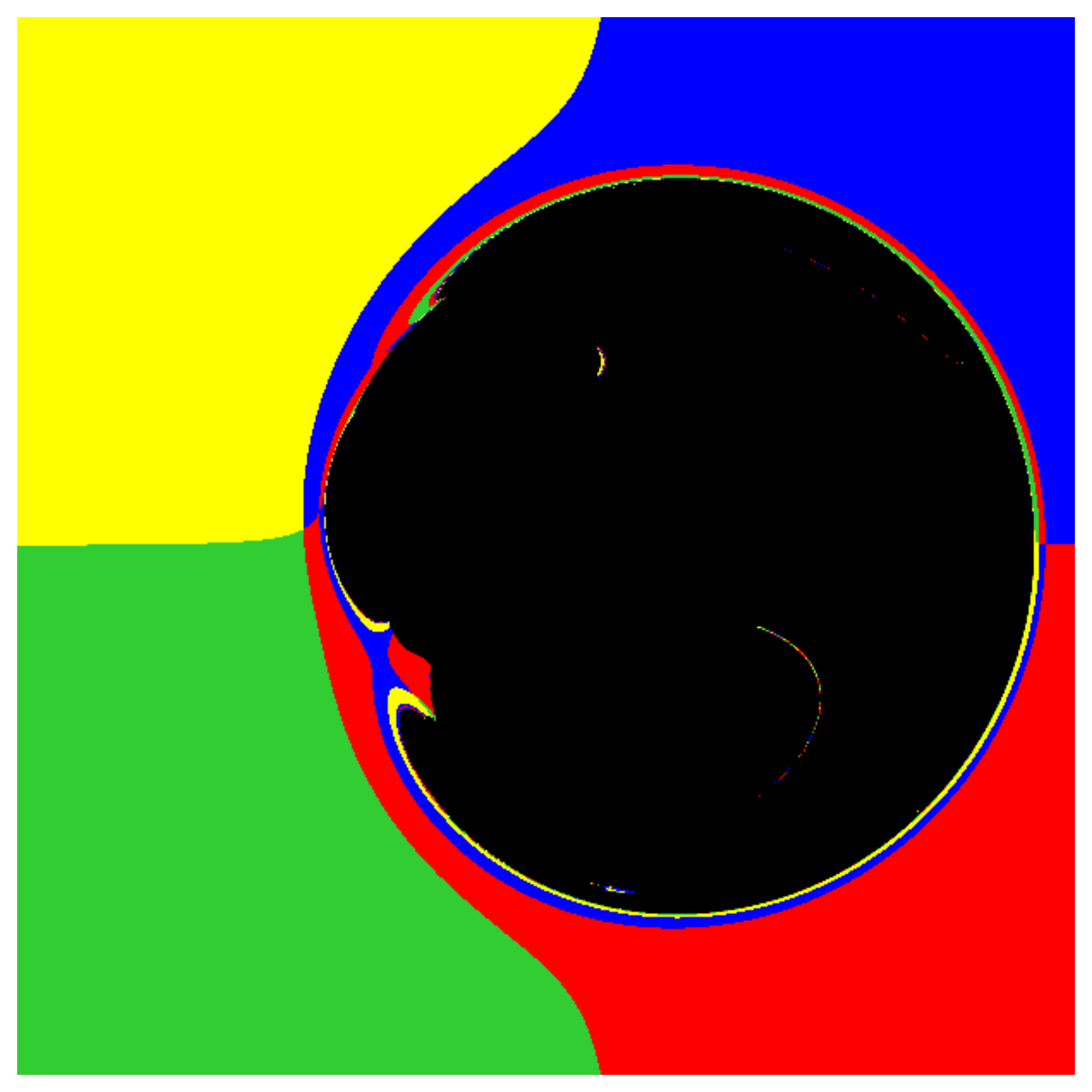}
\caption{Four-color screen ($\theta_0=90^\circ$)}
\end{subfigure}
\begin{subfigure}{0.48\textwidth}
\includegraphics[width=\textwidth]{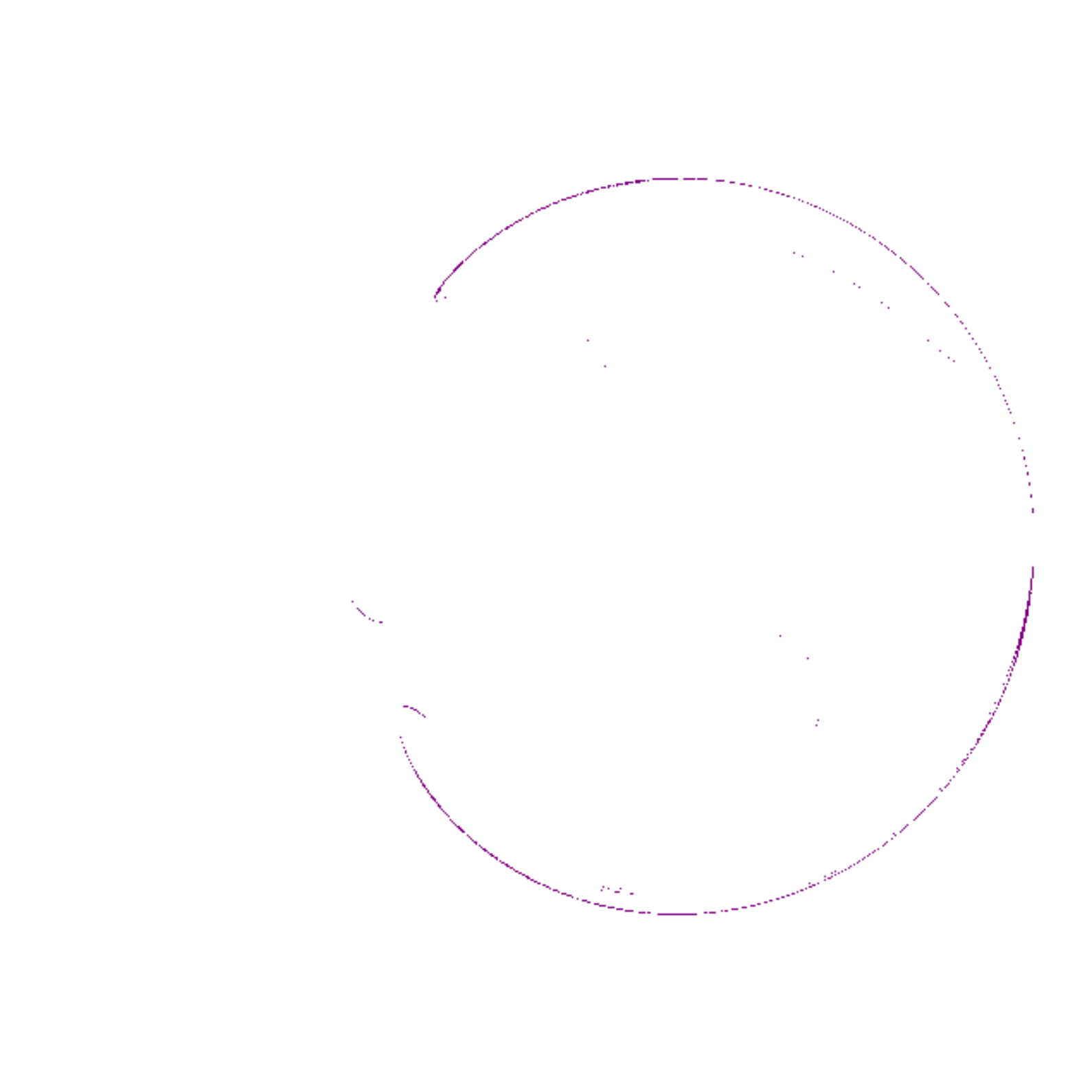}
\caption{$n=2$ photon ring ($\theta_0=90^\circ$)}
\end{subfigure}
\begin{subfigure}{0.48\textwidth}
\includegraphics[width=\textwidth]{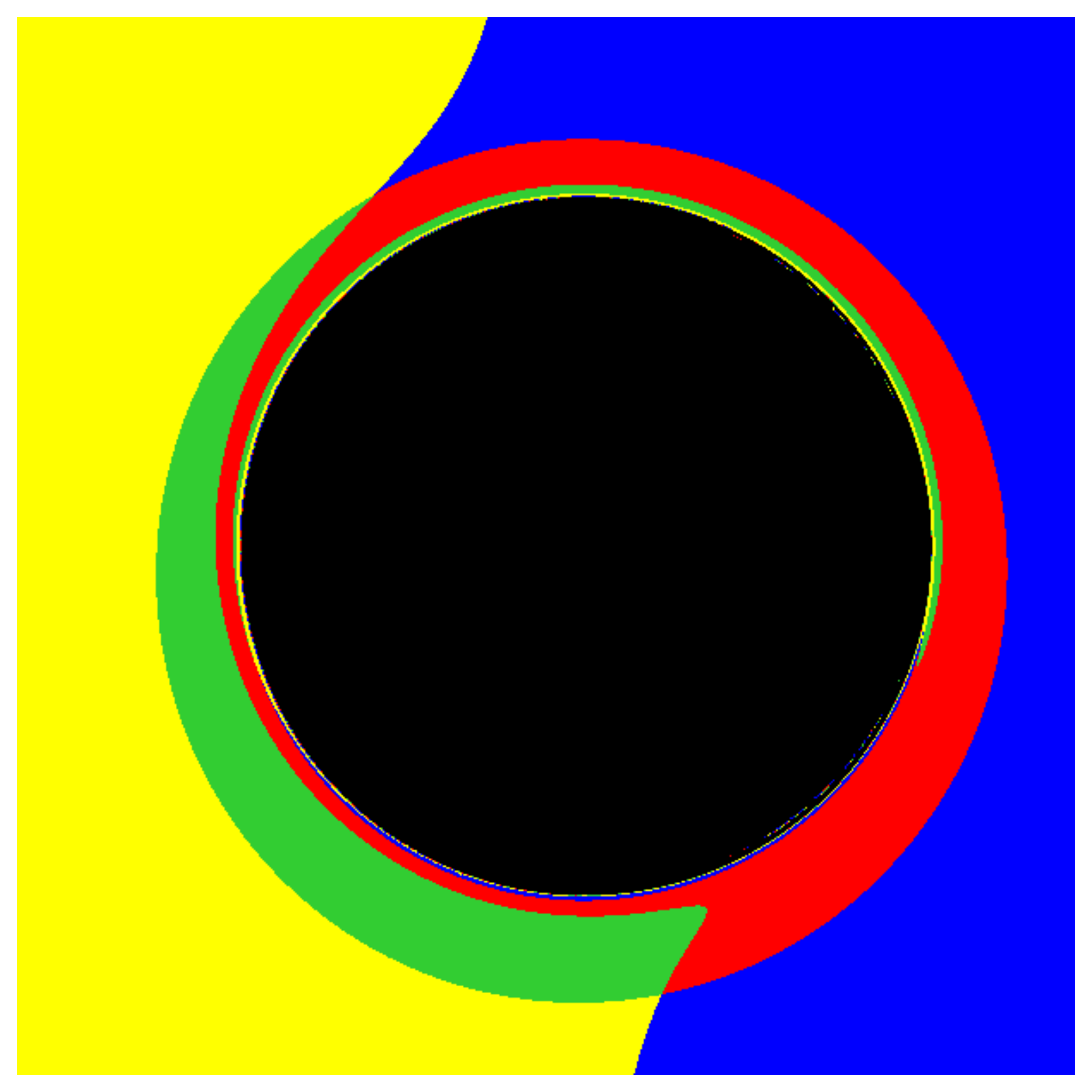}
\caption{Four-color screen ($\theta_0=15^\circ$)}
\end{subfigure}
\begin{subfigure}{0.48\textwidth}
\includegraphics[width=\textwidth]{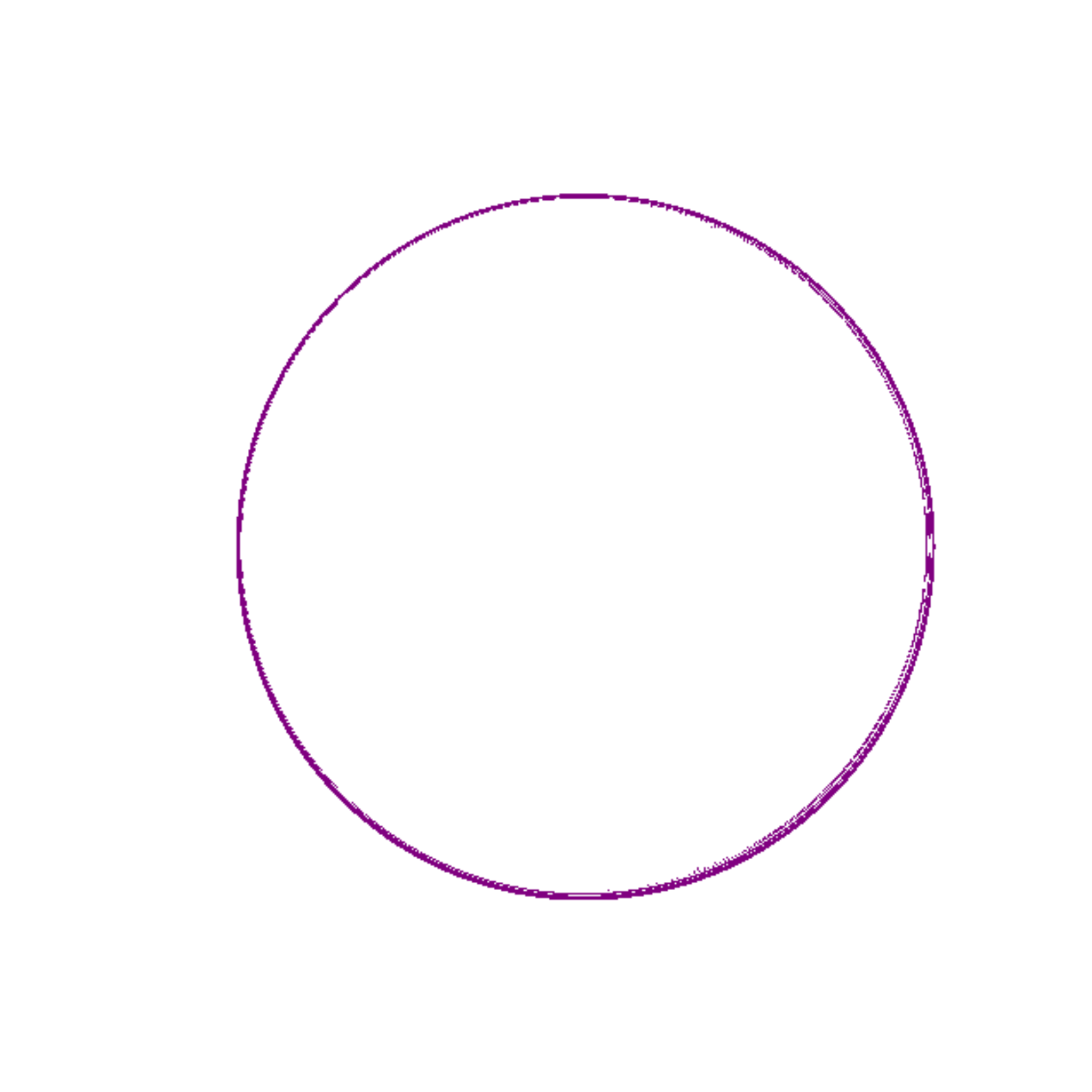}
\caption{$n=2$ photon ring ($\theta_0=15^\circ$)}
\end{subfigure}
\caption{Four-color screen images (left) of the Manko-Novikov black hole with $a/M = 0.94$ and the deviation parameter $\alpha_3 = 13$, and its $n=2$ photon ring (right), at inclinations $\theta_0=90^\circ$ (top) and $\theta_0=15^\circ$ (bottom).}
\label{fig:MNbasicfig}
\end{figure}

Null geodesics for the Manko-Novikov black hole are not separable or integrable, so the usual semi-analytical way of determining a critical curve as in Section \ref{sec: analytic shadow} is not applicable. Indeed, the lack of integrability leads to chaotic lensing phenomena \cite{Cunha:2015yba,Cunha:2016bjh,Bacchini:2021fig}, so that it is not even clear whether the notion of a critical curve (or photon ring)  is well-defined for images of such objects; see the four-color screens in Fig.~\ref{fig:MNbasicfig}.
Nevertheless, for relatively low inclinations $\theta_0\lesssim 35^\circ$, the photon rings can still be extracted from these images and their outer boundaries are to a very good approximation still connected, smooth curves. For higher inclinations, the photon ring starts to show ``holes'' and other disconnected phenomena, which are typical features for non-integrable metrics (see e.g.\ \cite{Cunha:2015yba,Cunha:2016bjh,Bacchini:2018zom}; this was also noticed for the Manko-Novikov metric when the quadrupole ($M_2$) bump is non-zero \cite{wang2018chaotic}); see Fig.~\ref{fig:MNbasicfig}. As can also been seen in Fig.~\ref{fig:MNbasicfig}, the Manko-Novikov metric breaks equatorial symmetry.

\begin{figure}[htpb]\centering
\begin{subfigure}{0.48\textwidth}
\includegraphics[width=\textwidth]{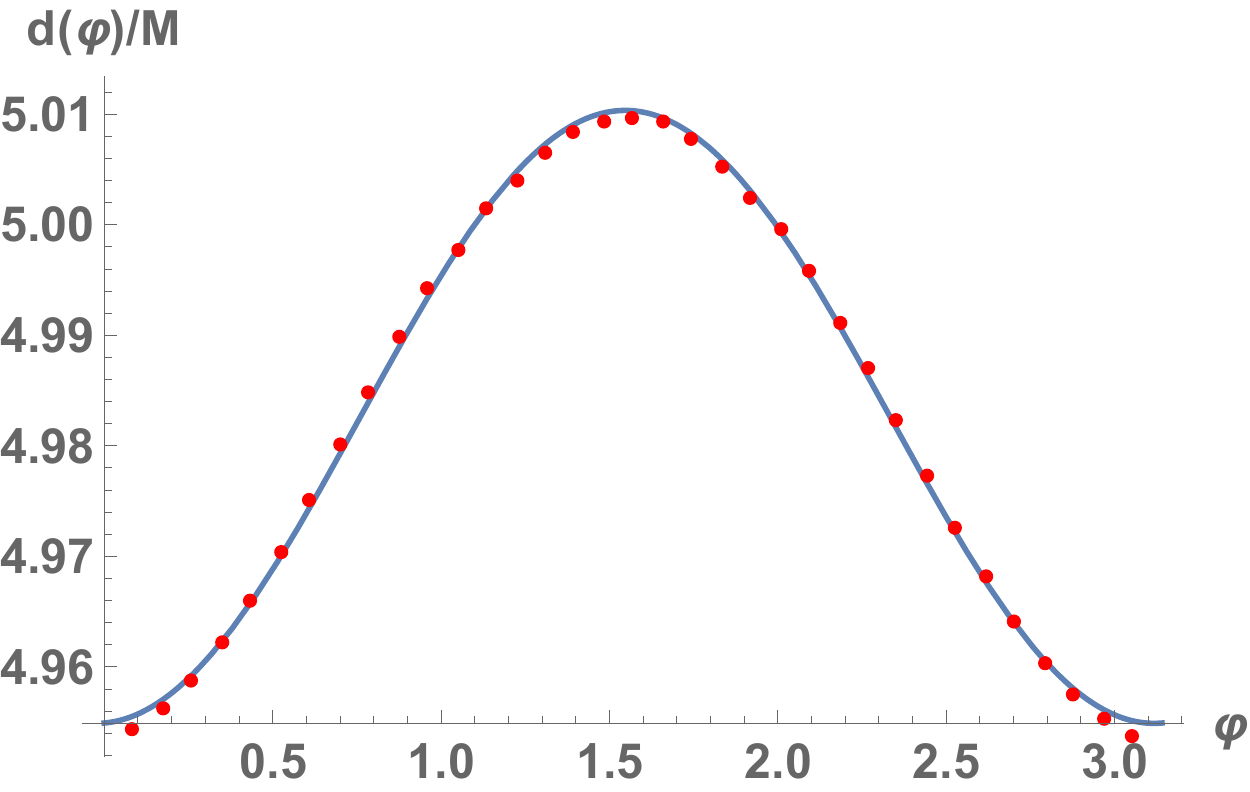}
\caption{Original $n=2$ ring data fit}
\end{subfigure}
\begin{subfigure}{0.48\textwidth}
\includegraphics[width=\textwidth]{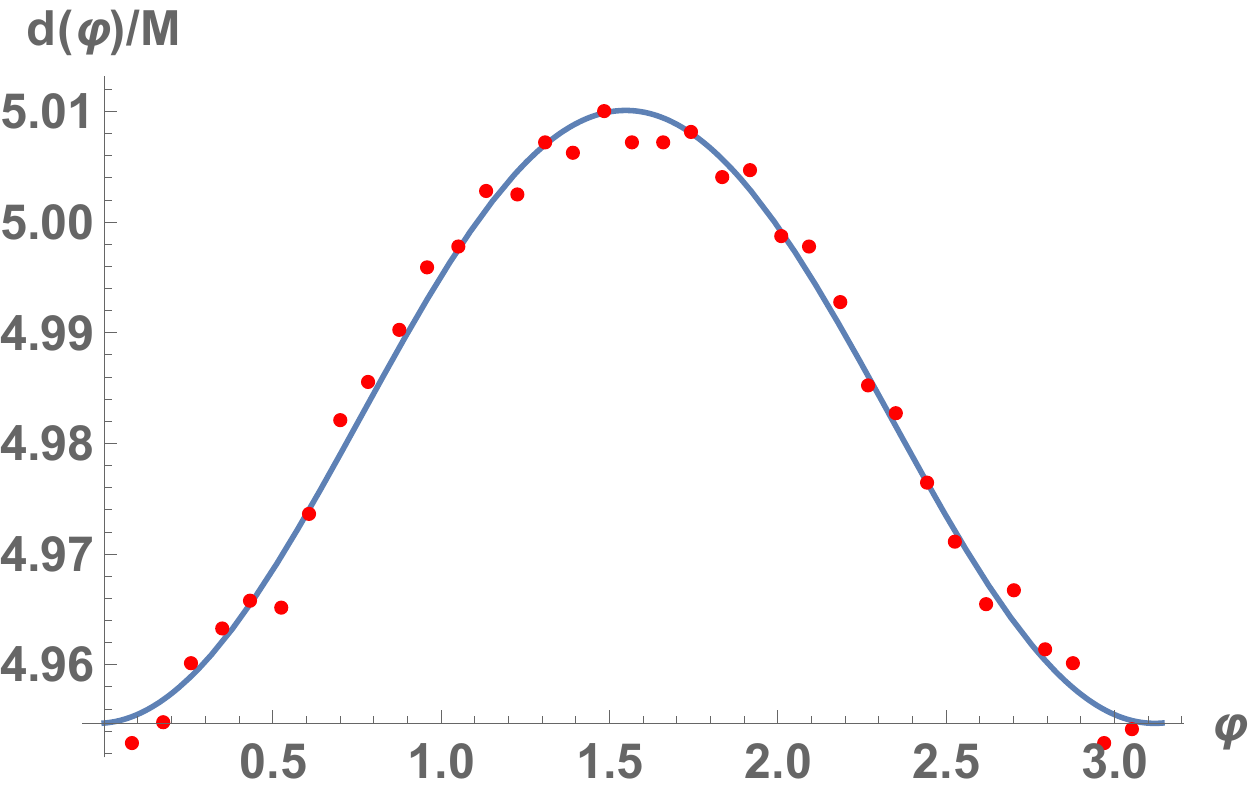}
\caption{Mock-noisy $n=2$ ring data fit}
\end{subfigure}
\begin{subfigure}{0.48\textwidth}
\includegraphics[width=\textwidth]{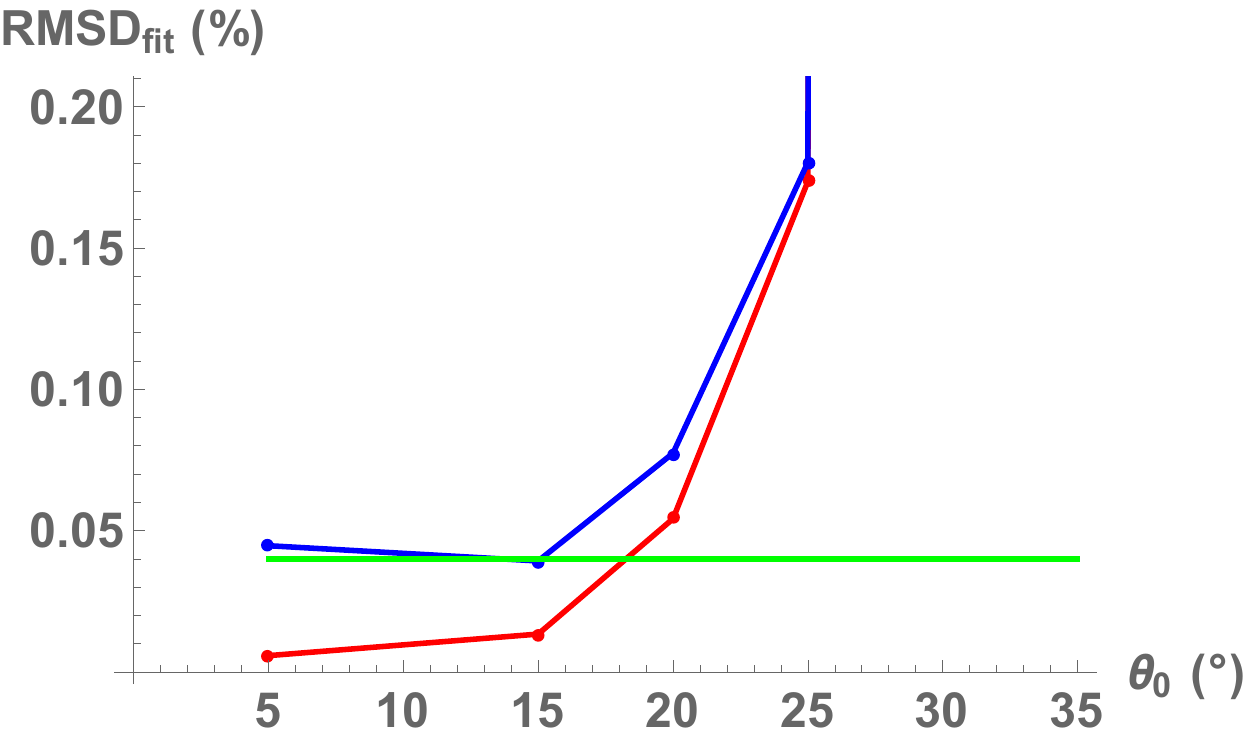}
\caption{RMSD varying $\alpha_3$ and fixed $\theta_0=17^\circ$}
\end{subfigure}
\caption{The circlipse fit for the $n=2$ rings of the Manko-Novikov black hole with parameters $a/M=0.94$. The top two panels show the projected diameter and circlipse fit for $\alpha_3=13$ at inclination $\theta_0=15^\circ$. The bottom panel shows the fit uncertainty $\RMSD$ of the circlipse fit for varying $\alpha_3$ and fixed $\theta_0=17^\circ$.
%, and for fixed $\alpha_{3}=13$ and varying $\theta_0$.
The red dots and line indicate the circlipse fit RMSD for the original data; the blue dots and line indicate the RMSD for the mock-noisy data. The green line indicates the pure Kerr baseline of $\RMSD=0.04\%$. Data points off the chart are $\RMSD = 16.1\%,22.0\%$ (original) and $\RMSD = 16.2\%,22.0\%$ (mock-noisy) for resp. $\theta_0 = 30^\circ,35^\circ$.}
\label{fig:MNshapefits}
\end{figure}

The projected diameters for the $n=2$ photon ring at inclination $\theta_0=15^\circ$ are still fit very well by the circlipse function ($\RMSD = 0.054\%,0.077\%$ for original and mock-noisy data, respectively); see the top panels Fig.~\ref{fig:MNshapefits}. However, at higher inclinations, the deviation from the circlipse quickly becomes more visible and the fit becomes worse; see the bottom panel of Fig.~\ref{fig:MNshapefits}. Once again, it is hard to distinguish the deviating black hole geometry from its Kerr limit at low inclinations.

\medskip

Our analysis in Section \ref{sec:shape} shows that the circlipse shape of the photon ring (and critical curve) is universal over a wide range of deviations from Kerr. A large inclination angle and/or a large deviation from Kerr is needed to have a measurable difference from the Kerr shape. Of course, such large deviations from Kerr would presumably show up (or have shown up already) in other observations as well. We can conclude that the photon ring does not distinguish very well between Kerr and other black holes. As we will see in the next section, the Lyapunov exponent is a more promising observable to see deviations from Kerr.

\clearpage

\section{Photon Ring Thickness \& Lyapunov Exponent}\label{sec:lyap}

The previous section shows how difficult it is to disentangle Kerr black holes from their more general counterparts, based on the shape of the critical curve or photon rings. Therefore, this section adresses an observable that could potentially help to distinguish the different metrics; the \textit{Lyapunov exponent} reflects the relative width (or total intensity) of subsequent photon rings, and strongly depends on the details of the metric. Furthermore, the Lyapunov exponent varies along the black hole critical curve, so that it represents an entire function worth of values that can be tested.

Consider a bound photon orbit at radius $r_B$. A \emph{nearly-}bound photon, initially at radius $r_B + \delta r_0$, will orbit around the black hole a number of times and evolve to radius $r_B+\delta r_n$ \cite{Johnson_2020}:
\begin{equation}\label{eq: exp growth of delta r}
    \delta r_n = e^{\gamma n}\delta r_0\,,
\end{equation}
where $n$ is the number of half-\textit{orbits} (defined below) the photon travels on. The exponent $\gamma$ is called the Lyapunov exponent. Equation (\ref{eq: exp growth of delta r}) holds as long as the photon remains on an orbit that is nearly bound, and is no longer true once the photon is well separated from the original bound orbit and flies off to infinity. The successive photon rings around the black hole, which correspond to photons travelling on an increasing number of orbits before escaping the black hole environment, then also decrease in width exponentially according to the Lyapunov exponent \cite{Johnson_2020}:
\begin{equation}\label{eq: exp decrease ring widths}
    \frac{\delta R_{n+1}}{\delta R_n} \approx e^{-\gamma} \quad \text{for } n \gg 1 \,.
\end{equation}

Just like for the critical curve, the Lyapunov exponent can be calculated analytically for metrics that admit separable null geodesics. We review this calculation in Section \ref{sec: analytic Lyap} and apply it to the Johannsen and Rasheed-Larsen metrics; in Section \ref{sec:lyapunovresults} we discuss the resulting Lyapunov exponents for these black holes. Note that since integrability of the null geodesic equations is necessary to calculate the analytic Lyapunov exponent, the formalism does not apply to the Manko-Novikov metric.

In Section \ref{sec: numerical estimates} we briefly show how this analytic Lyapunov exponent can be extracted (approximately) from ray-traced images, deferring details to Appendix \ref{app: num Lyap}. The work in this section is an extension of the results in the Master's thesis \cite{thesis}.

\subsection{Analytic formula for the Lyapunov exponent}\label{sec: analytic Lyap}

In this section we review the derivation of the analytic expression for the Lyapunov exponent (\ref{eq: Lyap general}), following the method of \cite{Johnson_2020}.

When null geodesics are separable, we can define a radial potential $\R(r)$ and angular potential $\Theta(\theta)$ as discussed in Section \ref{sec: analytic shadow}.
We can then write
\begin{equation}
    \pm_r \frac{1}{\sqrt{\R(r)}}\frac{\dd r}{\dd\sigma}=  \pm_\theta \frac{1}{\sqrt{\Theta(\theta)}}\frac{\dd\theta}{\dd\sigma}\,.
\end{equation}
The signs on both sides depend on the photon trajectory, and can change along the trajectory. We can then integrate this expression over the geodesic path from $(t_i, r_i, \theta_i, \phi_i)$ to $(t_f, r_f, \theta_f, \phi_f)$,
\begin{equation}\label{eq: relate R and Theta}
    \fint_{r_i}^{r_f} \pm_r \frac{1}{\sqrt{\R(r)}}\dd r= \fint_{\theta_i}^{\theta_f} \pm_\theta \frac{1}{\sqrt{\Theta(\theta)}}\dd\theta\,,
\end{equation}
where the slash denotes integration along the geodesic (meaning that the signs can change along the integration).

A photon on a bound geodesic at constant radius $r_B$ satisfies (\ref{eq: cond bound geod}). The motion in the $\theta$-direction is constrained between turning points $\theta_\pm$, i.e.\ the zeroes of the angular potential. We say that the photon on the bound geodesic has completed \emph{one orbit} when it makes one full oscillation in the $\theta$-direction.

For a photon on a nearly-bound geodesic at a radius $r_B + \delta r_0$, after $n$ half-orbits, the angular integral will be equal to $nG_\theta$ where $G_\theta$ is the angular integral over one single half-orbit,
\begin{equation}\label{eq: def Gtheta}
    G_\theta = \int_{\theta_-}^{\theta_+} \frac{1}{\sqrt{\Theta(\theta)}}\dd\theta\,.
\end{equation}
Meanwhile, the photon has advanced to the radius $r_B + \delta r_n$. We can approximate the radial potential of this nearly-bound photon as
\begin{equation}\label{eq:Rapprox}
    \R(r) \approx \frac12\R_B''(r_B)\left(r- r_B\right)^2\,,
\end{equation}
where $\R_B''(r_B)$ is the second derivative of the radial potential evaluated on the radius of the bound orbit.
This allows us to approximate the radial integral as
\begin{equation}
    \int_{r_B+\delta r_0}^{r_B + \delta r_n} \frac{\dd r}{\sqrt{\R(r)}} \approx \sqrt{\frac{2}{\R_B''(r_B)}}\ln \left(\frac{\delta r_n}{\delta r_0}\right)\,. 
\end{equation}
Combining the radial and angular parts gives us
\begin{equation}\label{eq: Obtain Lyap after approx}
    \frac{1}{n}\ln \left(\frac{\delta r_n}{\delta r_0}\right) =  \sqrt{\frac{\R_B''(r_B)}{2}}\ G_\theta(r_B)\,.
\end{equation}
Comparing this with (\ref{eq: exp growth of delta r}), we conclude that the Lyapunov exponent is given by
\begin{equation}\label{eq: Lyap general}
    \boxed{
    \gamma(r) = \sqrt{\frac{\R''(r)}{2}} G_\theta(r)\,.}
\end{equation}

We dropped the subscript $B$ here, but it should be noted that the constants of motion $\lambda,\chi$ in $\R$ are implicitly determined by the condition (\ref{eq: cond bound geod}).

Just like the black hole critical curve, the Lyapunov exponent is obtained from the conserved quantities associated with bound null geodesics, and therefore is imprinted with the spacetime metric properties. This implis the Lyapunov exponent is also parametrized by the radial coordinate in the interval where bound orbits exist. We can therefore associate a Lyapunov exponent value for every point on the critical curve. We introduce polar coordinates on the screen:
\begin{align}\label{eq: polar coord screen}
    R = r_0^{-1} \sqrt{\alpha^2 + \beta^2}\,,\qquad \cos \phi_R = \frac{\alpha / r_0}{R}\,,
\end{align}
where $\alpha, \beta$ are the impact parameters defined in (\ref{eq: alpha beta}). The Lyapunov exponent can then be parametrized as a function of the viewscreen polar angle $\phi_R$, i.e.\ $\gamma(\phi_R)$, making the map between values of the Lyapunov exponent values and the critical curve explicit. The Lyapunov exponent $\gamma(\phi_R)$ for the Kerr black hole, for various values of $a/M$, is shown in Fig.~\ref{fig: Ly Kerr}.

\subsubsection{Lyapunov exponent as a function of inclination angle}\label{sec:lyap-inclangle}

\begin{figure}[htpb]
    \centering
    \includegraphics[width = 0.7\textwidth]{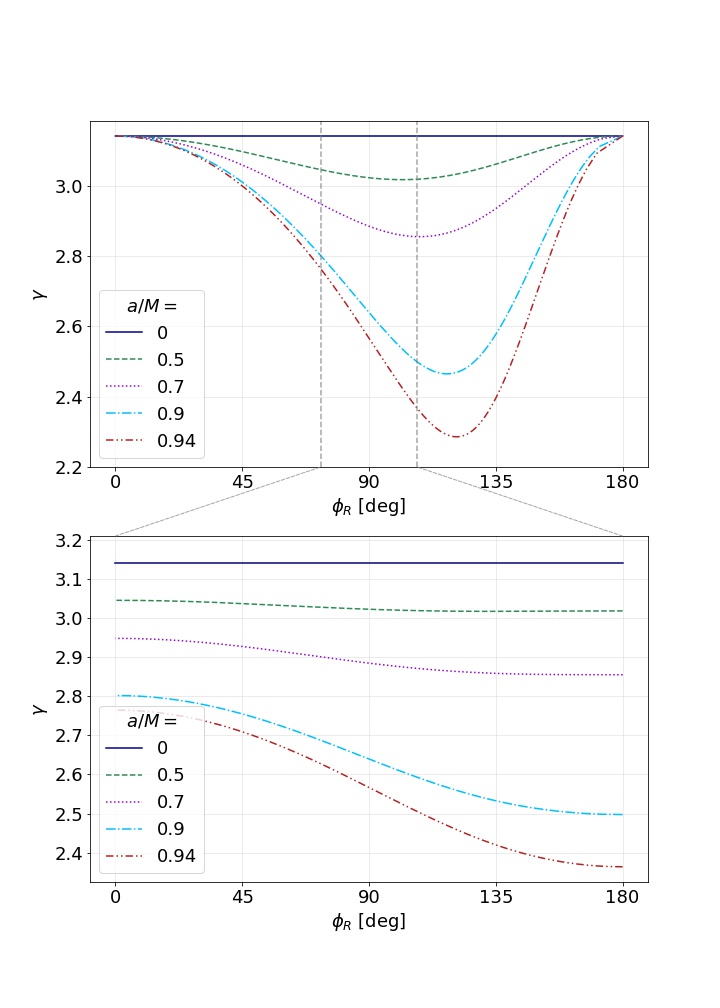}
    \caption{Influence of the spin $a$ on the Lyapunov exponent of a Kerr black hole critical curve. The figure also shows how the curve measured by an observer at $\theta_0 = 17^\circ$ (bottom) is contained within the curve measured by an equatorial observer (top).}
    \label{fig: Ly Kerr}
\end{figure}

The relation (\ref{eq: Lyap general}) does not depend on the coordinates of the observer. This makes sense, since $\gamma$ is a property of the bound null geodesics in the photon sphere which is independent of any observer. On the other hand, the range of $r$-values that contributes to the critical curve on the observer screen is dependent on the observer's inclination $\theta_0$ (through the condition that $\Theta(\theta_0)\geq 0$). This range of $r$ is `smeared out' over the black hole critical curve; every point on the critical curve corresponds to a different bound orbit radius. The same bound orbit radius determines the Lyapunov exponent at that point on the critical curve. Therefore, the information in the function $\gamma(\phi_R)$ for an off-equatorial observer will always be `contained' within the function $\gamma(\phi_R)$ as determined by an equatorial observer.\footnote{Actually, this is only true for spacetimes that are symmetric with respect to $\theta = \frac{\pi}{2}$. An observer at this inclination observes the 'maximal range' of radial coordinates. For spacetimes that do not have this symmetry, there is a different inclination angle $\theta^*$ that maximizes the range of $r$ over which the angular potential is positive for bound orbits. For the cases we studied, this angle turned out to deviate only mildly from $\theta = \frac{\pi}{2}$, much like the analysis in Appendix \ref{app: rings asymm spacetime}.} This is shown in Figure \ref{fig: Ly Kerr} by means of the dashed lines, which indicate which part of the function $\gamma(\phi_R)$ as determined by the equatorial observer is seen by an observer at an inclination $\theta_0 = 17^\circ$.

The position of these boundaries can be determined analytically; we will illustrate this for the Kerr metric. Consider two observers at a distance $r_0$, one in the equatorial plane and the other one at an inclination $\theta^*$. For the inclined observer, the geodesics that hit the horizontal axis on the observer screen at $\beta = 0$ obey the relation
\begin{equation}\label{eq: lambda edge obs}
    (\lambda^*)^2 = \eta^* \tan^2\theta^* + a^2 \sin^2\theta^*\,.
\end{equation}
Since $\lambda^*$ is a quantity associated with the geodesic itself and not with the observer, we can use this expression to determine where the equatorial observer registers this geodesic on their screen. Using the expressions (\ref{eq: polar coord screen}) with $\theta_0 = \pi /2$, we find that a geodesic subject to (\ref{eq: lambda edge obs}) hits the screen of the \emph{equatorial observer} at
\begin{align*}
    R^* & = r_0^{-1}\sqrt{a^2(1+\sin^2 \theta^*)+\eta^* \sec^2 \theta^*}\,, \\
    \phi_R^* & = \arccos\left( - \frac{\lambda^*}{R^*\, r_0 \sin\theta^*}\right)\,.
\end{align*}
We can then rewrite the second relation as
\begin{equation}\label{eq: phi bound Kerr}
    \sin\left(\frac{\pi}{2} - \phi^*_R\right) = \pm \sin(\theta^*) \sqrt{1-\frac{a^2 \sin^2 \theta^*}{a^2 (1+\sin^2 \theta^*)+\eta^* \sec^2 \theta^*}}\,.
\end{equation}
If we assume that the second term in the square root is small, we can approximate
$$ \sin\left(\frac{\pi}{2} - \phi^*_R\right) \approx \pm \sin(\theta^*) \,, $$
with the result that
\begin{equation}\label{eq: approx phi bound Kerr}
    \phi^*_R \approx  \left(\frac{\pi}{2}\pm \theta^* \right) \text{ mod } \pi,
\end{equation}
so an observer at inclination $\theta^*$ sees only a fraction $\sim 2\theta^*/\pi$ of the function $\gamma(\phi_R)$ that an equatorial observer sees.
This approximation (\ref{eq: approx phi bound Kerr}) is better\footnote{We have explicitly checked this approximation for a range of parameters. The largest deviation we have encountered between the exact value for $\phi_R^*$ and the approximation (\ref{eq: approx phi bound Kerr}) was on the order of 2$^\circ$.} for values of the spin close to zero or observers close to the poles, where it becomes exact. In the example of Figure \ref{fig: Ly Kerr}, this approximation tells us that the observer at $\theta_0 = 17^\circ$ observes the part of the plot for the equatorial observer that is contained between approximately 73$^\circ$ and $107^\circ$.

\subsection{Results}\label{sec:lyapunovresults}
The discussion of Section \ref{sec: analytic Lyap} applies for any stationary, axisymmetric metric with separable null geodesics. We have applied this formalism to the Johannsen and Rasheed-Larsen black holes; details are presented in Appendix \ref{app: aux calc Joh}. For the Johannsen metric, the integral $G_\theta$ can be analytically calculated in terms of an elliptic function. In the case of the Rasheed-Larsen metric, the angular integral has to be calculated numerically, as well as its turning points $\theta_\pm$.

\begin{figure}
    \centering
    \includegraphics[width = 0.6\textwidth]{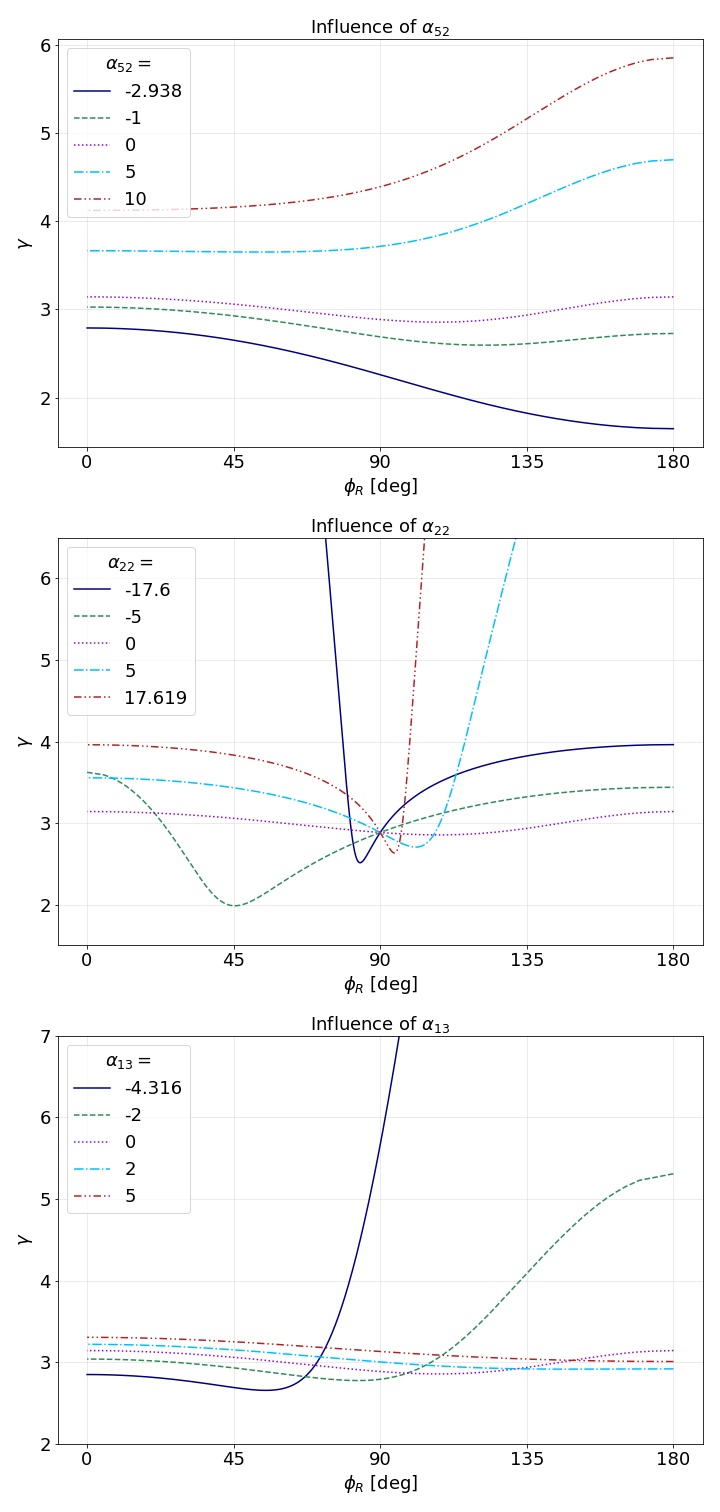}
    \caption{Influence of changing the different parameters of the Johannsen metric on the Lyapunov exponent, as measured by an equatorial observer. The spin parameter is $a/M=0.7$ and only the deviation parameter mentioned in each figure is non-zero. Note that $\alpha_{ij}=0$ corresponds to Kerr in each figure.}
    \label{fig: Ly Joh}
\end{figure}

\begin{figure}
    \centering
    \includegraphics[width = \textwidth]{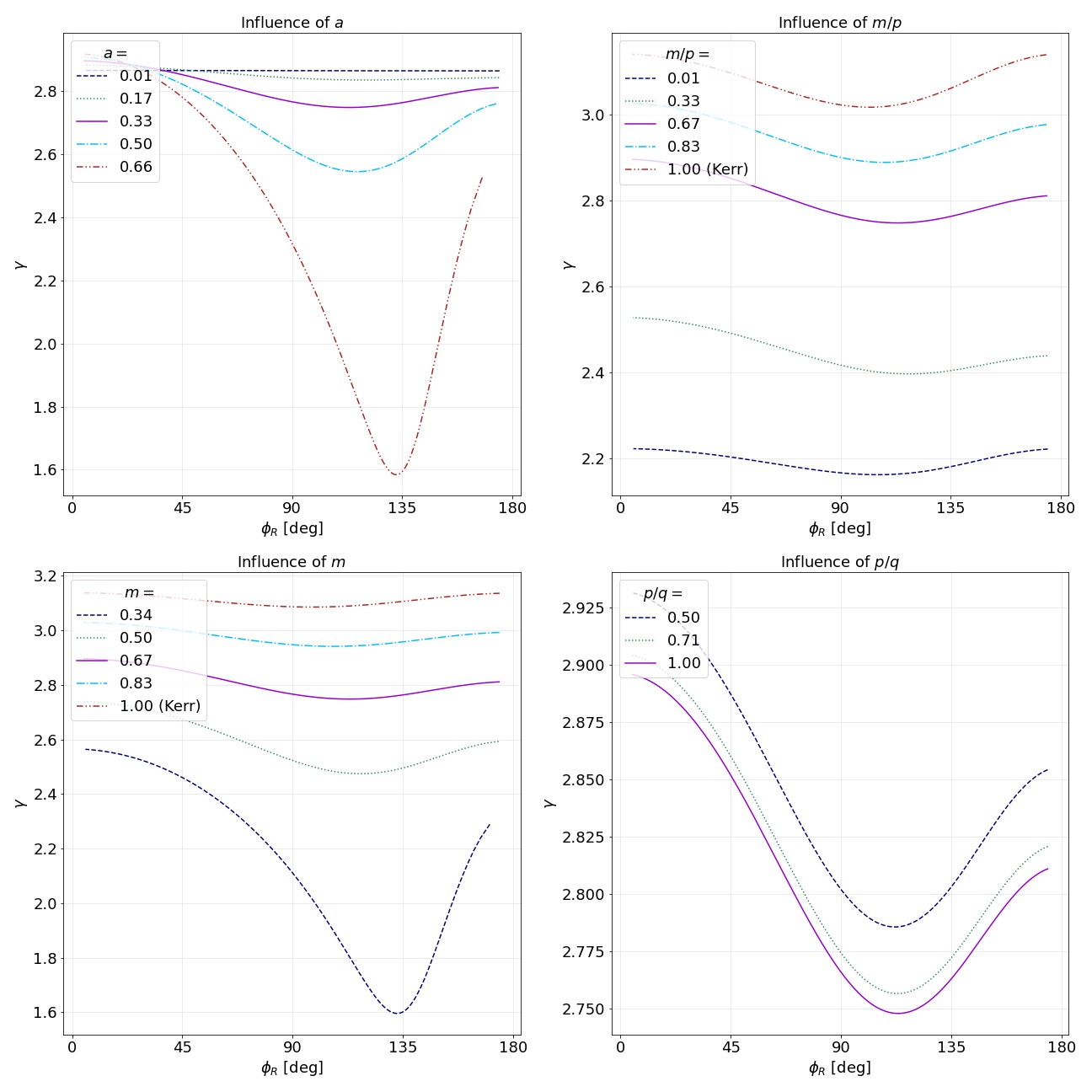}
    \caption{Influence of changing the different parameters of the Rasheed-Larsen metric on the Lyapunov exponent, as measured by an equatorial observer. The standard parameters are $a/M=1/3, m/M = 2/3, p/M=q/M=2$. In the two left panels we change $a$ and $m$, and in the top right panel we change $m/p$ while keeping $a/m$ and $p$ fixed. The bottom right panel shows the influence of changing $p/q$, keeping $a,m$ fixed. The two parameter values that correspond to Kerr black holes are indicated.}
    \label{fig: Ly RL}
\end{figure}

We plot the Lyapunov exponent $\gamma(\phi_R)$ for the Johannsen and Rasheed-Larsen black holes in Figs.~\ref{fig: Ly Joh} and \ref{fig: Ly RL}, as measured by an observer in the equatorial plane. For the Johannsen black hole (Fig.~\ref{fig: Ly Joh}), varying the spin parameter $a/M$ has an analogous effect as for Kerr (see Fig.~\ref{fig: Ly Kerr}). It is interesting to see that the parameter $\alpha_{52}$ does influence $\gamma$, and therefore the photon rings, even though the critical curve is not  affected by this parameter. Note that some of the curves in the plots for $\alpha_{22}$ and $\alpha_{13}$ are seen to diverge rapidly; this happens when the values for these parameters are close to the bounds given in (\ref{eq: single bounds a13 a22}).

For the Rasheed-Larsen metric (Fig.~\ref{fig: Ly RL}), the parameters $a$ and $m$ have a fairly complementary effect (see upper-left and lower-left panels): if the ratio $a/m$ approaches 1, the curve develops a strong dip, and gives lower values overall. This is again roughly the same effect as seen in Kerr and Johannsen when varying spin. If we keep the ratio $a/m$ fixed, increasing $m$ with respect to $p$ and $q$ (see upper-right panel) alters the overall curve, where we recover the Kerr limit for the limiting cases $p=q=2m$. The effect of increasing the imbalance between the charges (i.e.\ changing $p/q$, see lower-right panel) is to raise the overall curve while keeping its form roughly the same, and the results are symmetric in $p,q$.

In general, it is clear that the Lyapunov exponent varies strongly with the metric parameters. Measuring this exponent precisely at different polar angles would be a rigid test of the no-hair theorem, and could in principle distinguish Kerr from other black holes. This remains true at relatively low inclinations (e.g.\ $\theta_0=17^\circ$); even though observations at such inclinations only probe a part of the full function $\gamma(\phi_R)$ (see Fig.~\ref{fig: Ly Kerr}), this can still often suffice to distinguish Kerr from other black holes.

\clearpage

\subsection{On measuring the Lyapunov exponent in practice}\label{sec: numerical estimates}

From the discussion above, it is clear that the Lyapunov exponent depends strongly on the metric parameters and can identify deviations from Kerr. However, there are a number of important issues with this analysis which we touch upon here.

First of all, there is the question of measuring the Lyapunov exponent in practice. One possibility in principle is to compare the widths (or intensities) of successive photon rings \cite{Johnson_2020}. However, an accurate measurement of widths of photon subrings (even for $n=2$) currently seems unfeasible in the near future. A more exciting, realistic possibility may come from a recent proposal that the brightness autocorrelations of the photon ring --- which are certainly measurable with (an upgraded version of) EHT --- would encode information on the Lyapunov exponents and other photon trajectory critical exponents \cite{hadar2021photon}.

Besides the question of practical observability, there are also two issues with the analytical Lyapunov exponent:
\begin{enumerate}
    \item The analytic Lyapunov exponent, as calculated and discussed above, can only be determined for metrics that allow for integrable geodesic equations.
    \item The Lyapunov exponent is calculated as an approximation, using the first non-trivial order of the radial potential for geodesics that are only slightly deviating away from a bound orbit (see (\ref{eq:Rapprox})). It is not clear how good this approximation is, and in particular --- even if a hypothetical precise measurement of successive photon ring widths or intensities can be made --- whether (\ref{eq: exp decrease ring widths}) will indeed be satisfied for the Lyapunov exponent calculated in this approximation.
\end{enumerate}

The first question is relevant for e.g.\ the Manko-Novikov black hole we have discussed. It would seem that both the notions of critical curve and of photon ring as closed, connected objects break down for this black hole. Indeed, losing integrability for null geodesics invariably gives rise to chaotic phenomena in the image in general \cite{Cunha:2015yba,Cunha:2016bjh,Bacchini:2021fig}. It would be interesting to investigate how such chaotic ``photon ring-breaking'' features could translate into measurable interferometric signatures that can be looked for in observations (see also \cite{Bacchini:2021fig} for related discussions on chaotic image features and their observability).

\medskip

 The second issue, i.e.\ how well the calculated analytical Lyapunov exponent actually corresponds to the widths of successive photon subrings, seems to be largely unexplored territory --- this is mostly due to the prohibitively high resolutions needed to resolve subrings with $n\gtrsim 2$ in ray-traced images \cite{Cardenas-Avendano:2022csp}, and the uncertainty in the (claim of the in-)dependence of the underlying emission mechanisms and how well they are resolved in GRMHD simulations. 

 We take a first step towards addressing this issue in the analysis of Appendix \ref{app: num Lyap}. We use \texttt{FOORT} \cite{FOORT} to obtain ray-traced images at a high resolution, from which we extract the photon rings that are resolved in width.\footnote{Due to the (approximately) exponential decrease in width of successive subrings, the resolution needed to resolve an additional ring grows rapidly. The images considered here generally resolve 3-4 rings.} We compare the ratios of these widths to give a numerical estimate of the Lyapunov exponent for low-order rings. Our overall conclusion is that the relation (\ref{eq: exp decrease ring widths}) is not yet satisfied for the lowest-order rings, and that numerical noise (i.e.\ uncertainties on the small widths) is quick to dominate the estimate for increasing $n$. Using images that zoom in on the edge of the critical curve allows us to make better estimates (based on $n=4,5$ ring widths) that come close(r) to the theoretical predictions. However, even for these rings, there is still a clear difference between the ``measured'' Lyapunov exponent prediction and the ``theoretical'' value from the calculations of the previous sections.

 \medskip

 Our analysis in this section shows that the Lyapunov exponent is quite sensitive to metric deviations. Precise measurements of this observable could then be a strong differentiator between Kerr and other alternative black holes.
 A thorough study of the Lyapunov exponent and its measurability should also take into account the difference between the calculated analytical value and the value that can be extracted in practice from the particular observation (such as pointed out in this subsection for the observation of successive subring widths). We leave more comprehensive studies of the observability of the Lyapunov exponent to future work.

%%%%%%%%%%%%%%%%%%%%%%%%%%%%%%%%%%
%%%%%%%%%%%%%%%%%%%%%%%%%%%%%%%%%%
\section*{Acknowledgments}
We would like to thank A. C\'ardenas-Avenda\~{n}o, J. Davelaar, S. Gralla, T. Li, A. Lupsasca, H. Paugnat for interesting discussions and comments on an early draft of this work.
D.R.M.\ is supported by FWO Research Project G.0926.17N. 
L.K. acknowledges support from the ESA Prodex experiment arrangement 4000129178 for the LISA gravitational wave observatory Cosmic Vision L3.
F.B.\ acknowledges support from the FED-tWIN programme (profile Prf-2020-004, project ``ENERGY'') issued by BELSPO.
Support for this work was provided by NASA through the NASA Hubble Fellowship grant HST-HF2-51518.001-A awarded by the Space Telescope Science Institute, which is operated by the Association of Universities for Research in Astronomy, Incorporated, under NASA contract NAS5- 26555. Research at the Flatiron Institute is supported by the Simons Foundation.
This work is also partially supported by the KU Leuven C1 grant ZKD1118 C16/16/005. Computing resources were provided and supported by the VSC (Flemish Supercomputer Center), funded by the Research Foundation Flanders (FWO) and the Flemish Government -- department EWI.

%%%%%%%%%%%%%%%%%%%%%
\appendix

%%%% MN BH appendix
\section{Details of the Manko-Novikov Black Hole}
\label{app: MaNo}

The Manko-Novikov metric is originally given by the Weyl-Papapetrou line element in prolate spheroidal coordinates:
\begin{equation}\label{eq:MNWLlineelement}
 \dd s^2 = -f\left(\dd t - \omega \dd \phi\right)^2 + \frac{k^2}{f} e^{2\gamma} \left(x^2 - y^2\right) \left( \frac{\dd x^2}{x^2-1}+ \frac{\dd y^2}{1-y^2}\right) + \frac{k^2}{f} \left(x^2 - 1\right)\left(1-y^2\right) \dd \phi^2\,.
\end{equation}
In this metric, $k$ is a real constant and $f, \gamma, \omega$ are functions that depend only on the coordinates $x,y$. 
We only consider an octupole deformation, of which the strength is characterised by a parameter $\alpha_3$.
The series of auxiliary functions that go into (\ref{eq:MNWLlineelement}) are then given by
\begin{align*}
    f \, & = e^{2\psi} \frac{A}{B}\,, &
    \omega \, & = 2 k  e^{-2\psi} \frac{C}{A} - 4 k \frac{\alpha}{1-\alpha^2}\,, \\
    e^{2\gamma} \, & = e^{2\gamma'} \frac{A}{\left(x^2-1\right)\left(1-\alpha^2\right)^2} \,,&
    \psi \, & = \alpha_3\frac{P_3}{R^4}\,,
    \end{align*}
    and further
    \begin{align*}
    \gamma'\, & = \frac12 \ln \frac{x^2-1}{x^2-y^2} + 2\alpha_3^2\frac{P_4^2 - P_3^2}{R^8} + \alpha_3 \sum_{l=0}^3\left(\frac{x-y+(-1)^{3-l}(x+y)}{R^{l+1}}P_l\right) \,,\\
     & \\
    A \, & = \left(x^2 - 1\right) \left(1 + \tilde a b\right)^2 - \left(1-y^2\right) \left(b-\tilde a\right)^2 \,,\\
    B \, & = \left[x+1+\tilde a b\left(x-1\right)\right]^2 + \left[\left(1+y\right)\tilde a +\left(1-y\right)b\right]^2\,, \\
    C \, & = \left(x^2 - 1\right) \left(1+\tilde a b\right) \left[b-\tilde a-y\left(\tilde a+b\right)\right] + \left(1 - y^2\right)(b-\tilde a) \left[1+\tilde a b + x(1-\tilde a b)\right]\,,\\
     & \\
    \tilde a \, & = -\alpha \exp \left(2 \alpha_3\left[1 - (x-y)\sum_{l=0}^3 \frac{P_l}{R^{l+1}}\right] \right)\,, \\ 
    b \, & = \alpha \exp \left(2 \alpha_3\left[-1 + (x+y)\sum_{l=0}^3 \frac{(-1)^l P_l}{R^{l+1}}\right] \right)\,, \\ 
     & \\
    R \, & = \sqrt{x^2 + y^2 - 1}\,.
\end{align*}
The $P_l$ are the Legendre polynomials evaluated at $xy/R$, i.e.\ $P_l = P_l\left(xy/R\right)$. The expressions above correct two typos\footnote{One is in the expression for $M_2$: \cite{manko1992novikov} writes $\alpha_2$ at the end, which should be $\alpha^2$. The other is in the expression (13) of \cite{manko1992novikov}: the -1 and $(-1)^n$ at the end of the expressions should not be taken into the sum.} in \cite{manko1992novikov}, which were also noticed in  \cite{gair2008observable}. Finally, the parameters $\alpha, k$ are related to the mass and angular momentum $J=Ma$ as
\be k = \sqrt{M^2-a^2}, \qquad \alpha = \frac{k-M}{a}.\ee

The Manko-Novikov metric as described above can be brought to Boyer-Lindquist coordinates by the following transformation \cite{wang2018chaotic}:
\begin{equation}
    x = \frac{r-M}{k}\,, \qquad y = \cos\theta\,.
\end{equation}
The result of this transformation is the line element (\ref{eq: MaNo metric}).

%%%%% additional shadow/lyapunov calculations
\section{Black Hole Critical Curve \& Lyapunov Calculations}\label{app: auxiliary calcs}
In this Appendix, we present more details on the calculations of the analytic critical curves presented in Section \ref{sec: analytic shadow}, and the analytic Lyapunov exponents presented in Section \ref{sec: analytic Lyap}.

\subsection{Johannsen metric} \label{app: aux calc Joh}

The constants of motion determined by the bound geodesic condition (\ref{eq: cond bound geod}) with (\ref{eq: R Joh}) are 
\begin{align}
    \lambda & = \frac{\left[M(a^2-r^2)  + r \Delta\right]A_1 + (a^2+r^2)\Delta A_1'}{a(M-r)A_2 + a\Delta A_2'}  \,,\label{eq: lambda Joh}\\
    \chi & = \frac{\Delta\left[ (a^2+r^2)(A_2A_1' - A_1A_2') + 2rA_1A_2 \right]^2}{\left[(M-r)A_2 + \Delta A_2'\right]^2} \,.\label{eq: chi Joh}
\end{align}

\paragraph{Separability of the geodesic equations}

We start by demonstrating the separability of the geodesic equations. Starting from the Hamilton-Jacobi equation
\begin{equation}\label{eq: Ham Jac}
    -\frac{\partial S}{\partial \tau} = \frac12 g^{\alpha\beta}\frac{\partial S}{\partial x^\alpha}\frac{\partial S}{\partial x^\beta}\,,
\end{equation}
assuming a Hamilton-Jacobi function of the form
\begin{equation}\label{eq: Ham Jac func}
    S = \frac12 \mu^2 \tau - Et + L\phi +S_r(r) + S_\theta(\theta)\,,
\end{equation}
where $\mu$ is the mass of a test-particle on a geodesic. Plugging the metric (\ref{eq: Johannsen metric}) into (\ref{eq: Ham Jac}), we find that
\begin{equation}\label{eq: Ham Jac Joh}
    -\tilde{\Sigma}\mu^2 =  - \frac{1}{\Delta}\left[-A_1(r)\left(r^2+a^2\right)E + aA_2(r) L\right]^2 + \Delta A_5(r) p_r^2 + \frac{1}{\sin^2\theta}\left[L - aE\sin^2\theta\right]^2  + p_\theta^2\,,
\end{equation}
where we identified $\frac{\partial S_r}{\partial r} \equiv p_r\,, \frac{\partial S_\theta}{\partial \theta} \equiv p_\theta$. Using the expression (\ref{eq: dev functions Joh}) for $\tilde{\Sigma}$, we see that this equation can be separated in $r$ and $\theta$, so that we can introduce a constant $k$ to find
\begin{align}
    \Delta A_5(r) p_r^2 & = -\left(r^2+\epsilon_3 \frac{M^3}{r}\right)\mu^2 + \frac{1}{\Delta}\left[A_1(r)\left(r^2+a^2\right)E - a A_2(r)L\right]^2 - k \,, \\
    p_\theta^2 & = k- a^2\mu^2\cos^2\theta - \csc^2\theta\left[L - aE\sin^2\theta\right]^2  \,.
\end{align}
These expressions give the potentials (\ref{eq: Theta Joh}) and (\ref{eq: R Joh}), in which we have used the energy-rescaled constants of motion $\chi \equiv \frac{k}{E}$ and $\lambda \equiv \frac{L}{E}$, and set $\mu=0$ for null geodesics. Alternatively, we also use $\eta = \chi - (\lambda-a)^2$ for the Johannsen metric.

\paragraph{Analytic solution of angular integral}
We follow the method of  \cite{kapec2019particle} to calculate the integral (\ref{eq: G_theta Joh}).

Using the expression (\ref{eq: Theta Joh}) for the angular potential of the Johannsen metric, we make a change of variables $u=\cos^2\theta$ to find
\begin{equation}
    \Theta(u) = \frac{a^2}{1-u}(u_+ - u)(u-u_-)\,,
\end{equation}
where
\begin{equation}\label{eq: upm general}
    u_\pm = \Delta_\theta \pm \sqrt{\Delta_\theta^2 + \frac{\eta}{a^2}} \,, \quad \Delta_\theta  = \frac12 \left(1-\frac{\eta+\lambda^2}{a^2}\right)\,.
\end{equation}
Using (\ref{eq: lambda Joh}) and (\ref{eq: chi Joh}), this becomes an analytic expression for $u_\pm$ in terms of the radial coordinate $r$. When\footnote{We already mentioned that $\eta\geq 0$. However, for $\eta = 0$, the geodesic is bound to the equatorial plane, meaning that the angular integral is poorly defined. The result for the Lyapunov exponent will however be well-defined in the limit $\eta \to 0$.} $\eta > 0$ both roots are real. Furthermore, $u_- < 0 < u_+ \leq 1$. Therefore, we have $\theta_\pm = \cos^{-1}\left(\mp \sqrt{u_+}\right)$. One can now calculate that
\begin{equation}
    G_\theta= \frac{1}{\sqrt{a^2}}\int_0^{u_+} \frac{\dd u}{\sqrt{u\left(u_+ - u\right)\left(u-u_-\right)}}\,,
\end{equation}
and upon making the substitution $u = u_+ t^2$, we find
 \begin{equation}\label{eq: G_theta Joh}
     G_\theta = \frac{2}{\sqrt{-a^2u_-}} K\left(\sqrt{\frac{u_+}{u_-}}\right)\,.
 \end{equation}
 This is given in terms of a complete elliptic integral of the first kind:
 \begin{equation}
    K(k) = \int_0^1 \frac{\dd t}{\sqrt{\left(1-t^2\right)\left(1-k^2t^2\right)}}\,.
\end{equation}
Note that the argument for the elliptic function $K$ in (\ref{eq: G_theta Joh}) is imaginary, as $u_- <0$. However, the corresponding integral, and so the value of $G_\theta$, remains real.

\subsection{Rasheed-Larsen black hole} \label{app: aux calc RL}

\paragraph{Separability of the geodesics equations for null geodesics}

Again, we will show that the Hamilton-Jacobi equation (\ref{eq: Ham Jac}) is separable. However, this time the separability is only true in the case of null geodesics ($\mu = 0$) \cite{Keeler:2012mq}. Assuming (\ref{eq: Ham Jac func}), we get the equation
\begin{equation}
    -\mu^2 \sqrt{H_1 H_2} = E^2\frac{B^2 H_3^2 \csc^2\theta - \Delta H_1 H_2}{\Delta H_3} + 2EL \frac{BH_3 \csc^2\theta}{\Delta}+L^2 \frac{H_3\csc^2\theta}{\Delta} + \Delta p_r^2 + p_\theta^2\,,
\end{equation}
where we identified $\frac{\partial S_r}{\partial r} \equiv p_r\,, \frac{\partial S_\theta}{\partial \theta} \equiv p_\theta$. The goal is to show that this equation is separable in $r,\theta$. The term on the left-hand side depends on $r,\theta$ in a non-separable way, meaning that we need $\mu = 0$, i.e.\ massless particles, if we want separability. The terms on the right-hand side are separable:
\begin{itemize}
    \item The first term on the right-hand side is separable, as we find that
    \begin{equation}
        \frac{B^2 H_3^2 \csc^2\theta - \Delta H_1 H_2}{\Delta H_3} = T_r(r) + T_\theta(\theta) + T\,,
    \end{equation}
    where the auxiliary functions on the right-hand side are given by
    \begin{align}
    T = &\, -2m^2  -\frac32 pq + m(p+q) \,,\\
    4m^2(p+q)^2 \Delta T_r(r) = &\, (2m-p-q)r-r^2 - pq\bigg\{-a^2(4m^2+pq)^2 +\nonumber \\
    &\,m^2 \left((p-2m)^2(q-2m)^2+4(p+q)(4m^2+pq)r\right)\bigg\} \,,\\
    T_\theta(\theta) = & \,\frac{(p-q)\sqrt{(p^2-4m^2)(q^2-4m^2)}}{2m(p+q)}a\cos\theta - a^2\cos^2\theta\,.
    \end{align}
    \item The second term is easily seen to be separable, as the $\theta$-dependence of $B_3$ (see (\ref{eq: B RL})) gets canceled by the $\csc^2\theta$. We denote this term as
    \begin{equation}
        f(r) = \frac{B_3 \csc^2\theta}{\Delta}\,.
    \end{equation}
    \item The third term is also separable, as we can calculate that
    $$H_3 = \Delta - a^2\sin^2\theta\,,$$
    such that
    \begin{equation}
        \frac{H_3\csc^2\theta}{\Delta} = \csc^2\theta - \frac{a^2}{\Delta}\,.
    \end{equation}
\end{itemize}
The conclusion is that the Hamilton-Jacobi equation is indeed separable for null geodesics, meaning that we can introduce a constant $k$ such that
\begin{align}
    p^2_\theta & = k - L^2\csc^2\theta - E^2 T_\theta(\theta) \,, \\
    \Delta p_r^2 & = -(k+T) - E^2 T_r(r) - 2ELf(r) + \frac{a^2}{\Delta}L^2\,.
\end{align}
These expressions define the potentials (\ref{eq: Theta RL}) and (\ref{eq: R RL}), in which we have used the energy-rescaled constants of motion $\chi \equiv \frac{k}{E}$ and $\lambda \equiv \frac{L}{E}$. Alternatively, we also use $\eta = \chi - \lambda^2$.

%%%%%%%%%%%%%%%%%%%%%%
%%%%%%%%%%%%%%%%%%%%%%
%%%%%%%%%%%%%%%%%%%%%%
\section{Ray Tracing with \texttt{FOORT}}\label{app:FOORT}
In this paper, we use the Flexible Object-Oriented Ray Tracer (\texttt{FOORT}) developed by two of the authors \cite{FOORT}. Whereas most ray tracers are written with specifically the Kerr metric in mind, \texttt{FOORT} was constructed in first place to provide a flexible but fast framework for ray tracing geodesics in arbitrary spacetimes. The flexibility of \texttt{FOORT} allows us to implement the various metrics in this paper as well as easily adjust the geodesic integration to keep track of various diagnostical quantities along the geodesic trajectory --- such as the number of passes through the equator that the geodesics make.

The specific implementation details of \texttt{FOORT} will be presented elsewhere \cite{FOORT}; we present here only the features that are most important for our analysis. The ray tracing camera is placed at $r=1000M$, where $M$ is the mass of the black hole being considered, and $r$ is the Boyer-Lindquist-type coordinate (i.e.\ the radial coordinate which is presented for each metric in this paper). The null geodesics are then traced backwards from the camera position towards the object, following them until they either disappear into a horizon or escape outside the celestial sphere at $r=1000M$. We keep track of the quadrant of the celestial sphere in which the geodesic escapes, giving rise to four-color screen pictures such as Fig.~\ref{fig:Kerrexample}. We also keep track of the number of passes the geodesic makes through the equator, which tells us which order $n$ photon ring the geodesic belongs to (see below). We use \texttt{FOORT}'s adaptive mesh for deciding which pixels to ray trace, which allows us to achieve effective resolutions up to $19137^2$ pixels\footnote{We use an initial square equally spaced grid of $300^2$ pixels. \texttt{FOORT} then iteratively decides which ``squares'' of four pixels to subdivide in half (i.e.\ into nine pixels for each subdivided square). With a maximum subdivision level of 7, the resulting row and column size is $((300-1)\times 6^2)+1=19137$, where the $\pm 1$ takes into account the edge pixels (and the initial grid has subdivision level 1) \cite{FOORT}.}
on a $(15M)^2$-size viewscreen by only integrating ca.\ 1M pixels. This resolution is more than sufficient for our purposes; for example, the resulting (outer) $n=2$ photon ring that is extracted contains approximately a factor of 3-5 more than the number of pixels used in our analysis (see below in Appendix \ref{app:ringcurvetod}). 
Integrating all pixels on the viewscreen typically only takes $\mathcal{O}(10)$ minutes per image on an average personal laptop due to the adaptive mesh and an efficient C++ implementation that takes advantage of OpenMP parallelism to speed up the calculations.

\paragraph{Identifying photon rings}
The $n$-th photon ring is usually defined as the collection of photons that have completed $n$ half-orbits around the black hole before escaping. More precisely, one defines the $n$-th lensing band as the collection of photons that have passed through the equatorial plane ($\theta = \pi/2$) at least $n+1$ times (before either escaping to infinity or crossing the horizon) \cite{Gralla:2020srx}. The $n$-th photon ring then lies entirely inside the $n$-th lensing band; in particular, for a thin disc of isotropic emission on the entire equatorial plane, the $n$-th photon ring will precisely fill the $n$-th lensing band. 

One can distinguish (at least in principle, given high enough resolution) the $n=0$, $n=1$, $n=2$ peaks in brightness within the photon ring. This is intuitively clear: each time a geodesic passes through the equatorial plane, it ``picks up'' an additional photon and thus increases its brightness in the final image.
This peaked brightness profile provides a distinct interferometric signal in the Fourier transform of the (brightness) image, which is what the VLBI experiments such as the EHT actually measure \cite{Gralla:2020nwp,Johnson_2020}. The periodicity of the Fourier-transformed signal at a given angle and at a certain baseline in this Fourier transform precisely gives the \emph{projected diameter} of the $n=2$ photon ring; this is how the projected diameter and thus the shape of the $n=2$ photon ring is argued to be in principle observable with (near-)future VLBI experiments \cite{Gralla:2020srx,Cardenas-Avendano:2022csp,Paugnat:2022qzy}.

With \texttt{FOORT}, we identify the $n$-th lensing bands by keeping track of the number of passes through the equatorial plane that each geodesic takes. As \emph{thickness} of the $n$-th photon ring, we simply take the thickness of this $n$-th lensing band (in pixels). To define the \emph{shape} of the $n$-th photon ring, we consider the pixels on the \emph{outer} bounding curve of the $n$-th lensing band.

%%%%%%%%%%%%%%%%%%%%%%
%%%%%%%%%%%%%%%%%%%%%%
%%%%%%%%%%%%%%%%%%%%%%
\section{Additional information on the ring data generation}\label{app:moreshapemethod}

\subsection{From photon ring curve to projected diameter}\label{app:ringcurvetod}
Since the $n=2$ ring has a finite size in this image, when determining the $n=2$ ring shape, we take the outermost boundary of the $n=2$ photon ring pixels; note that this choice is arbitrary as the size of the $n=2$ ring is insignificant compared to the length scale of the ring itself \cite{Gralla:2020srx}. We truncate this collection further to about 5,000-8,000 points for manageability. The critical curve, if we consider it, is always computed semi-analytically as discussed in Section \ref{sec: analytic shadow}.

We then extract the projected diameter (\ref{eq:defdphi}) from the photon ring or critical curve as a function of the $\varphi$ angle (\ref{eq:phifromtheta}) in the following way.
Instead of building an interpolating function $(x(\theta),y(\theta))$ so that (\ref{eq:phifromtheta}) can be used, we use a simple linear regression on a number of neighboring points to determine the local tangent to the ring --- the $\varphi$ angle is then (by definition) the angle that the vector normal to this tangent makes with the $x$-axis. The optimal number of neighboring points for this linear regression is about 10 --- i.e.\ roughly 0.1\% of the total pixels in the ring; we find this gives the most optimal balance between smoothness and accuracy. Once the angle $\varphi$ has been determined at every point on the curve, we simply calculate the projected position $f(\varphi)$ using (\ref{eq:deffphi}) and $d(\varphi)$ using (\ref{eq:defdphi}).
This generates a large amount of data points $(\varphi, (d(\varphi))$. We then truncate this collection to the collection of 35 datapoints for $\varphi=5^\circ, \cdots, 175^\circ$, where in practice we take as the datapoint the point in our original collection with $\varphi$ closest to the sought-after value. We exclude $\varphi = 0^\circ$ as it is often the least well-behaved and most prone to extraneous numerical errors --- this is due to the ring pixels being more sparse for the part near the equator.

\subsection{Finding the rings in equatorially asymmetric spacetimes}\label{app: rings asymm spacetime}
The $n$-th photon subring for Kerr is defined as the collection of photons that have passed through the equatorial plane $n$ times (or, equivalently, made $n$ half-complete oscillations in the $\theta$ direction) \cite{Johnson_2020}. To determine the subring that a given pixel in an image of a Kerr black hole belongs to, we can then simply keep track of the number of times the corresponding null geodesic passes through the equatorial plane $\theta = \pi/2$.

The same methodology easily generalizes to any equatorially symmetric metric. However, when the metric is no longer equatorially symmetric (e.g.\ the Rasheed-Larsen or Manko-Novikov black holes), although one can still keep track of the number of equatorial passes, it is not clear this is the correct way to ``count'' the photon subring that the pixel belongs to. Indeed, the idea behind this counting is that a null geodesic belonging to the $n$-th subring picks up $n$ photons from an isotropically emitting optically thin \emph{equatorial} accretion disc. If the metric is no longer equatorially symmetric, the equator is no longer a stable place where the particles in this hypothetical accretion disc can orbit. Instead, this (thin) accretion disc should have a non-trivial profile $\theta(r)$ (expressed in some asymptotic Boyer-Lindquist or spherical coordinates) which only asymptotically satisfies $\lim_{r\rightarrow\infty}\theta(r) = \pi/2$; the behavior closer to the horizon will be non-trivial \cite{Chen:2021ryb}.

\begin{figure}[htbp]\centering
\begin{subfigure}{0.48\textwidth}
\includegraphics[width=\textwidth]{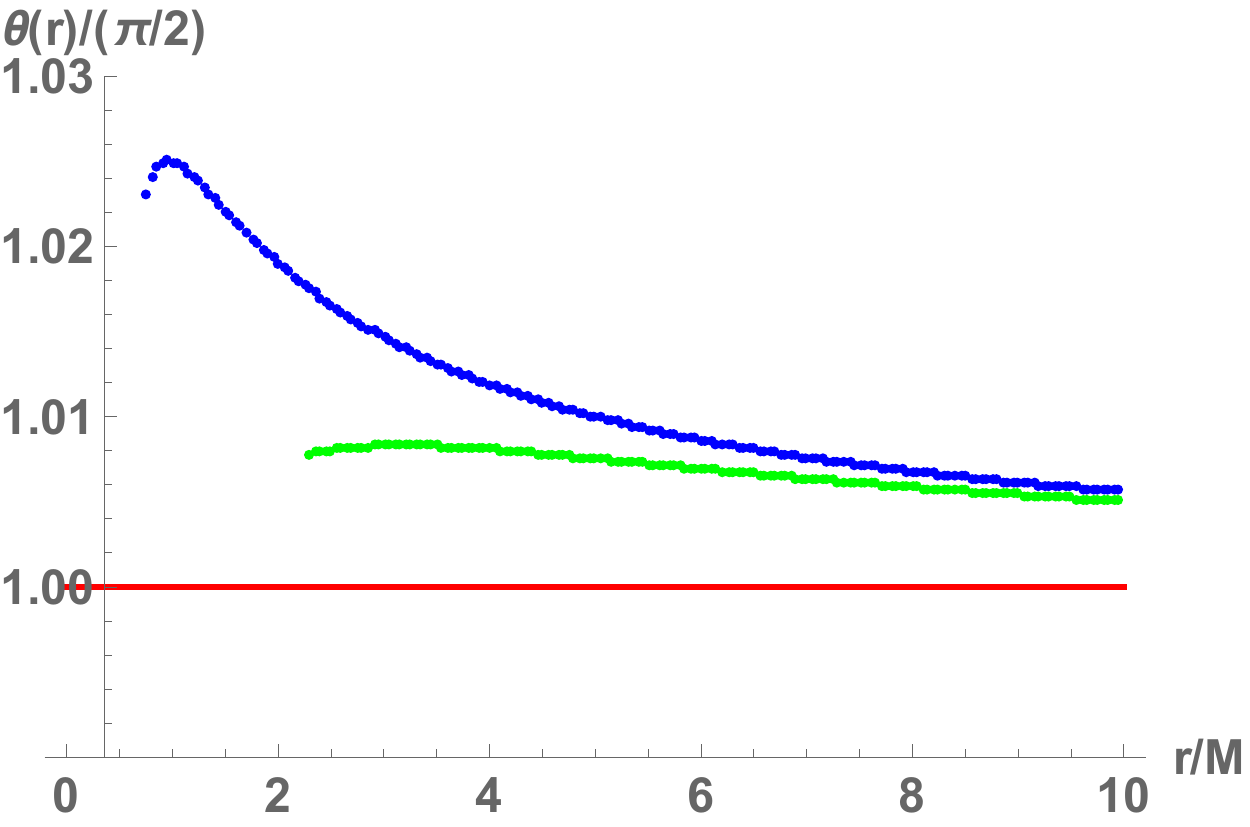}
\caption{Rasheed-Larsen $\theta(r)$ planes}
\end{subfigure}
\begin{subfigure}{0.48\textwidth}
\includegraphics[width=\textwidth]{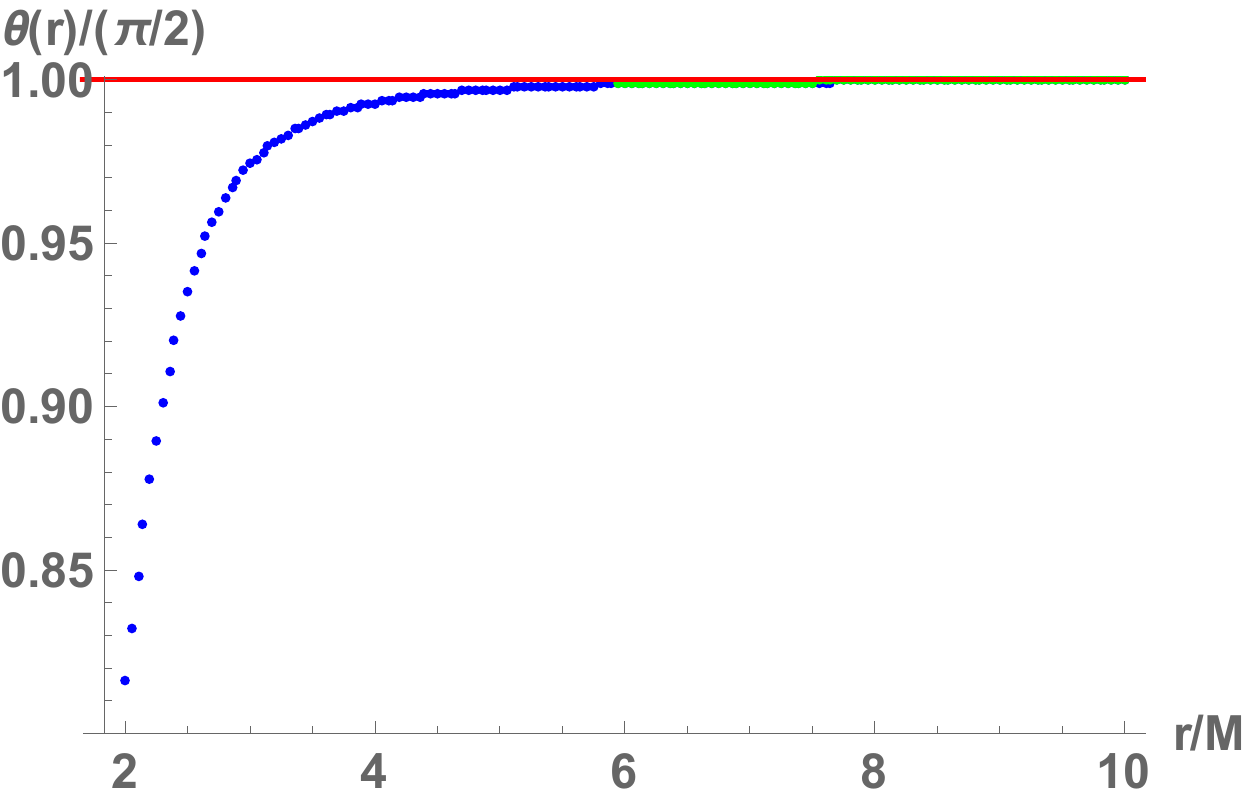}
\caption{Manko-Novikov $\theta(r)$ planes}
\end{subfigure}
\caption{The three possible ``accretion disc planes'' $\theta(r)$ as defined in the text, for the Rasheed-Larsen black hole with parameters (\ref{eq:RLmaxasymparams}) and the Manko-Novikov black hole with $a/M=0.94$ and $\alpha_3=13$. The red line is $\theta=\pi/2$ (definition 1); the blue line is definition 2, and the green line is definition 3. Note that the Rasheed-Larsen horizon is at $r_H\approx 0.55M$, and the Manko-Novikov horizon is at $r_H\approx1.995M$. All lines converge to the asymptotically flat value $\lim_{r\rightarrow\infty}\theta(r) = \pi/2$.}
\label{fig:thetarforRL}
\end{figure}

To model this profile $\theta(r)$ for the equatorially asymmetric metrics used in this paper (Rasheed-Larsen and Manko-Novikov), we can consider three different definitions of the ``accretion disc plane'' $\theta(r)$ through which null geodesics have to pass $n$ times in order to be counted as part of the $n$-th subring:
\begin{enumerate}
\item $\theta(r) = \pi/2$, which assumes that the deviation from equatorial symmetry does not significantly change the equilibrium disc position.
\item $\theta(r)$ is determined at a given $r=r_0$ by demanding that there exists a circular timelike geodesic at this constant $r_0$ and $\theta(r_0)$ that is \emph{co-rotating} with the black hole.
\item $\theta(r)$ is determined at a given $r=r_0$ by demanding that there exists a circular timelike geodesic at this constant $r_0$ and $\theta(r_0)$ that is \emph{contra-rotating} with the black hole.
\end{enumerate}
In the two latter cases, $\theta(r)$ will cease to exist for small enough $r$ close to the horizon. We illustrate these surfaces in Fig.~\ref{fig:thetarforRL} for the Rasheed-Larsen and Manko-Novikov black holes that feature in Section \ref{sec:shape}. 

We confirmed that all three of these $\theta(r)$ ``accretion disc planes'' give equivalent results for the photon rings. To do this in practice, \texttt{FOORT} can be configured to only register an equatorial pass once a certain threshold $\epsilon$ has been passed beyond the equator, i.e.\ $|\theta-\pi/2|>\epsilon$. After changing $\epsilon$ (whose default value is $\epsilon = 0.01$) such that $\epsilon > \max\{|\theta(r)-\pi/2|\}$ (i.e.\ $\epsilon = 0.026$ for Rasheed-Larsen and $\epsilon = 0.187$ for Manko-Novikov), we find that the analysis of Section \ref{sec:shape} in completely unchanged (in particular, Figs.~\ref{fig:RLshapefits} and \ref{fig:MNshapefits} are unchanged), which confirms that the precise location of the accretion disc plane is irrelevant for the photon ring images.

Our conclusion is that extracting the photon rings using equatorial passes remains correct for the equatorial asymmetric black holes and parameters we have considered.

\section{Numerical estimates of the Lyapunov exponent}\label{app: num Lyap}

The relation (\ref{eq: exp decrease ring widths}) is an approximation that is expected to be valid in the large-$n$ limit. In this Appendix, we investigate how well the relative widths of the photon subrings obey this relation when $n$ is small. We extract estimates for the Lyapunov exponent based on ray-traced images of the Johannsen and Rasheed-Larsen black holes, by comparing the relative widths from the photon rings.

Note also that the method presented in Section \ref{sec: analytic Lyap} to calculate the analytic Lyapunov exponent only works for metrics that have integrable geodesic equations. Therefore, the work in this section is also relevant to assess the feasibility of numerically determining the Lyapunov exponent for more complicated metrics that have non-integrable geodesic equations\footnote{Of course, this is based on the assumption that such a Lyapunov exponent even exists, as the relation (\ref{eq: exp growth of delta r}) is also determined theoretically. However, it is in any case still sensible to compare the widths of different photon rings.} like the Manko-Novikov metric (\ref{eq: MaNo metric}).

\begin{figure}[htbp]
    \centering
    \includegraphics[width = 0.49\textwidth]{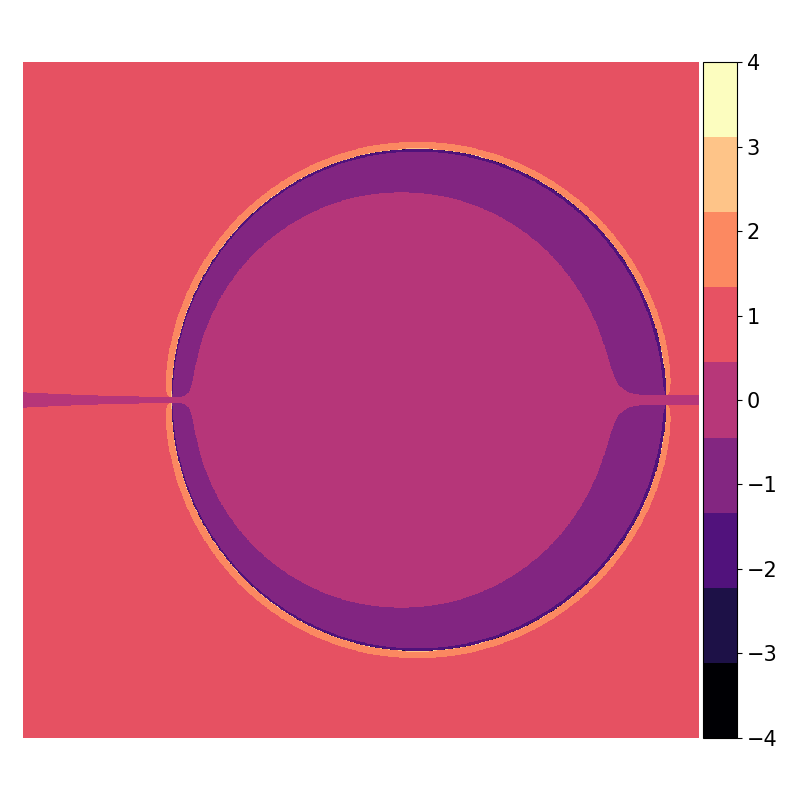} \includegraphics[width = 0.49\textwidth]{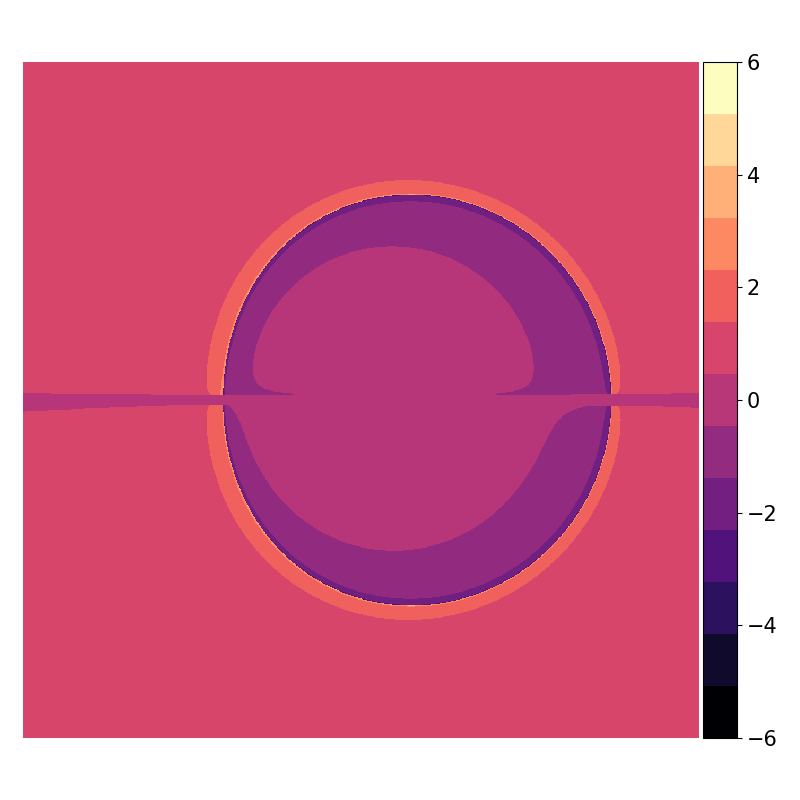}
    \caption{(left) Ray-traced image of a Johannsen black hole with non-zero parameters $a/M = 0.7$, $\alpha_{13} = 2$, as seen by an equatorial observer. (right) Ray-traced image of a Rasheed-Larsen black hole with parameters (\ref{eq:RLmaxasymparams}), as seen by an equatorial observer. The color denotes the number of equatorial passes that a geodesic has experienced; the absolute value of the number of passes, while a negative number indicates the geodesic terminated inside the horizon.}
    \label{fig: ex Eq Passes}
\end{figure}

\begin{figure}[htbp]
    \centering
    \includegraphics[width = \textwidth]{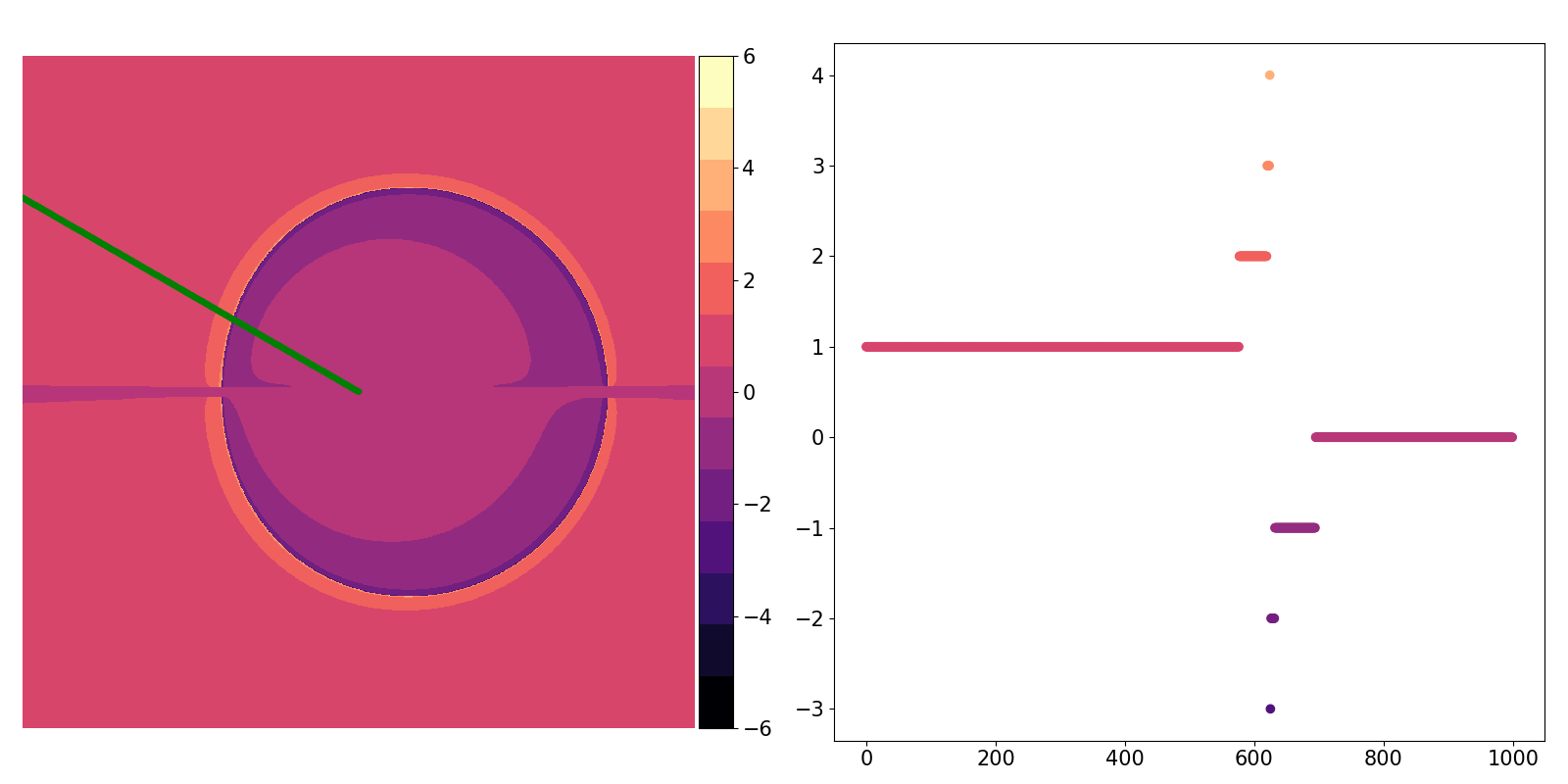}
    \caption{Example of an interpolated line at angle $150^\circ$. The left image corresponds to the Rasheed-Larsen black hole of Fig.~\ref{fig: ex Eq Passes}. (right) Number of equatorial oscillations (vertical) per pixel along the line (horizontal). A low resolution is used to make this plot, but for the analysis the number of pixels on the line is on the order of $\sim$60{,}000.}
    \label{fig: ex line}
\end{figure}

We use \texttt{FOORT} to obtain ray-traced images of the black hole metrics. Fig.~\ref{fig: ex Eq Passes} shows two examples of the number of equatorial passes that a geodesic (characterized by its impact parameters) experiences. From such images, we extract the thickness of each resolvable ring at angles $\phi_R \in [5^\circ, 15^\circ, \cdots, 155^\circ, 165^\circ]$.\footnote{We avoid $\phi_R\approx 0^\circ,180^\circ$ as there are numerical artifacts at those angles due to the viewscreen being placed on the equator.}  The thickness is obtained by considering a straight line from the center of the black hole at the given angle, and interpolating the number of equatorial passes from the neighbouring pixels (using the same resolution for the line as the pixel grid). An example is shown in Fig.~\ref{fig: ex line}.

The Lyapunov exponent can be estimated from the ratios of successive ring widths determined by the method above,
\begin{equation}
    \hat{\gamma}_n = -\ln \frac{w_{n+1}}{w_n}\,.
\end{equation}
It is illustrative to consider the measurement error of this estimate by accounting for the resolution of the image. As explained before, the resolution of pixels on the line is the same as the resolution of the image itself. The width of the $n$-th and $(n+1)$-th ring are determined as $w_n = x_{n}-x_{n-1}$ and $w_{n+1} = x_{n+1}-x_n$, where $x_n$ is the outermost pixel on the line that has performed $n$ equatorial passes. For each $x_n$, the measurement error due to the resolution is $\delta x_n=1$ pixel. The measurement error on the Lyapunov exponent is then
\begin{equation}
\delta \hat{\gamma}_n = \delta x_n \sqrt{\frac{2}{w_n^2} + \frac{2}{w_{n+1}^2}+\frac{2}{w_n w_{n+1}}}\,. 
\end{equation}
Note that, as the error $\delta x_n$ is always equal to 1, extracting estimates from zoomed-in images will decrease the error, as the ring widths (in terms of number of pixels) are larger.

\begin{figure}[htbp]
    \centering
    \includegraphics[width=0.49\textwidth]{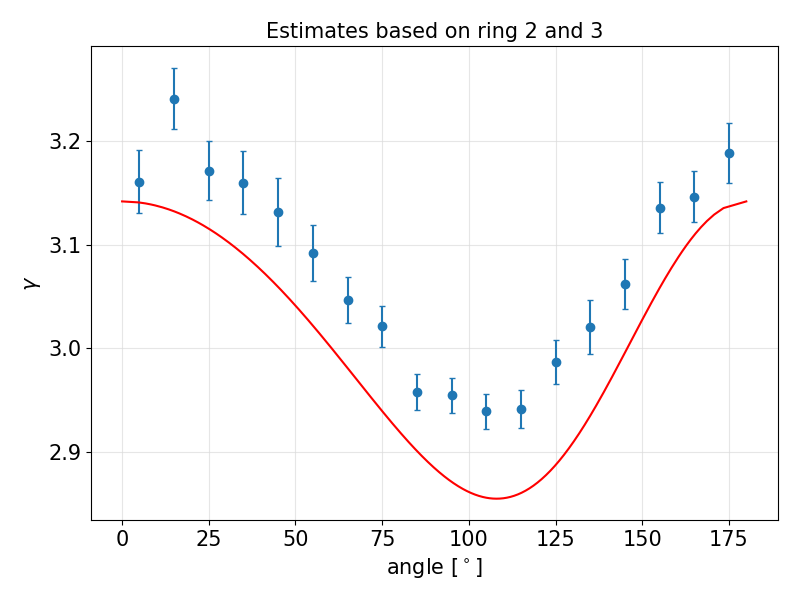}
    \includegraphics[width=0.49\textwidth]{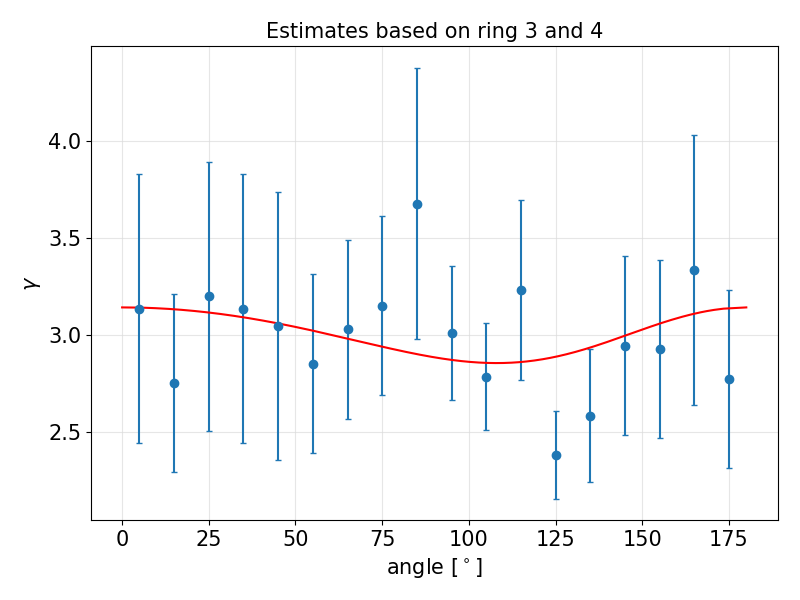}
    \caption{Estimates for the Lyapunov exponent as function of the on-screen angle for the Kerr metric with $a/M=0.7$ as determined by an equatorial observer.}
    \label{fig: Kerr estimates}
\end{figure}

\begin{figure}[htbp]
    \centering
    \includegraphics[width=0.49\textwidth]{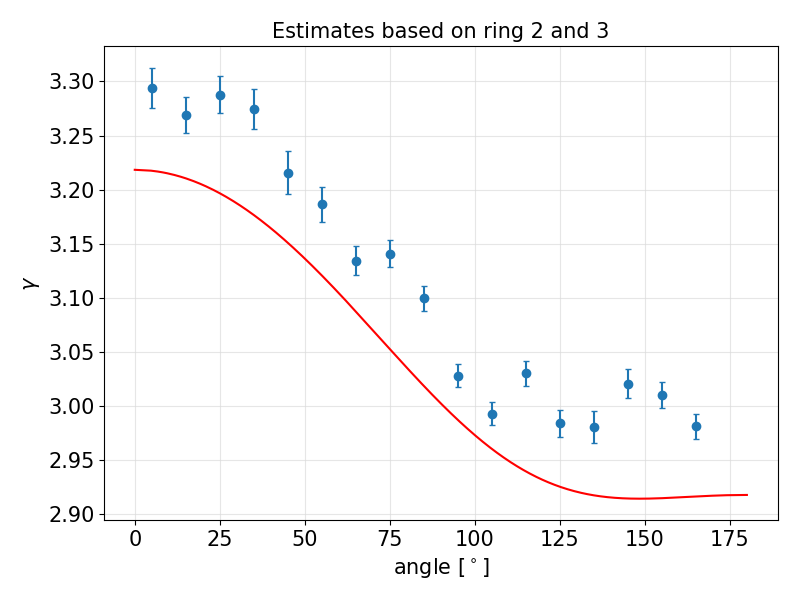}
    \includegraphics[width=0.49\textwidth]{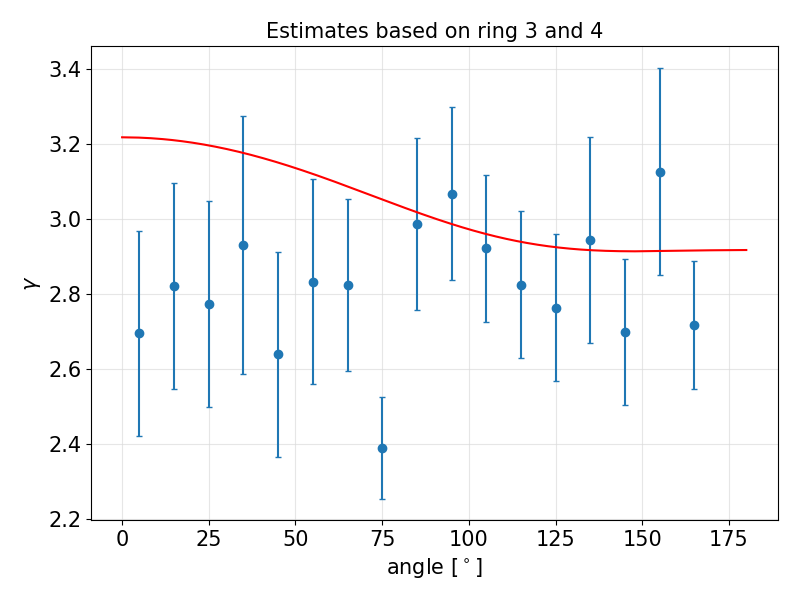}
    \caption{Estimates for the Lyapunov exponent as function of the on-screen angle for the Johannsen metric with non-zero parameters $a/M=0.7, \alpha_{13} = 2$ as determined by an equatorial observer.}
    \label{fig: Joh estimates}
\end{figure}

\begin{figure}[htbp]
    \centering
    \includegraphics[width=0.49\textwidth]{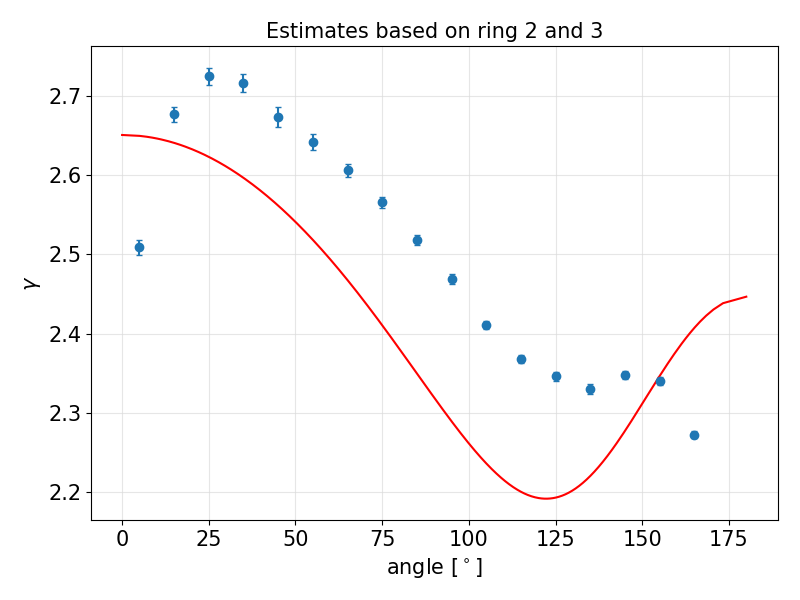}
    \includegraphics[width=0.49\textwidth]{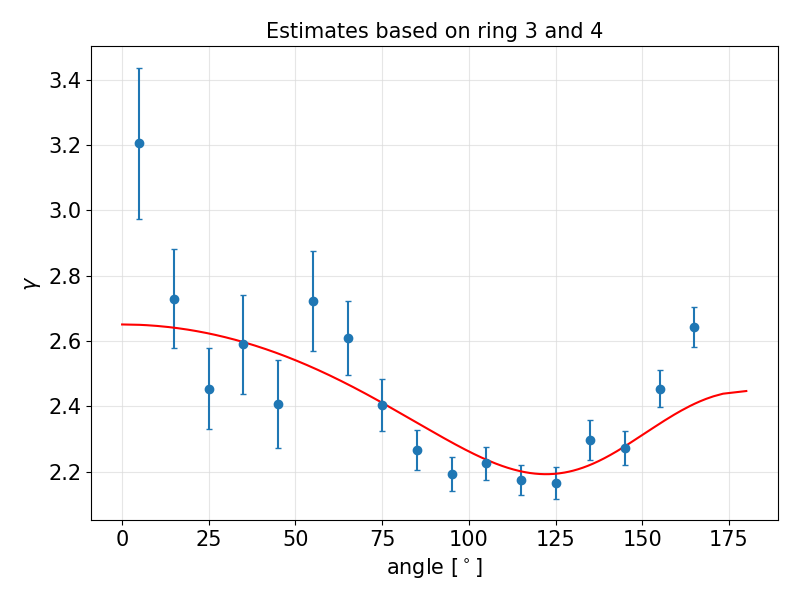}
    \includegraphics[width=0.49\textwidth]{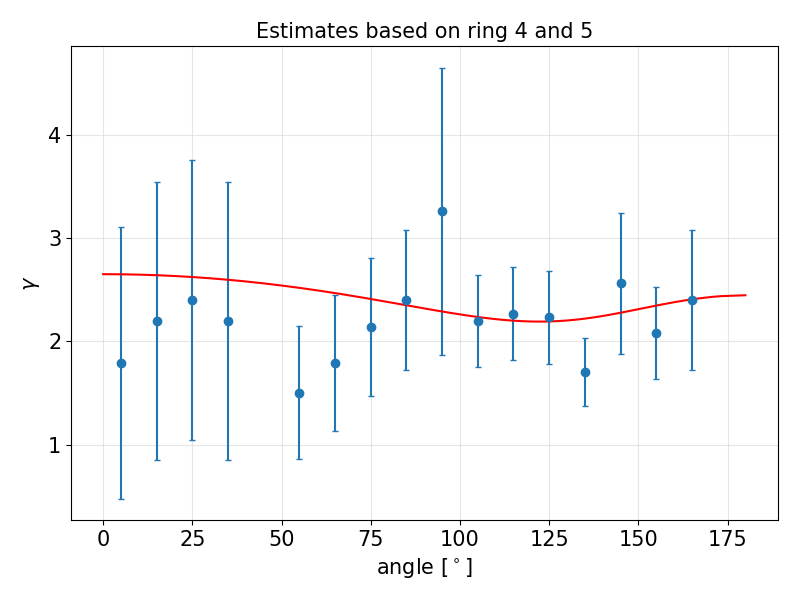}
    \caption{Estimates for the Lyapunov exponent as function of the on-screen angle for the Rasheed-Larsen metric with parameters (\ref{eq:RLmaxasymparams}) as determined by an equatorial observer.}
    \label{fig: RL estimates}
\end{figure}

The estimates for the Lyapunov exponent, based on different rings in the images in Fig.~\ref{fig: ex Eq Passes}, are shown in Figs.~\ref{fig: Kerr estimates}, \ref{fig: Joh estimates}, and \ref{fig: RL estimates}. The error is smaller for estimates based on low-$n$ rings, as they are wider so that the estimate suffers less from the resolution. We note that in both cases, the numerical estimate based on the lowest-order rings overestimates the theoretical Lyapunov exponent, but still generally follows the theoretical Lyapunov function shape. 
As expected, the estimates based on the smaller rings show larger measurement errors and appear to be more scattered. Note that not all of the error-bars include the theoretical prediction. The estimates $\hat{\gamma}_3$ (top right in Fig.~\ref{fig: RL estimates}) does still roughly follow the general functional shape of $\gamma(\phi_R)$. Finally, we notice that generically, estimates close to $\phi_R\approx0^\circ,180^\circ$ are worse; the region around the equator is more sensitive to numerical issues, as can also be seen in Fig.~\ref{fig: ex Eq Passes}.
It is interesting to note that the estimates for the Johannsen and Rasheed-Larsen black holes considered in Figs.~\ref{fig: Joh estimates} and \ref{fig: RL estimates} differ significantly from the estimates for the Kerr black hole. Even though the theoretical values are not accurately recovered, the shape of the curve as determined by the point estimates resembles the one obtained from theory, meaning that these estimates can really distinguish non-Kerr effects. This remains true for observers at e.g.\ lower inclinations $\theta_0=17^\circ$ --- these only have access to a smaller ``window'' of $\gamma(\phi_R)$ (see Fig.~\ref{fig: Ly Kerr}), but this nevertheless contains enough information to distinguish deviations from the Kerr Lyapunov values.

\begin{figure}[htbp]
    \centering
    \includegraphics[width=0.49\textwidth]{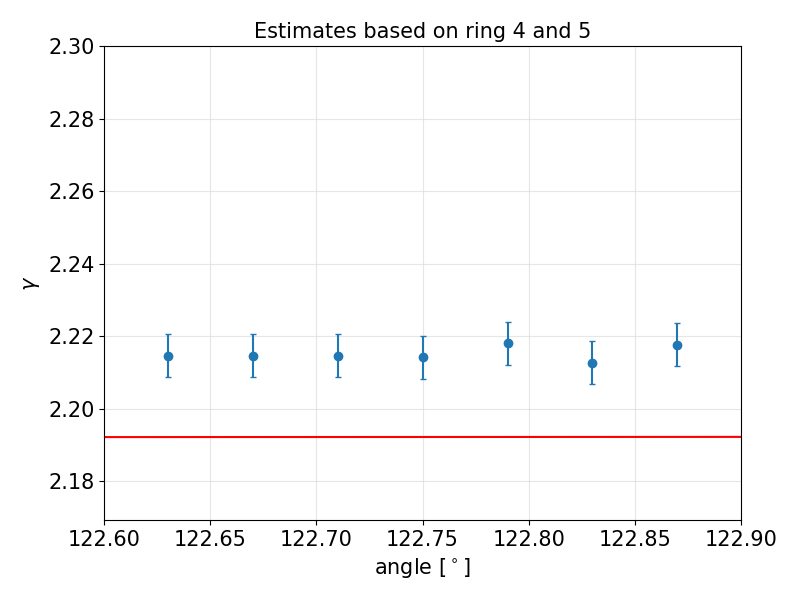}
    \includegraphics[width=0.49\textwidth]{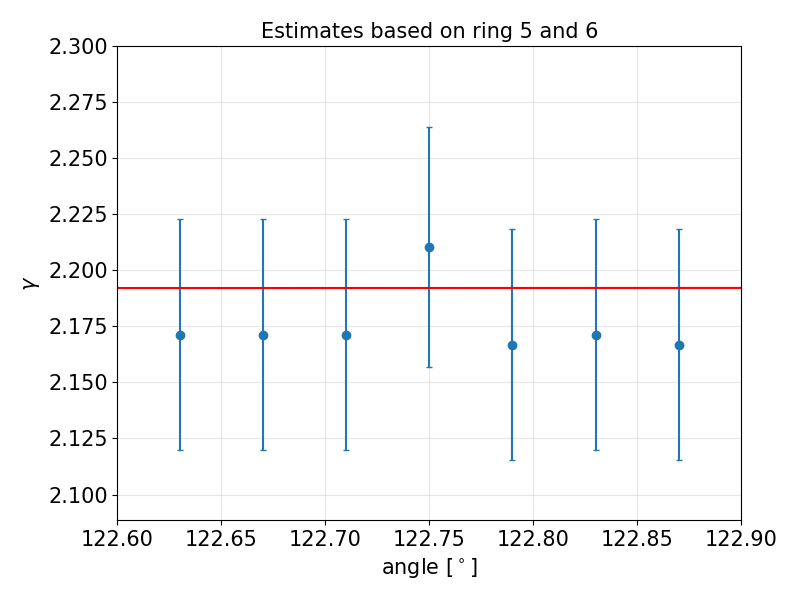}
    \includegraphics[width=0.49\textwidth]{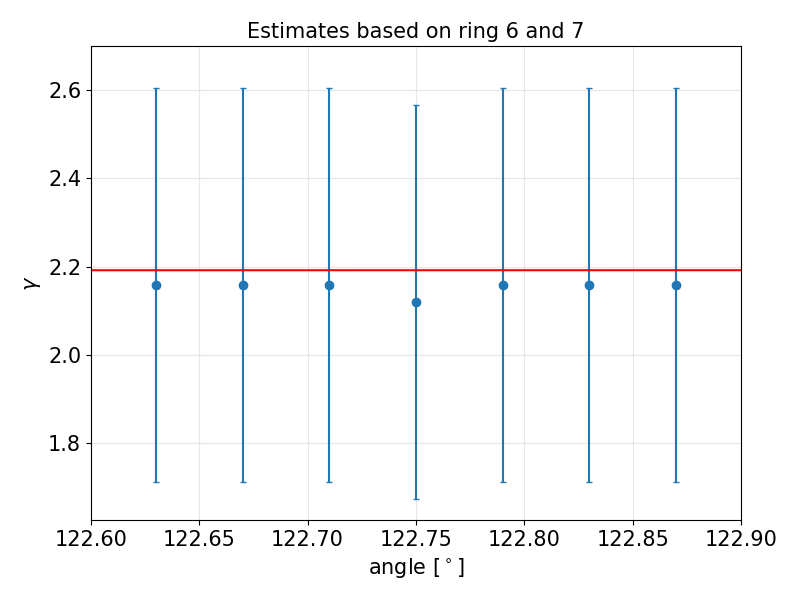}
    \caption{Estimates for the Lyapunov exponent as function of the on-screen angle around $122.75^\circ$ for the Rasheed-Larsen metric with parameters (\ref{eq:RLmaxasymparams}) as determined by an equatorial observer.}
    \label{fig: RL zoom estimates}
\end{figure}

Given that the rings decrease exponentially in width, an extremely high resolution is needed to see high-$n$ rings. To resolve these rings, we use images that zoom in on parts of the critical curve edge at a given angle. Fig.~\ref{fig: RL zoom estimates} shows estimates based on a zoomed-in image (by a factor 600) around the critical curve edge at the fiducial angle $\phi_R=122.75^\circ$. This image resolves rings up to $n\in\{4,5,6,7\}$. The lowest-order rings in the image now give consistent estimates around this angle that differ only by about 0.02 from the theoretical value; the higher-order ring widths approach the theoretical value better. (However, note that the error bars still do not encompass the analytic value.) Note also that the estimates based on the subsequent rings stay fairly consistent and are mostly limited by the resolution.

%%%%%%%%%%%%%%%%%%%%%%
%%%%%%%%%%%%%%%%%%%%%%
%%%%%%%%%%%%%%%%%%%%%%
\bibliographystyle{toine}
\bibliography{fuzzballshadows,extra_ref}

\end{document}